\documentclass[tightenlines,aps,showpacs,floatfix,twocolumn,pre,noeprint]{revtex4-1}
\usepackage[dvips]{graphicx}
\graphicspath{{figures/}}

\usepackage[bbgreekl]{mathbbol}
\usepackage{booktabs}
\usepackage{bigstrut}
\usepackage{tabularx}
\usepackage[caption=false,subrefformat=subsimple,labelformat=simple,listofformat=subsimple]{subfig}

\usepackage{amsmath}   
\usepackage{amssymb}
\usepackage{verbatim}
\usepackage[usenames,dvipsnames]{color} 
\usepackage[colorlinks=true,linkcolor=blue]{hyperref}
\usepackage{cleveref} 
\crefname{figure}{Fig.}{Figs.}
\Crefname{figure}{Figure}{Figures}

\crefname{equation}{Eq.}{Eqs.}
\Crefname{equation}{Equation}{Equations}

\crefname{table}{Table}{Tables}
\Crefname{table}{Table}{Tables}

\crefname{section}{Sec.}{Secs.}
\Crefname{section}{Section}{Sections}

\usepackage{multirow}

\usepackage{makecell}
\usepackage{pbox}

\usepackage{bbm}

\usepackage[american]{babel}

\DeclareMathOperator{\Tr}{Tr}

\begin{document}

\newcommand{\br}{\mathbf{r}}
\newcommand{\bbr}{\mathbbm{r}}
\newcommand{\bbbr}{\bar{\bbr}}
\newcommand{\cbbbr}{\check{\bbbr}}

\newcommand{\dex}{\Delta_{\textrm{ext}}}
\newcommand{\din}{\Delta_{\textrm{int}}}
\newcommand{\bba}{\mathbbm{a}}
\newcommand{\bbb}{\mathbbm{b}}
\newcommand{\bbc}{\mathbbm{c}}

\newcommand{\bR}{\mathbf{R}}
\newcommand{\bRt}{\bR_\theta}

\newcommand{\cbba}{\check{\bba}}
\newcommand{\cbbr}{\check{\bbr}}
\newcommand{\ct}{\check{\theta}}

\newcommand{\bbs}{\mathbb{\Sigma}}
\newcommand{\bbbs}{\bar{\bbs}}
\newcommand{\ml}{\mathbb{m}_\ell}
\newcommand{\mxl}{\mathbb{m}_{\xi\ell}}
\renewcommand{\sl}{\Sigma_\ell}
\newcommand{\sxl}{\Sigma_{\xi\ell}}

\newcommand{\cbbs}{\check{\bbs}}
\newcommand{\tbbs}{\tilde{\bbs}}
\newcommand{\cbbbs}{\check{\bbbs}}

\newcommand{\CR}{\frac{c_-}{c_+}}

\newcommand{\nc}{N_\text{core}}
\newcommand{\no}{N_-}
\renewcommand{\ni}{N_+}

\newcommand{\ro}{r_-}
\newcommand{\ri}{r_+}

\newcommand{\nch}{\hat{N}_\text{core}}
\newcommand{\noh}{\hat{N}_-}
\newcommand{\nih}{\hat{N}_+}
\newcommand{\roh}{\hat{r}_-}
\newcommand{\rih}{\hat{r}_+}

\title{Feasibility of rational shape design of single-polymer micelle using spontaneous surface curvature}
\date{\today} 
\author{Brian Moths}

\begin{abstract}
Polymeric micelles are used in a variety of applications, with the micelle's shape often playing an important role.
Consequently, a scheme to design micelles of arbitrary shape is desirable.
In this paper, we consider micelles formed from a single, linear, multiblock copolymer, and we study how easily the micelle's shape can be controlled by altering the copolymer block lengths.
Using a rational design scheme, we identify a few aspects of the multiblock composition that are expected to have a well-behaved, predictable effect on micelle shape.
Starting from a reference micelle composition, itself already exhibiting a nonstandard shape having a moderately sized dimple, we alter these aspects of the multiblock composition and observe the regularity of the micelle shape response. 
The response of the shape is found to be somewhat smooth, but significantly nonlinear and sometimes nonmonotonic, suggesting that sophisticated techniques may be required to aid in micelle design.
\end{abstract}

\maketitle 

\section{Introduction}
\label{sec:introduction}
Micelles are self-organized aggregates occurring in a solvent, and consisting of two chemically incompatible regions: the exterior of the micelle, occupied by solvophilic material which is miscible with the solvent, and the interior of the micelle, containing solvophobic, immiscible material.
The chemical dissimilarity between the micelle interior and exterior allow the micelle to transport material which would normally be immiscible in the solvent.
This is what makes micelles effective in their perhaps best-known role as detergents.
A related application is to use micelles as a drug carrier, with the drug payload residing in the interior of the micelle.
It has been found that a drug carrier's shape affects aspects of the drug carrier's performance such as where in the body (e.g., into which organ) the drug payload is deposited \cite{Decuzzi10,Devarajan09,Muro08,Patil08,Gillies04}.
A second application where micelle shape may be important concerns micelles aggregating together to form higher-order structures of various shapes such as cubes, pyramids, or long chains \cite{Bose2017,dule2015,Li2015,Qiu2015}.
The shape of these aggregates might be controlled through the shape of the constituent micelles.

Because of the importance of micelle shape, it would be useful to have a rational design scheme to create micelles of a precisely tailored shape.
A good rational design scheme would identify a few key control parameters governing how the micelle is synthesized, and these control parameters would have a well-behaved effect on the micelle shape.
Ideally, the effect of the control parameters would be so regular that, given the observed shapes from a small number of control parameter values, the relationship between shape and control parameters could be accurately determined by a naive linear model.

In this work, we characterize the performance of such a scheme wherein the micelle consists of a single, linear, multiblock copolymer (i.e., a polymer containing solvophilic and solvophobic monomers segregated into multiple homogeneous blocks), and the number of these blocks and their lengths, which we collectively refer to as the ``micelle composition", are used as control parameters to set the micelle shape.
We study the design scheme by simulation, which, for simplicity, is performed in two dimensions, a choice we will justify in \cref{sec:discussion}.
In a previous paper~\cite{firstPaper}, we demonstrated that this scheme can indeed be used to produce a micelle of a nonstandard dimpled shape, and we showed, by varying two aspects of the micelle composition, that the micelle shape could be controlled.
In this paper, we go beyond merely demonstrating that it is possible to control the micelle shape: we select several control parameters governing the micelle composition, and we assess the regularity of the micelle's shape dependence on these parameters.
We seek to determine if the micelle's shape dependence can be explained by a straightforward rationale and whether this dependence is simple enough that it may be represented by a naive linear model.

\begin{figure} 
	\centering
	\includegraphics[width=\linewidth]{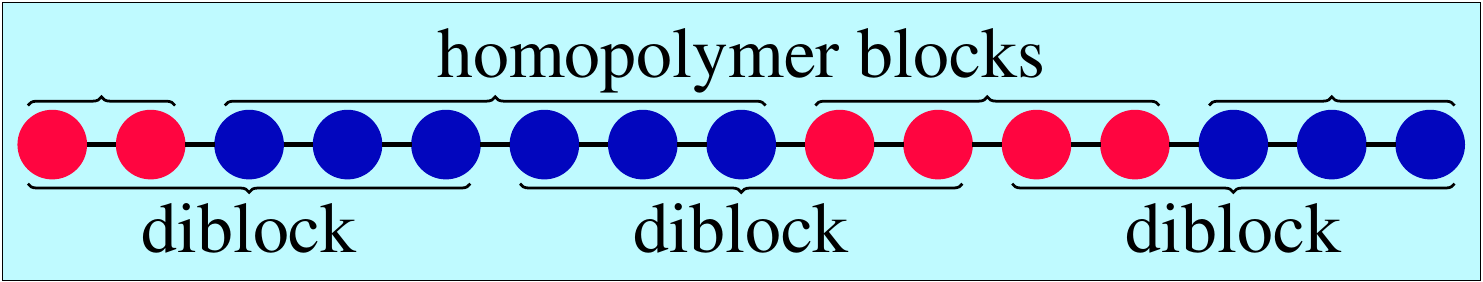}
	\caption{Two different views of a linear multiblock copolymer composed of two species of monomer, shown in red (light gray) and blue (dark gray).
	In the first view, the multiblock is considered a collection of homopolymer segments.
	In the second view, it is considered a sequence of diblocks joined end to end.
    Figure reproduced from \cite{firstPaper}.}
	\label{fig:multiblockDiblock}
\end{figure}

\begin{figure}
	\centering
	\includegraphics[width=.7\linewidth]{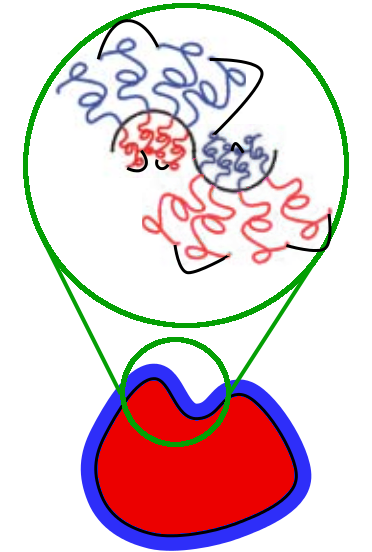}
	\caption{
		Illustration of shape-design rationale.
		The multiblock, viewed as a collection of diblocks, exhibits a configuration where the junction points of the diblock lie on the micelle surface.
		Diblocks of different composition, and therefore different spontaneous curvatures, cause a nonuniform surface curvature, giving the micelle its desired shape.
		The relative positioning of the diblocks is enforced by joining them end to end.
		Figure reproduced from \cite{firstPaper}.
	}
	\label{fig:rationale}
\end{figure}

To better motivate which aspects of the micelle composition we vary in this work, we now give a more detailed description of the rationale underlying our shape-design scheme.
The key idea is to view the multiblock copolymer not as a sequence of homopolymer blocks joined together, but rather as a sequence of diblocks.
Thus two homopolymer blocks are joined at a diblock junction point, and adjacent diblocks are joined to each other in the middle of a homopolymer segment, as illustrated in \cref{fig:multiblockDiblock}.
With this view in mind, we now give an explanation, illustrated in \cref{fig:rationale}, of how the diblocks' block lengths may affect the micelle shape.
In solution, the solvophilic blocks are located at the exterior of the micelle, while the solvophobic blocks compose the micelle interior.
Thus, the diblock junction points occupy the boundary separating the two regions.
It is well-known that such an interface containing diblocks has a spontaneous curvature depending on the diblocks' compositions and chemical properties~\cite{Wang91}.
By this reasoning, we may judiciously choose the diblock lengths at each point on the micelle surface to imprint a spontaneous curvature profile giving rise to the desired shape.

We have now explained why the multiblock is viewed as a collection of diblocks, but we have not explained why these diblocks must be joined together as opposed to simply being allowed to aggregate as in a typical self-assembled micelle composed of diblock amphiphiles.
The diblocks are joined in order to prevent them from diffusing across the micelle surface, since such diffusion would erase the intended spontaneous curvature profile.
Nevertheless, as will be indicated in \cref{sec:methods}, the multiblock structure of the polymer often fails to ensure the intended diblock arrangement on the surface.
To eliminate these failures something further must be done, but an in-depth study of this issue is beyond the scope of the present work.
Instead, we simply discard the problematic micelles.
A justification for discarding the problematic micelles and proposals for how they may be completely eliminated in future work are given in \cref{sec:discussion}.

In the rest of this paper, we describe an assessment of the performance of this design scheme.
In \cref{sec:methods}, we describe how we simulate single-polymer micelles: we identify a micelle composition that assumes a nonstandard shape; we select five aspects of the micelle composition as suitable control parameters; and we choose two features of the micelle shape whose dependences on the control parameters are to be assessed.
In \cref{sec:results}, the results of varying our chosen aspects of micelle composition are presented, and the effect on the shape features is examined.
In \cref{sec:discussion}, we discuss the implications of our results for the practicality of the shape-design scheme presented in this paper, and we revisit unresolved issues mentioned in \cref{sec:introduction} and \cref{sec:methods}.
In \cref{sec:conclusion}, we conclude.

\section{Methods}
\label{sec:methods}
To assess our shape-design scheme, we simulate micelles of various compositions and compare the resulting shapes.
However, before simulations can be performed, a physical model for the micelles must be selected.
A detailed description of our model and simulation method is given in \cite{firstPaper}.
We now present the most relevant features starting with the model.

\begin{figure}
\centering
\includegraphics[width=.4\linewidth]{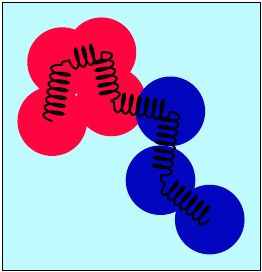}
\caption{Schematic of a short diblock copolymer as represented in our model.
This copolymer consists of seven beads: four solvophobic beads, shown in red (light gray), and three solvophilic beads, shown in blue (dark gray).
Harmonic springs connect beads adjacent along the polymer.
The light blue background represents the implicit solvent.}
\label{fig:beadSpring}
\end{figure}

We choose a simple coarse-grained bead-spring model with implicit solvent (similar to those of~\cite{detcheverry2009,Binder11,Dimitrakopoulos04,Hsieh06}) because our shape-design mechanism should not depend on details of the interactions of the polymer constituents.
A polymer molecule is represented as a linear sequence of beads with consecutive beads joined by harmonic springs, as illustrated in \cref{fig:beadSpring}.
There are two species of beads: solvophilic beads, which interact with other beads through only a short-range repulsion, and solvophobic beads, which experience an additional longer-range attraction with other solvophobic beads because of their immiscibility with the solvent.
The particular values of the interaction parameters and the simulation's temperature are chosen to replicate macroscopic behavior of real polymer.

This model is simulated at constant temperature using the LAMMPS molecular dynamics package \cite{lammps95}.
At regular intervals of the simulated time, the simulation records the junction points' positions, defined as the midpoint between adjacent beads of opposite species.
Because we are interested only in the overall shape of the micelle and not the individual positioning of each bead, these junction points provide sufficient data for our purposes.

We refer to each list of junction point positions as a ``shape", denoting it by a blackboard bold symbol (e.g., $\bbr$).
Therefore, each shape $\bbr$ has the form 
\begin{equation}
\left(\br_1,\br_2,\dots,\br_i,\dots,\br_{N_j}\right),
\label{eq:formOfR}
\end{equation} 
where $N_j$ is the number of junction points in the micelle and each $\br_i$ is a two dimensional junction point position.
The output of the simulation is a time sequence of such shapes: $\bbr_\alpha$, $\alpha=1,2,\dots,N_s$, where $N_s$ is the number of sampled shapes.

After the simulation runs are complete, the shape sequences are further analyzed.
We summarize the resulting sequence $\bbr_\alpha$ of shapes by its average $\bbbr$, the shape variance matrix $\bbs$ of dimension $2 N_j \times 2 N_j$ characterizing the variance shape's thermal fluctuations, and another $2 N_j \times 2 N_j$ variance matrix $\bbbs$ representing the uncertainty in the mean shape $\bbbr$.

For a given micelle composition, we run several simulations.
Despite the bonds joining adjacent diblocks, roughly half of the simulations result in poorly formed micelles where the diblocks do not keep their intended positioning on the surface.
Since we are interested in the behavior of our shape-design scheme, which depends on the diblocks maintaining their intended positioning, we exclude any poorly formed micelles from further analysis.
Concretely, we reject any simulation run whose average micelle either has two neighboring junction points separated by more than forty percent of the median distance between adjacent junction points, or whose the shortest closed path connecting all the junction points does not have the intended ordering.
In \cref{sec:discussion}, we explain why the exclusion of these poorly formed micelles is justified.
We combine the results of the remaining simulations to make a best estimate of $\bbbr$, $\bbs$, and $\bbbs$.

Having described how the micelles are simulated, we now describe which micelle compositions to simulate in order to examine their effect on micelle shape.
We start with a reference micelle composition previously shown in~\cite{firstPaper} to produce a micelle of nonstandard dimpled shape.
Then we select several parameters of this micelle composition to be changed.

\begin{figure}
\centering 
\includegraphics[width=\linewidth]{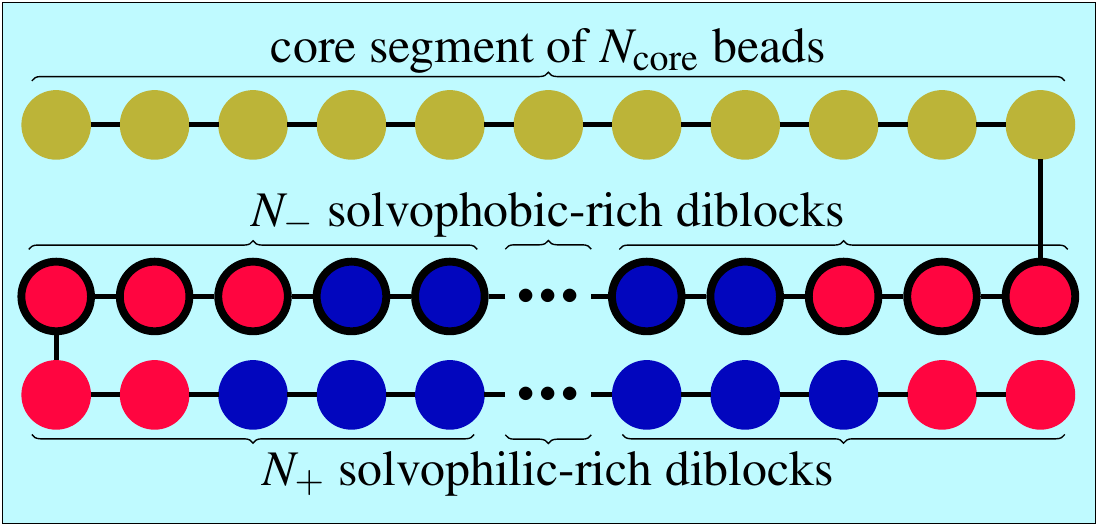}
\caption{
Schematic of multiblock bond architecture.
Tan (top row) and red (light gray, bottom rows) disks represent solvophobic beads, while blue disks (dark gray, bottom rows) represent solvophilic beads.
Black segments represent bonds between beads.
The multiblock begins with a core segment composed of solvophobic beads shown in tan and occupying the top row.
To this segment solvophobic-rich diblocks, outlined in black as in \cref{fig:endPicture}, are successively attached end to end.
Lastly solvophilic-rich diblocks are attached end to end.
The ``$\bullet \bullet\bullet$'' symbols represent further diblocks not shown.
Figure adapted from \cite{firstPaper}.
}
\label{fig:multiblockdiblockwithcore} 
\end{figure}
\begin{figure}
\centering %
\includegraphics[width=\columnwidth]{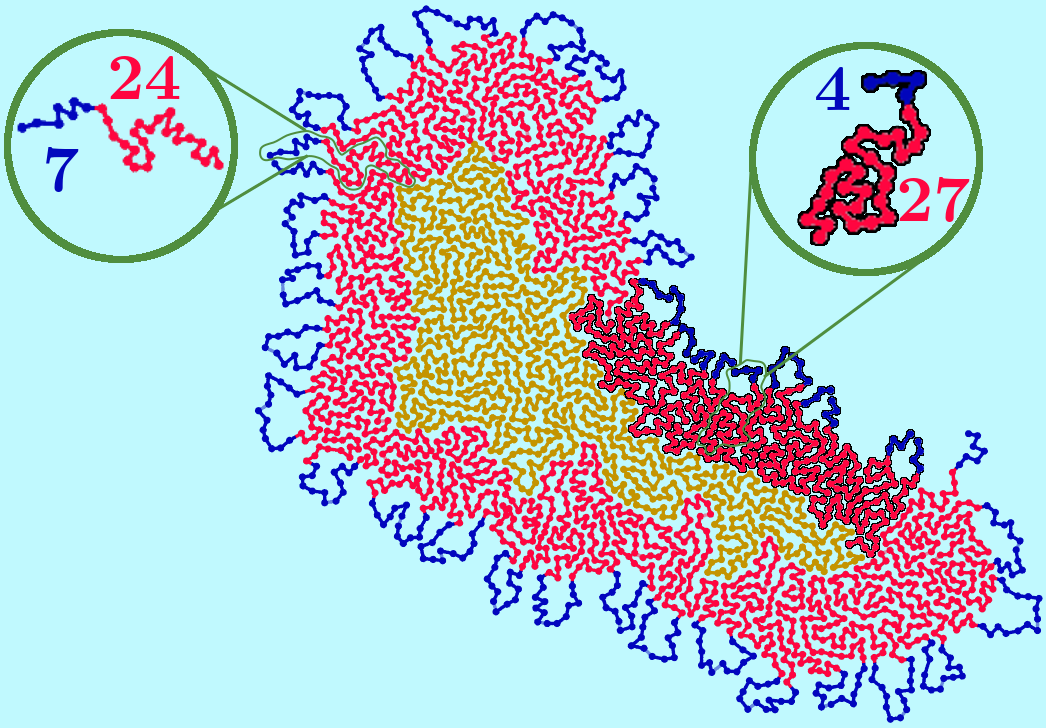} %
\caption{
A typical configuration of the reference micelle during the course of a simulation.
The red (light gray) beads are solvophobic; the blue (dark gray beads on micelle exterior), solvophilic.
The tan beads (dark gray beads in micelle interior), which constitute the micelle core, are also solvophobic.
The micelle is constructed from two types of diblocks, termed solvophobic-rich and solvophilic-rich.
The solvophobic-rich diblocks, outlined in black, are located near the dimple.
}
\label{fig:endPicture}
\end{figure}

The basic design of the co\-polymers simulated in this work is shown in \cref{fig:multiblockdiblockwithcore}.
The key feature of the design is that the micelle contains two species of diblock having a common length but distinguished by their composition: a ``solvophobic-rich'' species of diblock, having relatively more solvophobic beads and therefore favoring a more negative, concave curvature, and a ``solvophilic-rich'' species of diblock, having relatively more solvophilic beads and therefore favoring a more positive, convex curvature.
This contrast in preferred curvature is designed to cause the formation of a dimple.
The linear co\-polymer begins with a long sequence of solvophobic beads, which forms a ``core" to be situated in the micelle's interior.
To one end of this core segment are joined end-to-end a sequence of solvophobic-rich diblocks.
To the free end of this sequence of solvophobic-rich diblocks, we attached a sequence of solvophilic-rich diblocks.
The micelle has $700$ core beads, $12$ solvophobic-rich diblocks each containing $27$ solvophobic beads and $4$ solvophilic beads, and $55$ solvophilic-rich diblocks each containing $24$ solvophobic beads and $7$ solvophilic beads.
A chemical formula representing this monomer sequence is $R_{700}(R_{27}B_4B_4R_{27})_6(R_{24}B_7 B_7 R_{24})_{27} R_{24}B_7$, where $R$ represents a solvophobic monomer and $B$ represents a solvophilic monomer.
A typical thermal configuration of a micelle having the reference composition is illustrated in \cref{fig:endPicture}. 

\begin{figure}
\hspace*{\fill}%
\subfloat[][]{%
\label{fig:asymmetryOne}%
\centering%
\includegraphics[width=.4\columnwidth]{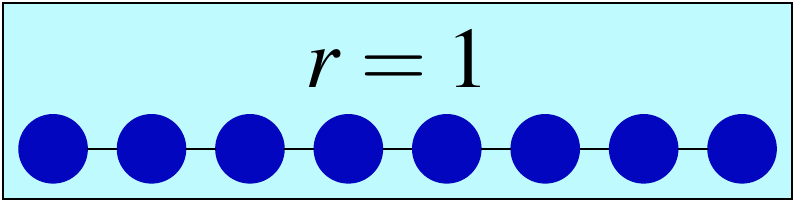}%
}\hspace*{\fill}%
\subfloat[][]{%
\label{fig:asymmetryHalf}%
\centering%
\includegraphics[width=.4\columnwidth]{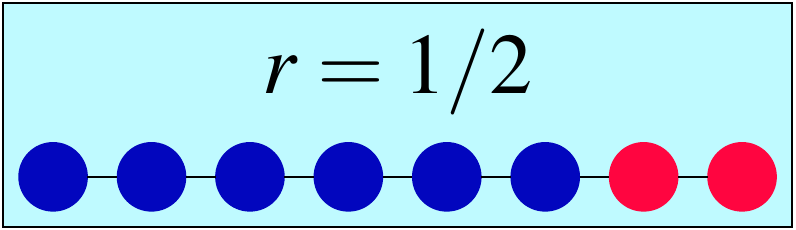}%
}\hspace*{\fill}

\hspace*{\fill}%
\subfloat[][]{ %
\label{fig:asymmetryZero}%
\centering%
\includegraphics[width=.4\columnwidth]{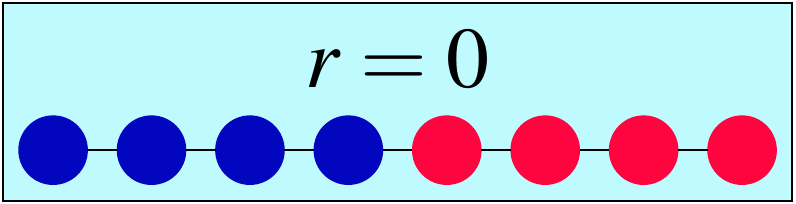}%
}\hspace*{\fill}%
\subfloat[][]{%
\label{fig:asymmeryMinusOne}%
\centering%
\includegraphics[width=.4\columnwidth]{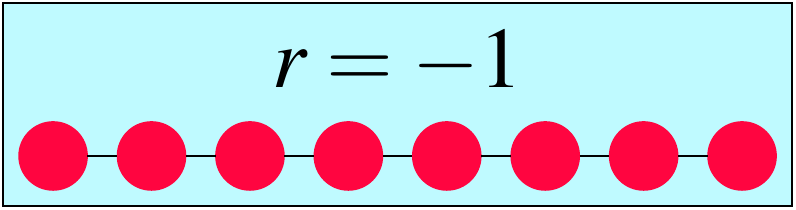}%
}\hspace*{\fill}
\caption{Several diblocks (or, in the extreme cases where only one species of bead is present, homopolymers) and their asymmetry ratios $r$ defined in \cref{eq:asymmetryRatio}.
Solvophobic beads are shown in red (light gray), and solvophilic beads, in blue (dark gray). }
\label{fig:asymmetryRatio}
\end{figure}

Next, we describe our chosen control parameters---parameters of the micelle composition that we alter to control the micelle's shape.
A good control parameter must have a well-behaved effect on the micelle shape, and, further, its effect on micelle shape ought to be predictable using a simple rationale.
Accordingly, we will describe each parameter's anticipated effect as it is introduced.
The first parameter we define is the number of core beads, $\nc$, having a value of $700$ for the reference micelle composition; it can be used to set the enclosed volume of the micelle without affecting the surface properties.
Two additional parameters concerning the number of beads in the micelle are the numbers of solvophobic-rich ($\no$, the ``$-$'' reflecting that these diblocks prefer a relatively negative, concave curvature) and solvophilic-rich ($\ni$, the ``$+$'' reflecting that these diblocks prefer a relatively positive, convex curvature) diblocks in the micelle, which we expect to set the preferred perimeter of their respective regions of the micelle surface without directly affecting either region's preferred curvature.
These parameters have the values of $\no=12$ and $\ni=55$ for the reference micelle composition.
The two final parameters concern the compositions of the solvophobic-rich and solvophilic-rich diblocks.
We keep the length of either species of diblock fixed at $31$, changing only the relative amount of the two species of beads (that is, the asymmetry of the diblock).
The asymmetry of a diblock containing $n_\text{phobic}$ solvophobic beads and $n_\text{philic}$ solvophilic beads is quantified by the ``asymmetry ratio" $r$ defined by 
\begin{equation}
r=\frac{n_\text{philic} - n_\text{phobic}}{n_\text{philic} + n_\text{phobic}}.
\label{eq:asymmetryRatio}
\end{equation}
The asymmetry ratio of several model diblocks is illustrated in \cref{fig:asymmetryRatio}.
We denote the asymmetry ratio of the solvophobic-rich diblocks and solvophilic-rich diblocks by $\ro$ and $\ri,$ respectively.
By definition, solvophilic-rich diblocks have a larger asymmetry ratio, so that $\ro$ and $\ri$ satisfy $\ro < \ri$.
These two parameters provide a way to control the spontaneous curvature of their respective regions of the micelle surface while only weakly changing the preferred perimeter.
Specifically, we expect that the more positive a diblock's asymmetry ratio, the more positive its associated preferred curvature.

\begin{figure}
\centering %
\includegraphics[width=.8\columnwidth]{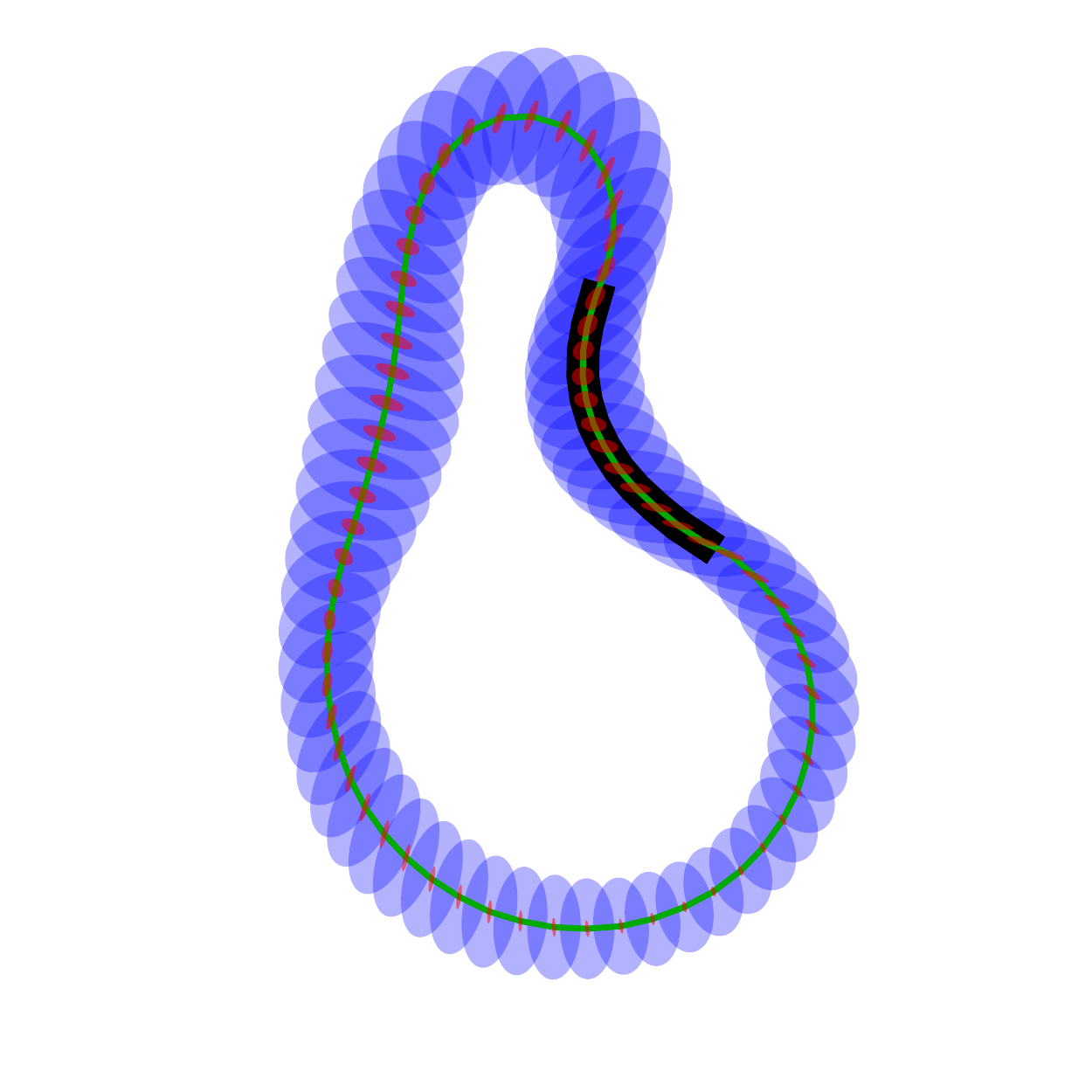} %
\caption{A graphical representation of the average shape $\bbbr$, the shape fluctuations $\bbs$, and the uncertainty in the average shape $\bbbs$ of the reference micelle.
The average shape $\bbbr$ is represented as a green curve (passing through the midline of shaded region) connecting the average position of the junction points.
The curve segments connecting solvophobic-rich diblocks are outlined in black.
The shape fluctuations $\bbs$ are represented by a large blue ellipses surrounding each junction point indicating the $40\%$ confidence region (corresponding to one standard deviation) for the junction point's position during the course of the simulation.
The uncertainty in the mean shape is represented similarly with smaller red ellipses indicating the confidence region for the mean junction point position.
}
\label{fig:baseCaseAverage}
\end{figure}
Having described how the micelle compositions are changed, we now describe what features of the resulting thermal micelle shape distribution we study.
A graphical representation of the average shape $\bbbr$, shape sample variance $\bbs$, and variance in the mean shape $\bbbs$ characterizing the micelle shape distribution is shown in \cref{fig:baseCaseAverage}.
In previous work \cite{firstPaper}, we have validated that the mean shapes are reproducible and the errors in the mean shape are indeed consistent with the variability in the mean shape between simulation runs.
However, since the average shape $\bbbr$, shape sample variance $\bbs$, and variance in the mean shape $\bbbs$ are high-dimensional objects, we choose, for the sake of concreteness, to look at only two scalar shape features summarizing these quantities, which we soon define: the curvature ratio, characterizing the strength of the average shape's dimple, and the normalized fluctuation, characterizing the size of thermal shape fluctuations. 

\begin{figure}
\includegraphics[width=.7\columnwidth]{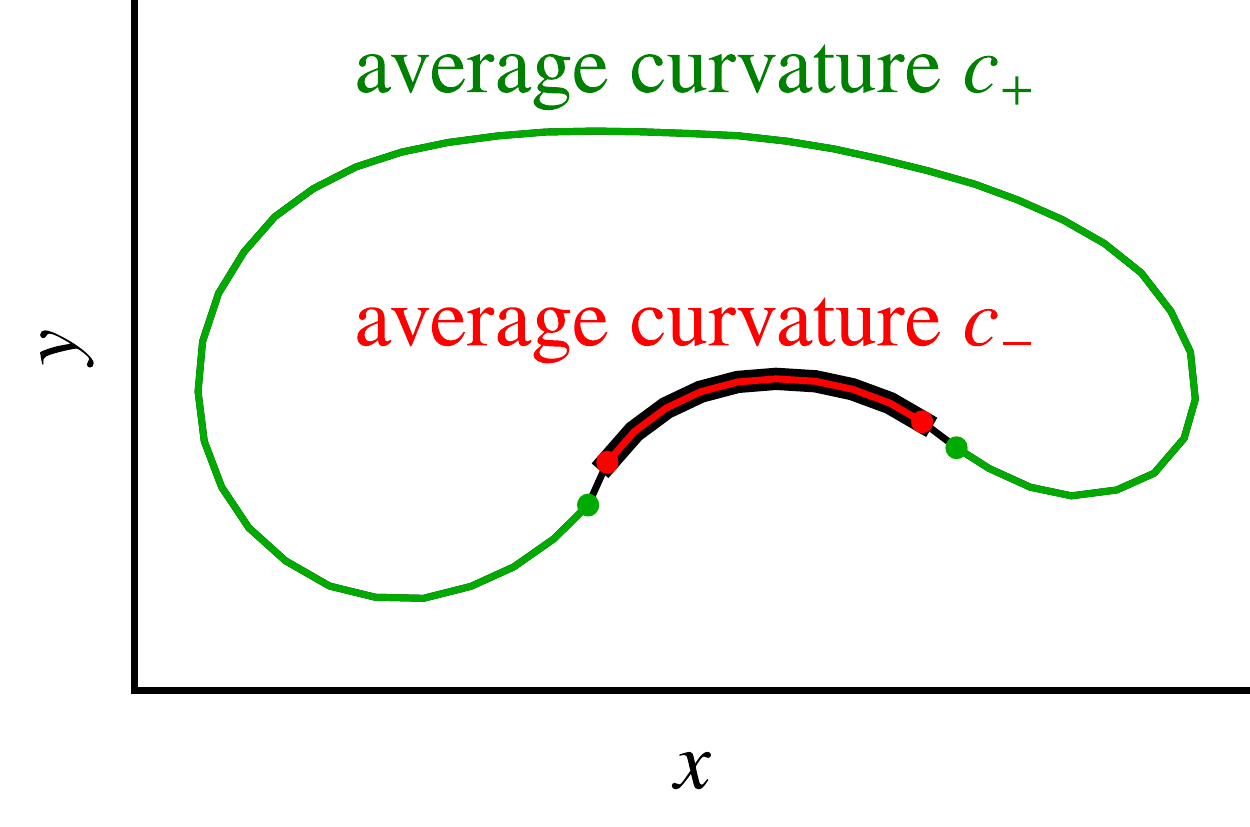}
\caption{Illustration of curvature ratio definition.
The region occupied by solvophobic-rich diblocks is shown in red, and its average signed curvature (having a negative value in this case) is denoted by $c_-$.
By contrast, the region occupied by solvophilic-rich diblocks is shown in green, and its average curvature is denoted $c_+$.
The curvature ratio is defined as the ratio $c_-/c_+$.
Figure reproduced from \cite{firstPaper}.
}
\label{fig:curvatureRatio}
\end{figure}

The curvature ratio $\CR$, illustrated in \cref{fig:curvatureRatio}, is defined as the shape's average signed curvature $c_-$ in the region occupied by the solvophobic-rich diblocks divided by the average signed curvature $c_+$ in the region occupied by the solvophilic-rich diblocks.
Thus a circle has a curvature ratio of one, and negative curvature ratios indicate the presence of a concave dimple, with increasingly negative curvature ratios indicating stronger dimples.

The normalized fluctuation $\delta$ is defined by the formula
\begin{equation}
\delta = \sqrt{\frac{\Tr \bbs}{2 N_j R_g^2}},
\label{eq:normalizedFluctuation}
\end{equation}
where $R_g$ is the radius of gyration of the average shape $\bbbr$.
Intuitively, the factor $\sqrt{\frac{\Tr \bbs}{2N_j}}$ may be interpreted as the root-mean-square length of the semi-axes of the blue ellipses shown, for example, in \cref{fig:baseCaseAverage} (the blue ellipses being the one standard deviation confidence regions for the sampled position of the junction points).
The normalized fluctuation is a scalar measure of the size of the shape fluctuations, normalized so as not to scale with the number of junction points or overall spatial extent of the micelle shape.

The uncertainties in these two shape features can, like the values themselves, be estimated from the micelle shape distribution statistics $\bbbr$, $\bbs$, and $\bbbs$.
Since the curvature ratio depends only on the mean shape $\bbbr$, its uncertainty can easily be inferred from the error in the mean $\bbbs$.
However, our estimate for the uncertainty in the normalized fluctuation is more subtle; we refer the reader to \cite{firstPaper} for a description and validation of this uncertainty estimate.

\section{Results}
\label{sec:results}

In this section, we show the dependence of the two shape features $\CR$ and $\delta$ on the five micelle composition parameters $\nc$, $\no$, $\ni$, $\ro$, and $\ri$ introduced in \cref{sec:methods}.
To speak to the question we raised in \cref{sec:introduction} of whether this observed dependence is explained by a straightforward rationale, we give simple arguments accounting for the observed behavior in terms of the micelle surface's tension and bending energy.
The adequacy of our proposed explanations, as well as what these results imply about the feasibility of a naive design strategy will be discussed in \cref{sec:discussion}.

We are not so much concerned with the exact numerical values of the composition parameters as we are with how the micelle shape qualitatively depends on them.
Therefore, to simplify discourse, we normalize the composition parameters by their values for the reference micelle, and we denote the normalized values with a hat (\,$\hat{}$\,).
For example, the normalized amount of core $\nch$ is given by $\nc/700$, since the reference micelle composition has $700$ core beads.
Similarly, the normalized number of solvophilic-rich chains is given by $\nih=\ni/55$, since the reference micelle composition has $\ni=55$, etc.

To frame the explanation of our observed results, we first explain what one might naively expect.
The shape dependence can be thought of as a function from the five-dimensional space of micelle composition parameters to the two-dimensional shape feature space.
In this work, we start from a base micelle composition and change different aspects of the micelle composition (in other words, moving in different directions in micelle composition space) and observe the resulting change in the shape features (in other words, how the resulting shape changes in shape feature space).
A naive expectation, which must be borne out for small changes in the micelle composition, 
is that the micelle shape change depends linearly on the change in micelle composition.
In the typical case, we expect the map to have full rank so that it is possible to change the curvature ratio without changing the normalized fluctuation and vice-versa through appropriate changes to the micelle composition.
Then by the rank-nullity theorem~\cite{artin}, there must be three directions in the micelle composition space (typically not corresponding to a change in any single composition parameter) that lead to no change in the shape features.
Since we expect the three null directions to have no relationship to the axes defined by the five composition parameters, we expect the five composition parameters to each change the shape features in a unique direction in the two-dimensional shape feature space.
We will compare our results to these expectations after presenting the results.

\begin{figure}
\subfloat[][]{%
\label{fig:coreCurvature}%
\centering%
\includegraphics[width=\columnwidth]{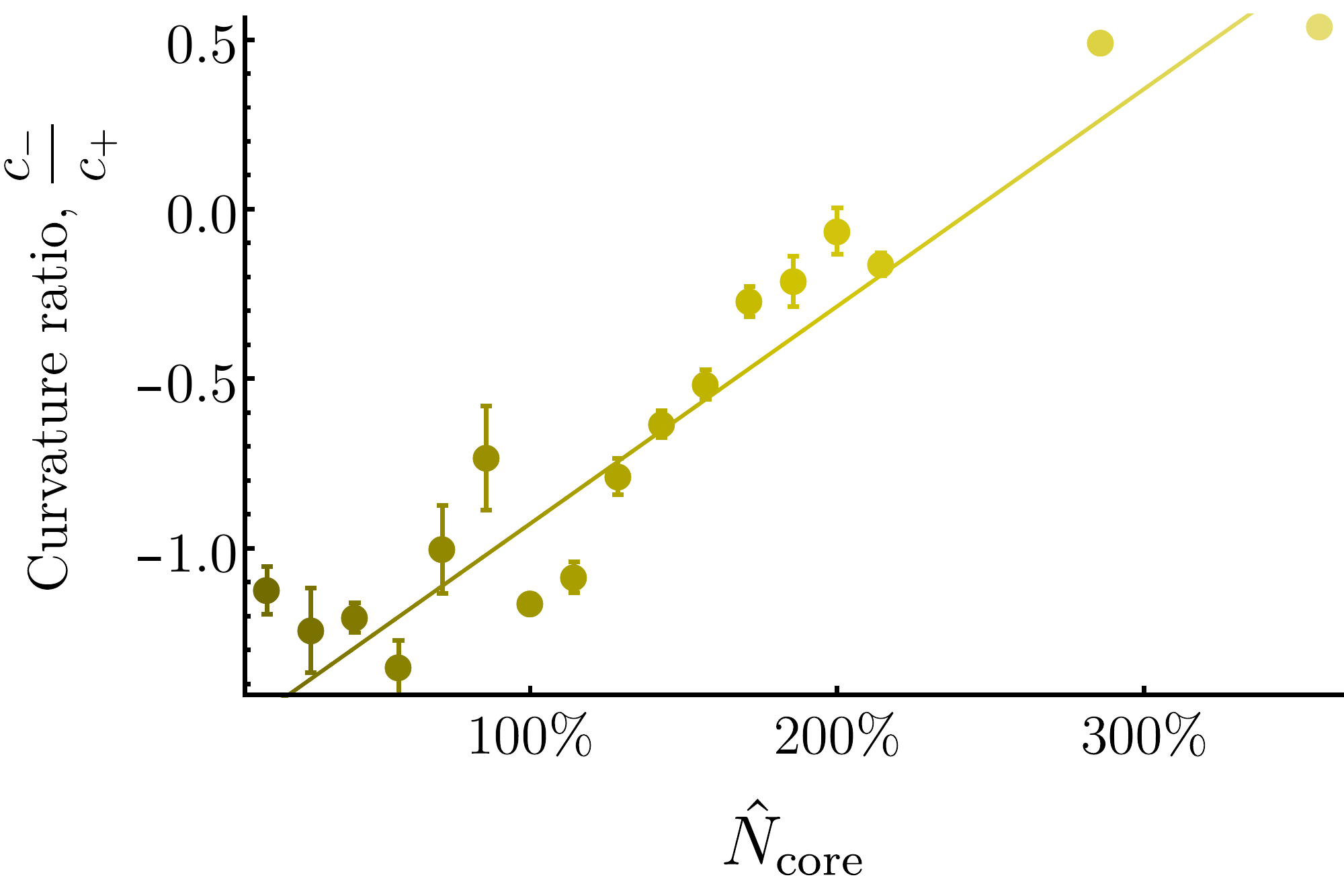}%
}

\subfloat[][]{%
\label{fig:coreFluctuation}%
\centering%
\includegraphics[width=\columnwidth]{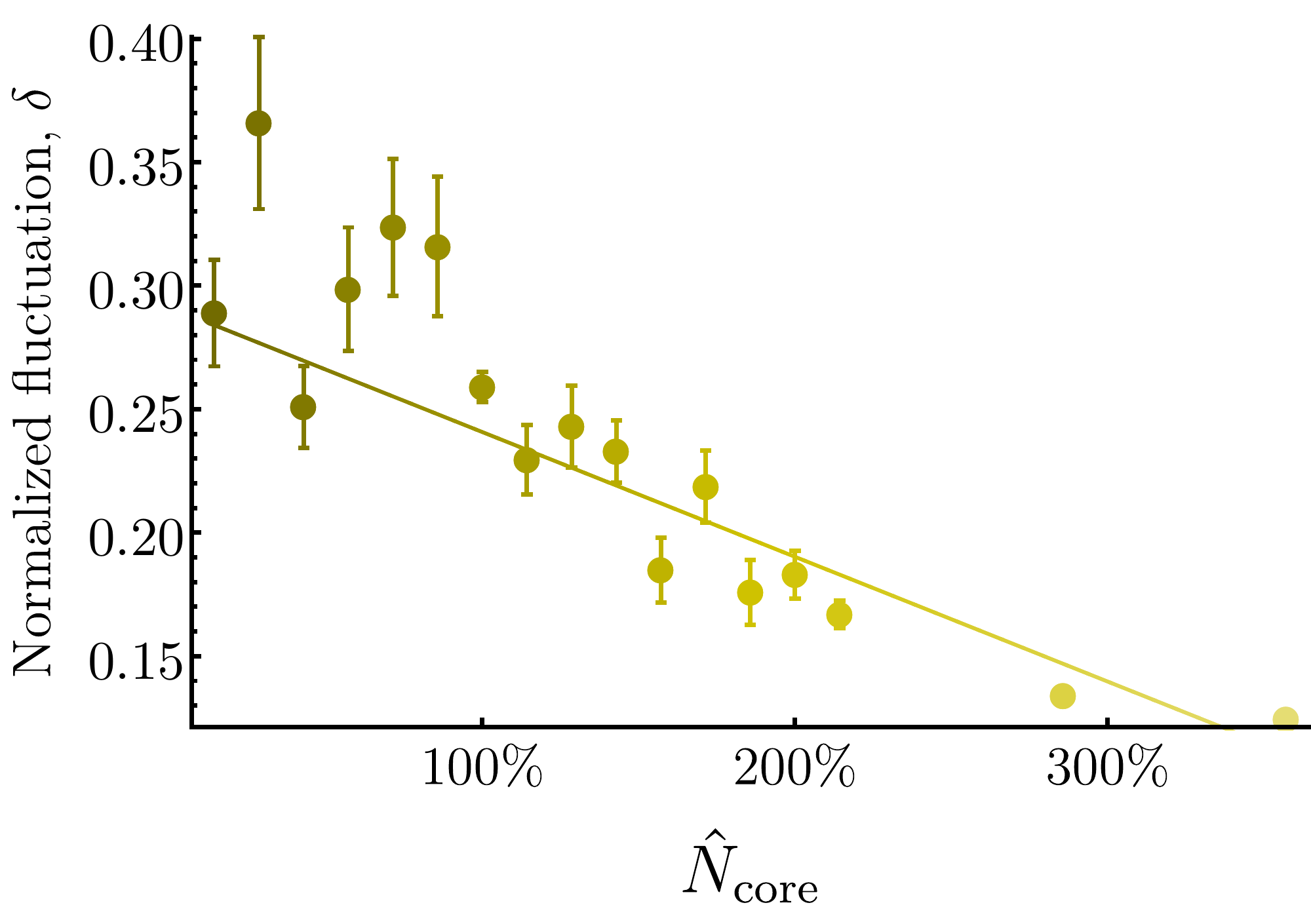}%
}

\caption{Plots of \protect\subref{fig:coreCurvature} the curvature ratio $\CR$ and \protect\subref{fig:coreFluctuation} the normalized  fluctuation $\delta$, with standard errors indicated, as a function of $\nch$, the number of core beads $\nc$ divided by the number of core beads in the reference micelle.
Also plotted is a line of best fit to the data. 
The lightness of the points and the fit line is set by the value of $\nch$.
(While the color is redundant in this plot, it is included to introduce the pattern used in \cref{fig:combinedResults}.)
The curvature ratio shows a fairly regular increasing trend (meaning the micelle shape becomes more circular) with the number of core beads, while the normalized fluctuation shows a decreasing trend.
To asses the linearity of these trends, linear fits are performed to both sets of data.
}
\label{fig:corePlots}
\end{figure}

We begin by examining the shape features' dependence on the number of core beads $\nc$ while holding the other four composition parameters fixed.
\newcommand{\figureSize}{.12}
\begin{table*}
\centering
\newcommand\T{\rule{0pt}{2.6ex}}
\begin{tabular*}{.85\textwidth}{|c|cccccc|}
\cline{1-7}
$\nch$ \T
 &  14\% & 29\% & 43\% & 57\% & 71\% & 86\%  \\
 &
\raisebox{-.5\height}{\includegraphics[width=\figureSize \linewidth]{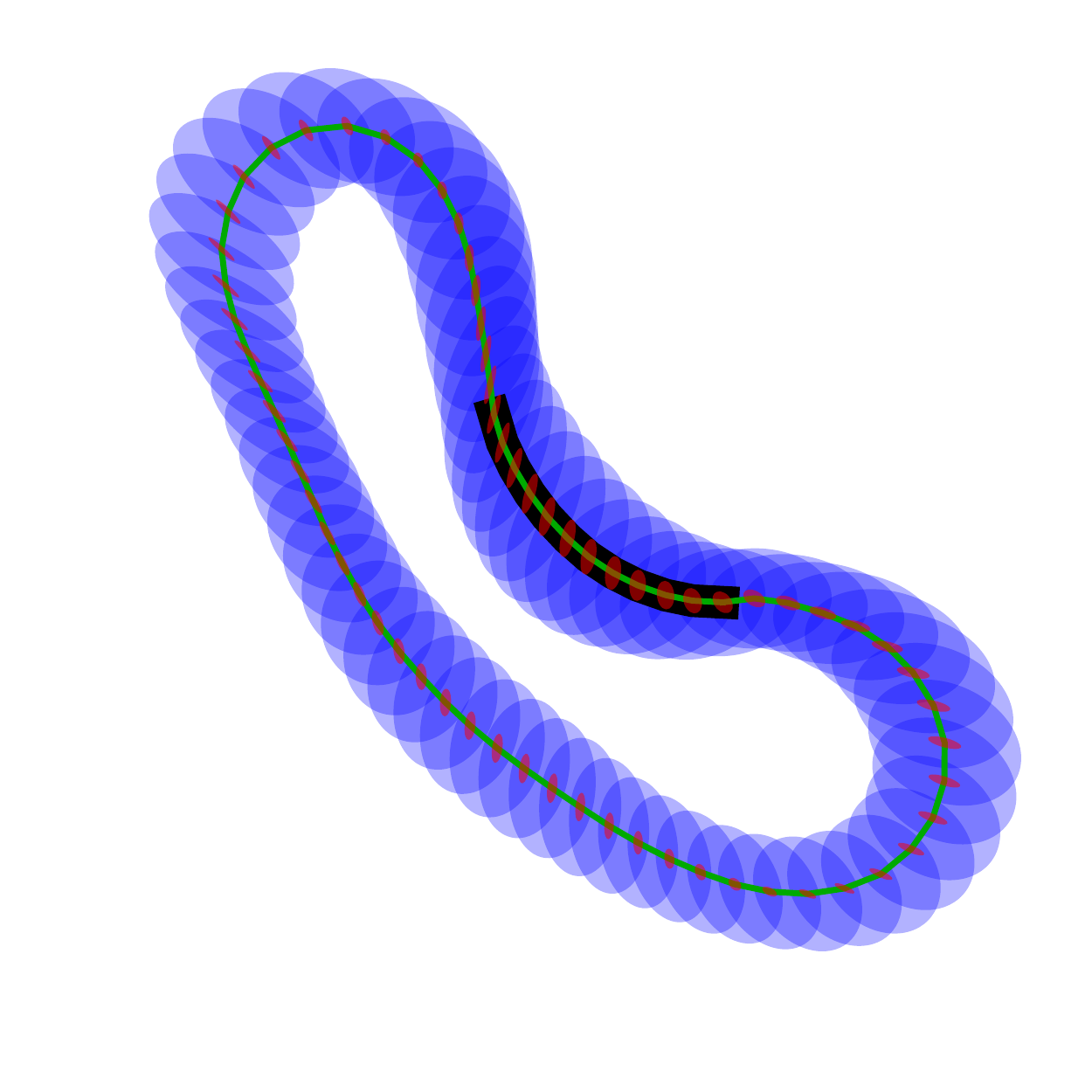}}
&
\raisebox{-.5\height}{\includegraphics[width=\figureSize \linewidth]{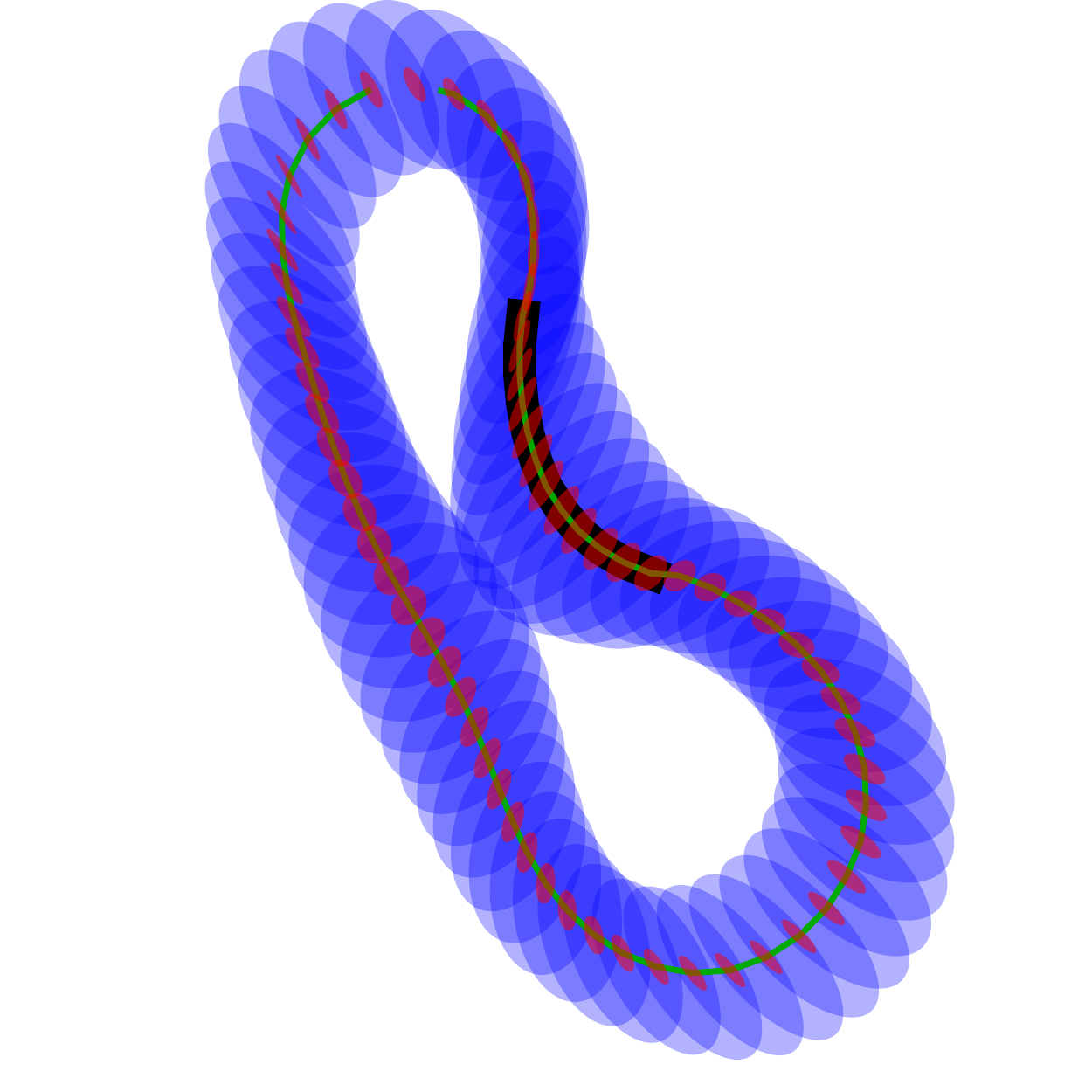}}
&
\raisebox{-.5\height}{\includegraphics[width=\figureSize \linewidth]{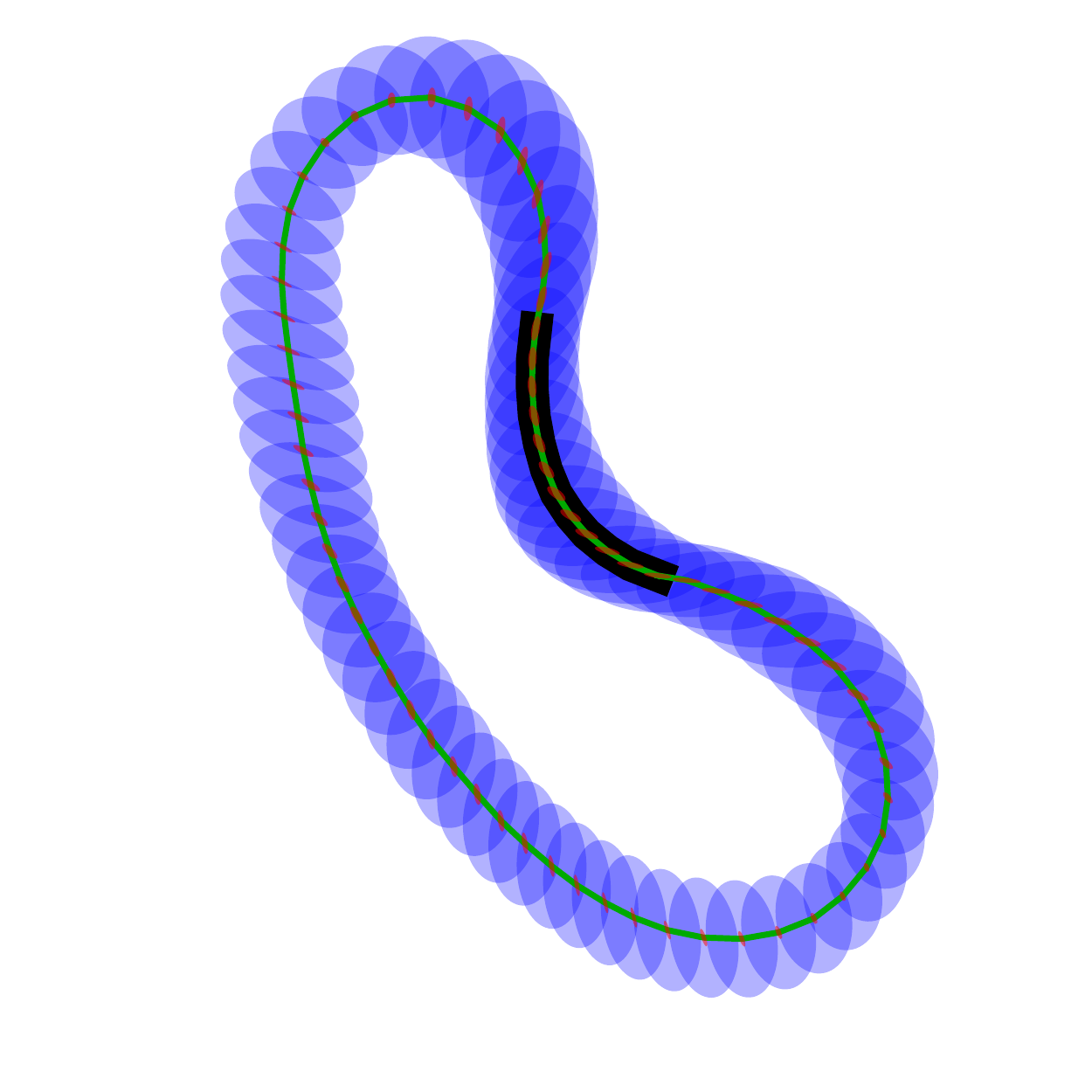}}
&
\raisebox{-.5\height}{\includegraphics[width=\figureSize \linewidth]{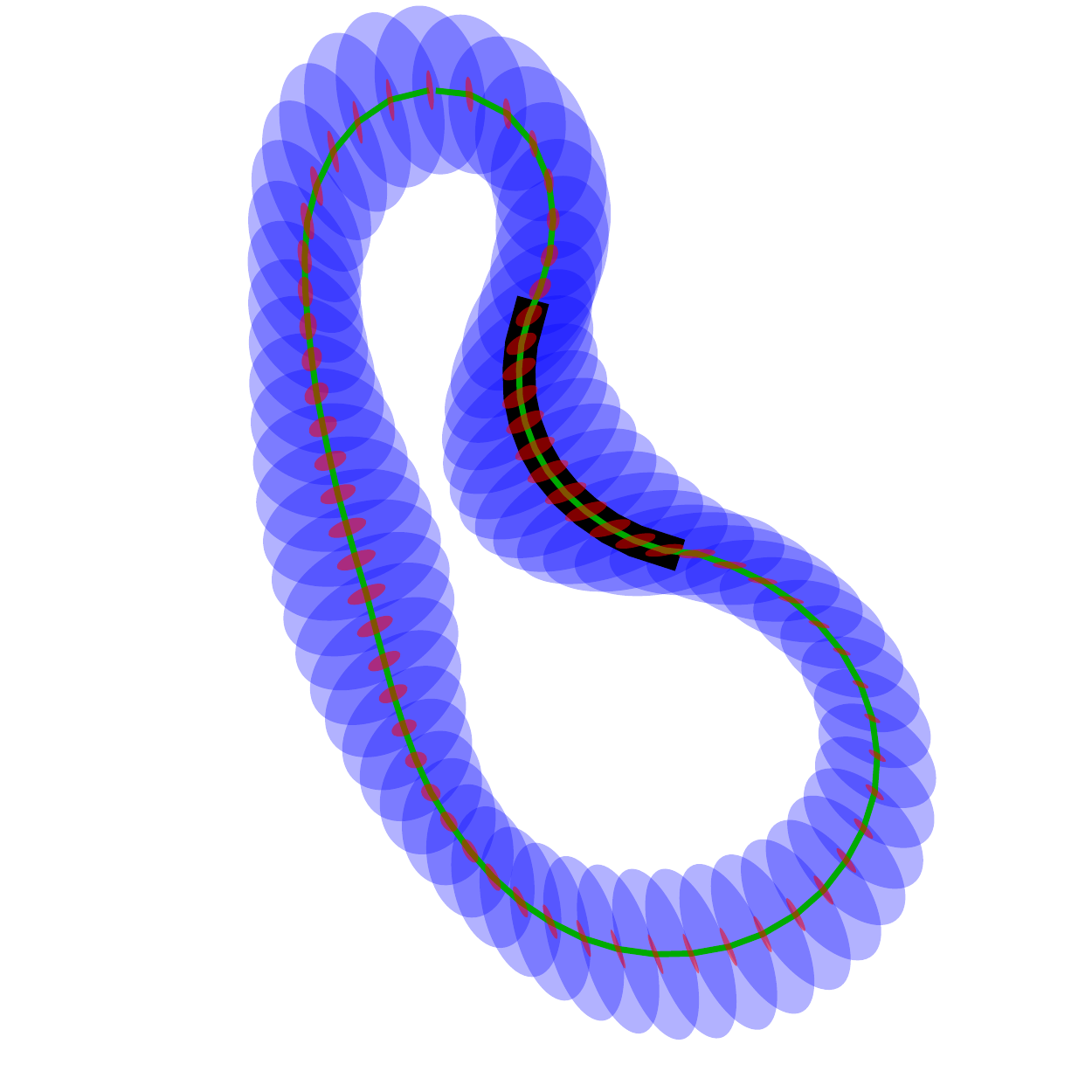}}
&
\raisebox{-.5\height}{\includegraphics[width=\figureSize \linewidth]{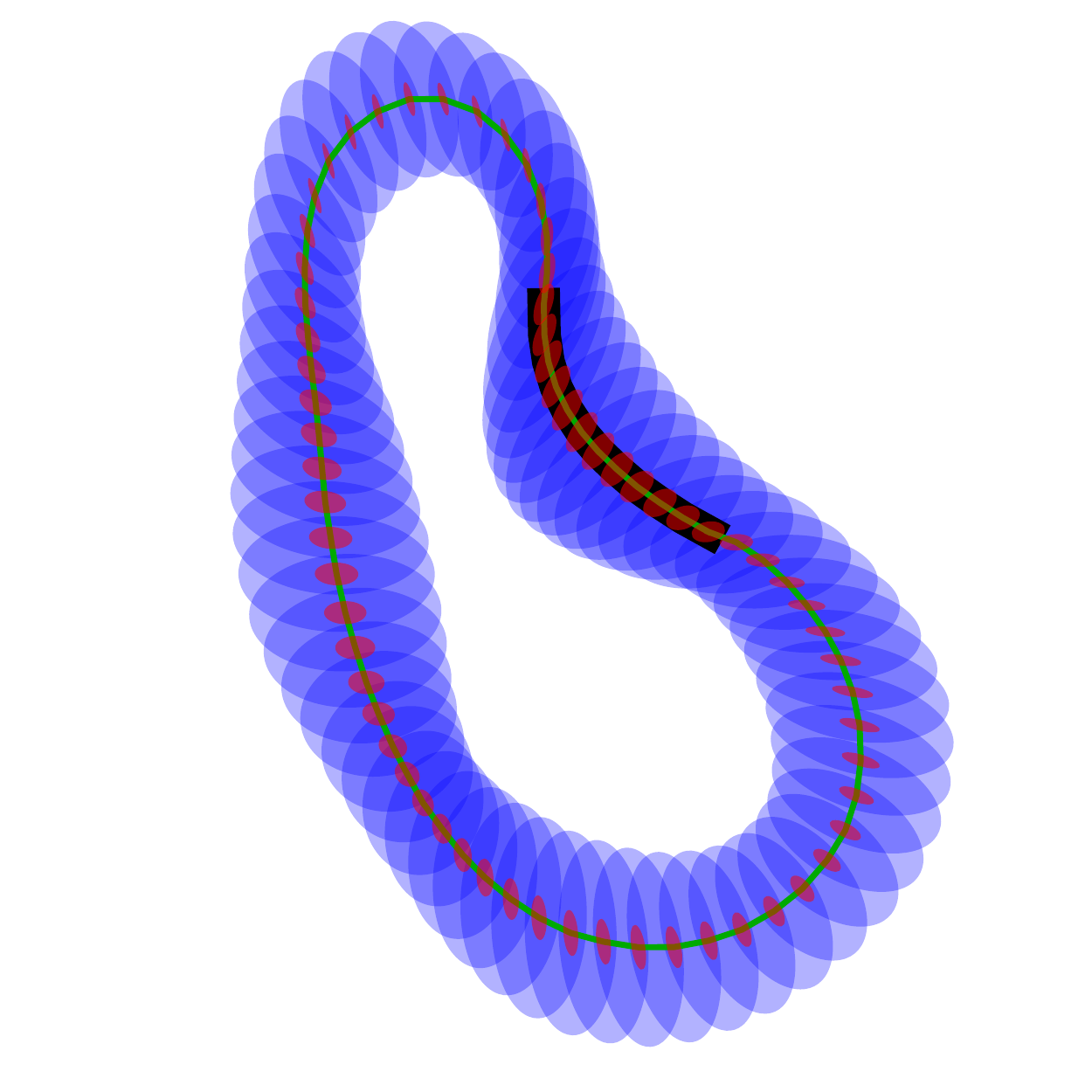}}
&
\raisebox{-.5\height}{\includegraphics[width=\figureSize \linewidth]{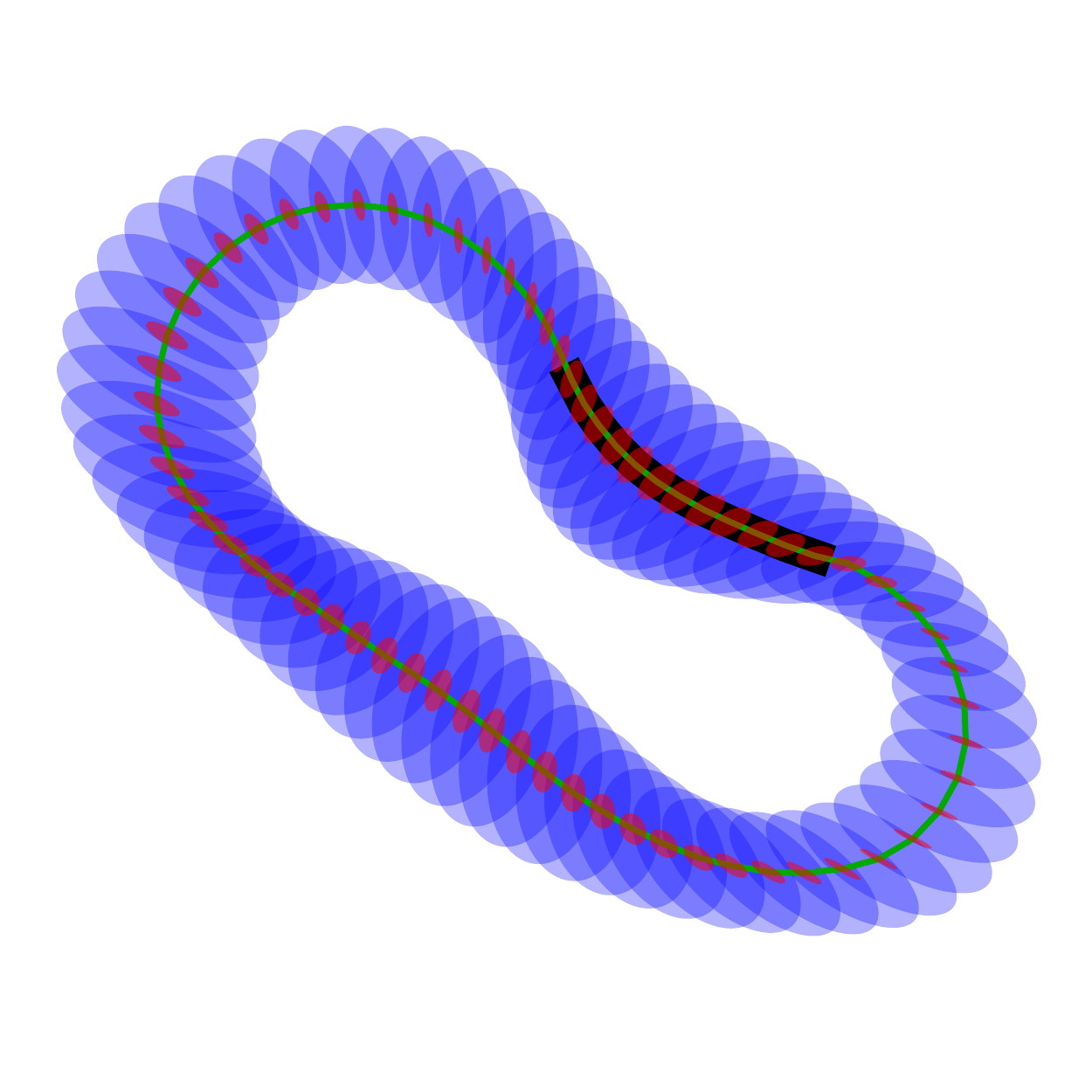}}
\\
\cline{1-7}
$\nch$ \T& 100\%&\small 114\% &\small 129\% &\small 143\% &\small 157\%  &\small 171\%  \\
&
\raisebox{-.5\height}{\includegraphics[width=\figureSize \linewidth]{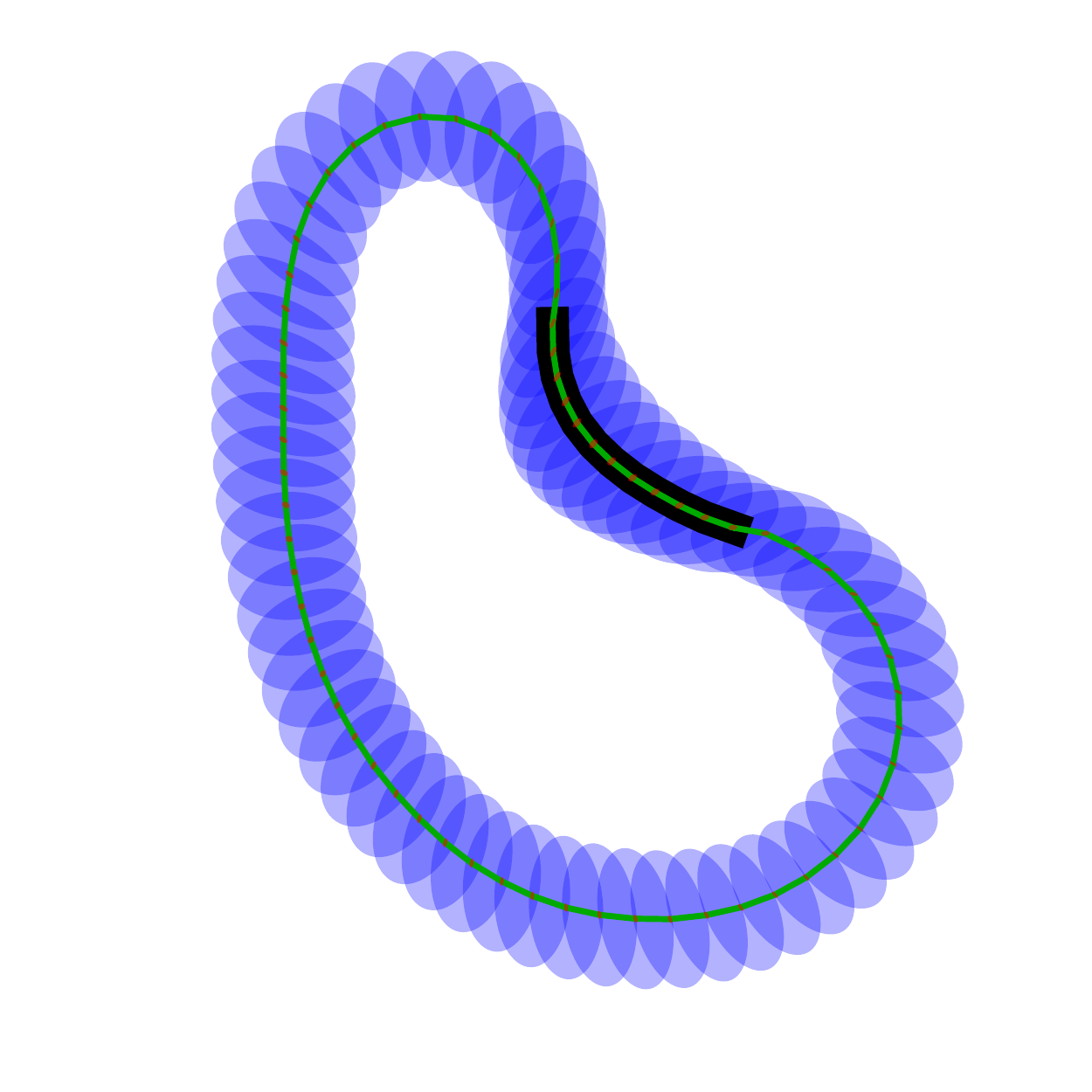}}
&
\raisebox{-.5\height}{\includegraphics[width=\figureSize \linewidth]{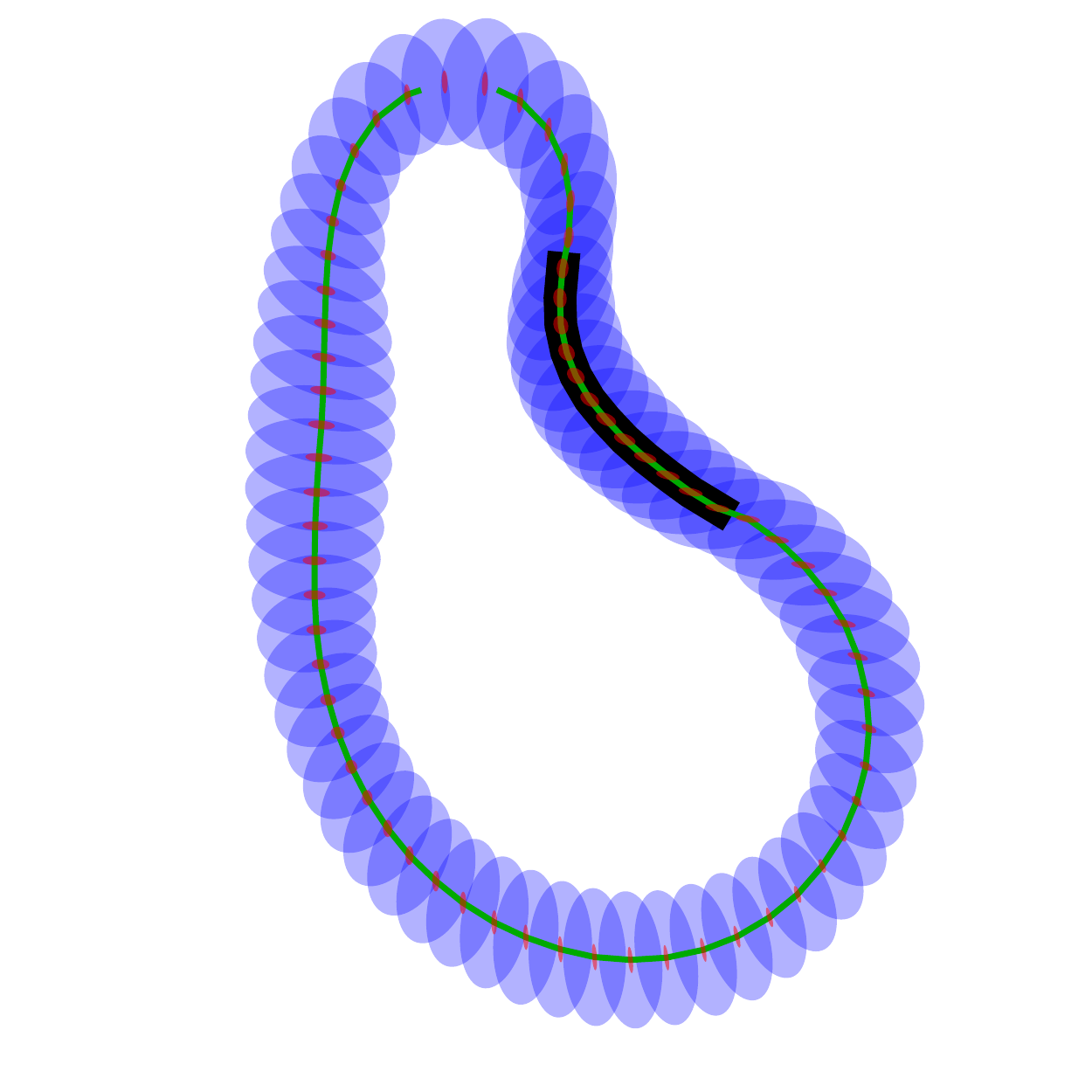}}
&
\raisebox{-.5\height}{\includegraphics[width=\figureSize \linewidth]{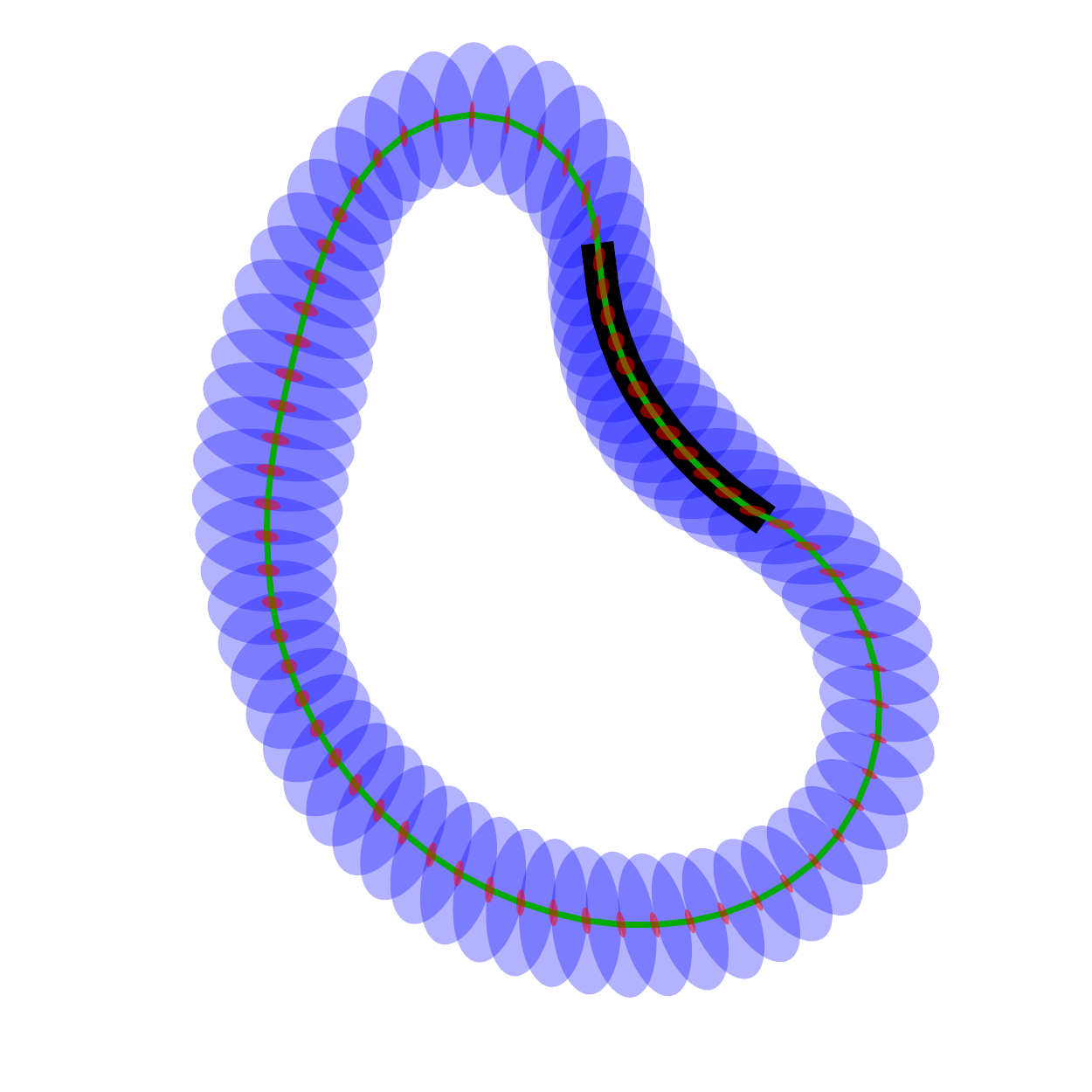}}
&
\raisebox{-.5\height}{\includegraphics[width=\figureSize \linewidth]{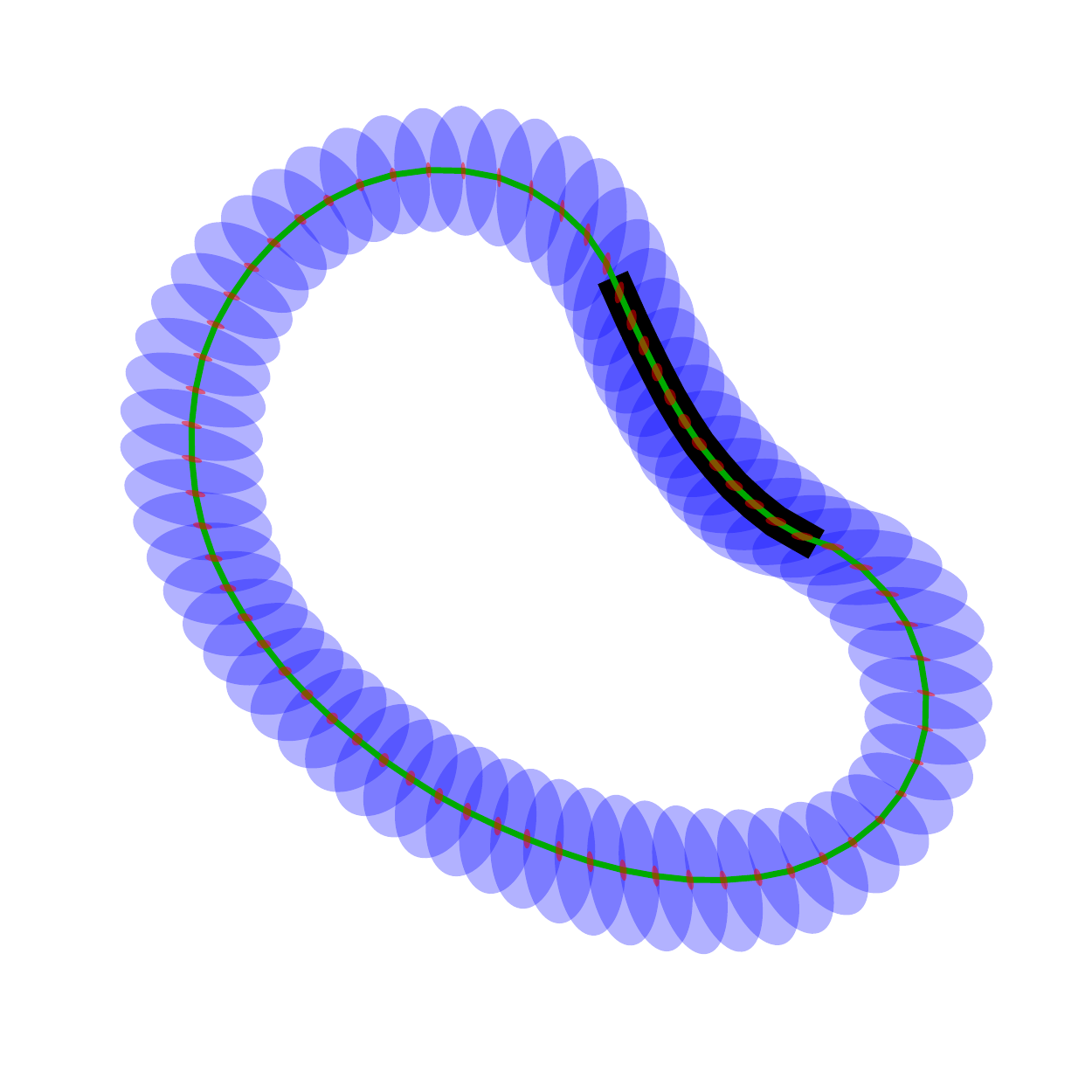}}
&
\raisebox{-.5\height}{\includegraphics[width=\figureSize \linewidth]{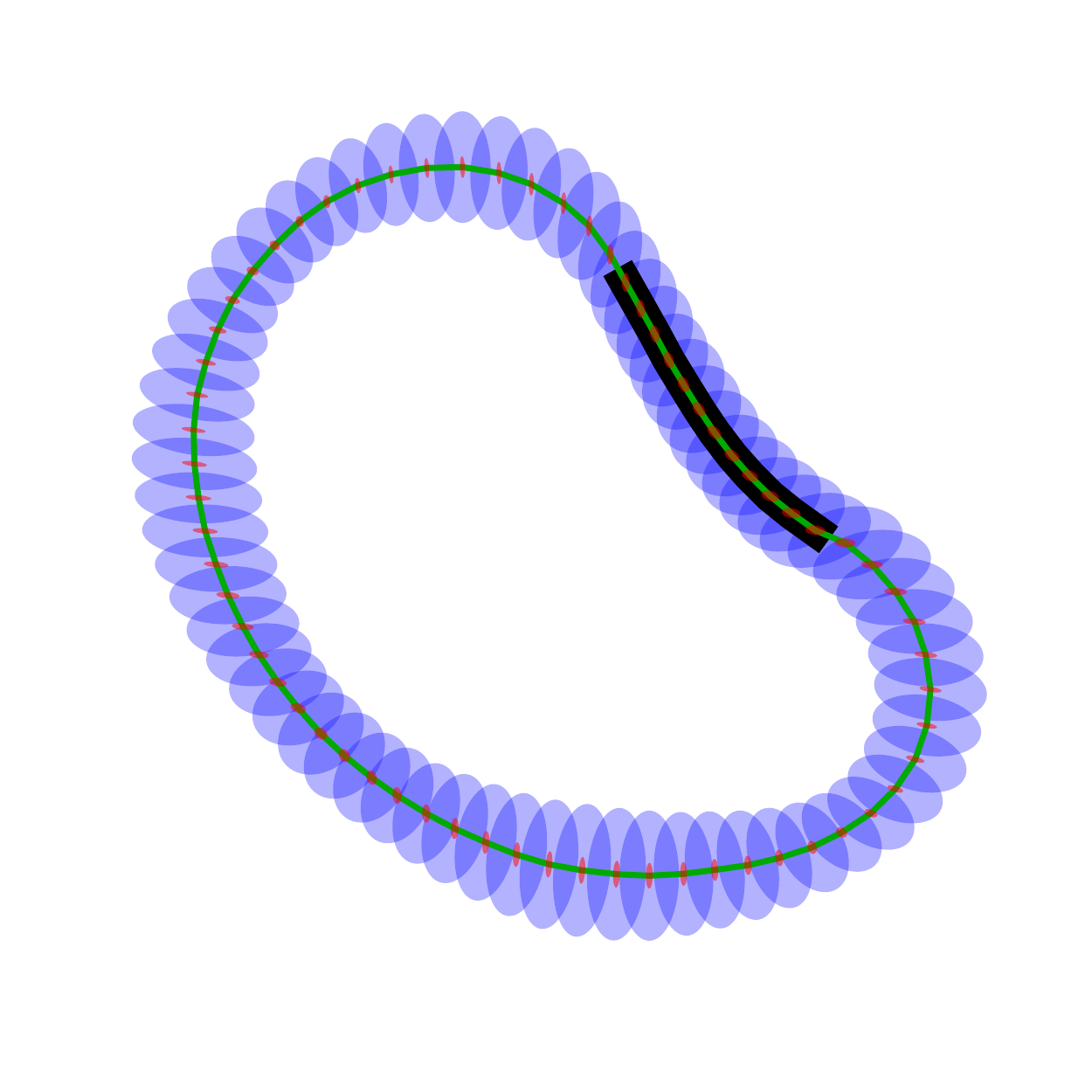}}
&
\raisebox{-.5\height}{\includegraphics[width=\figureSize \linewidth]{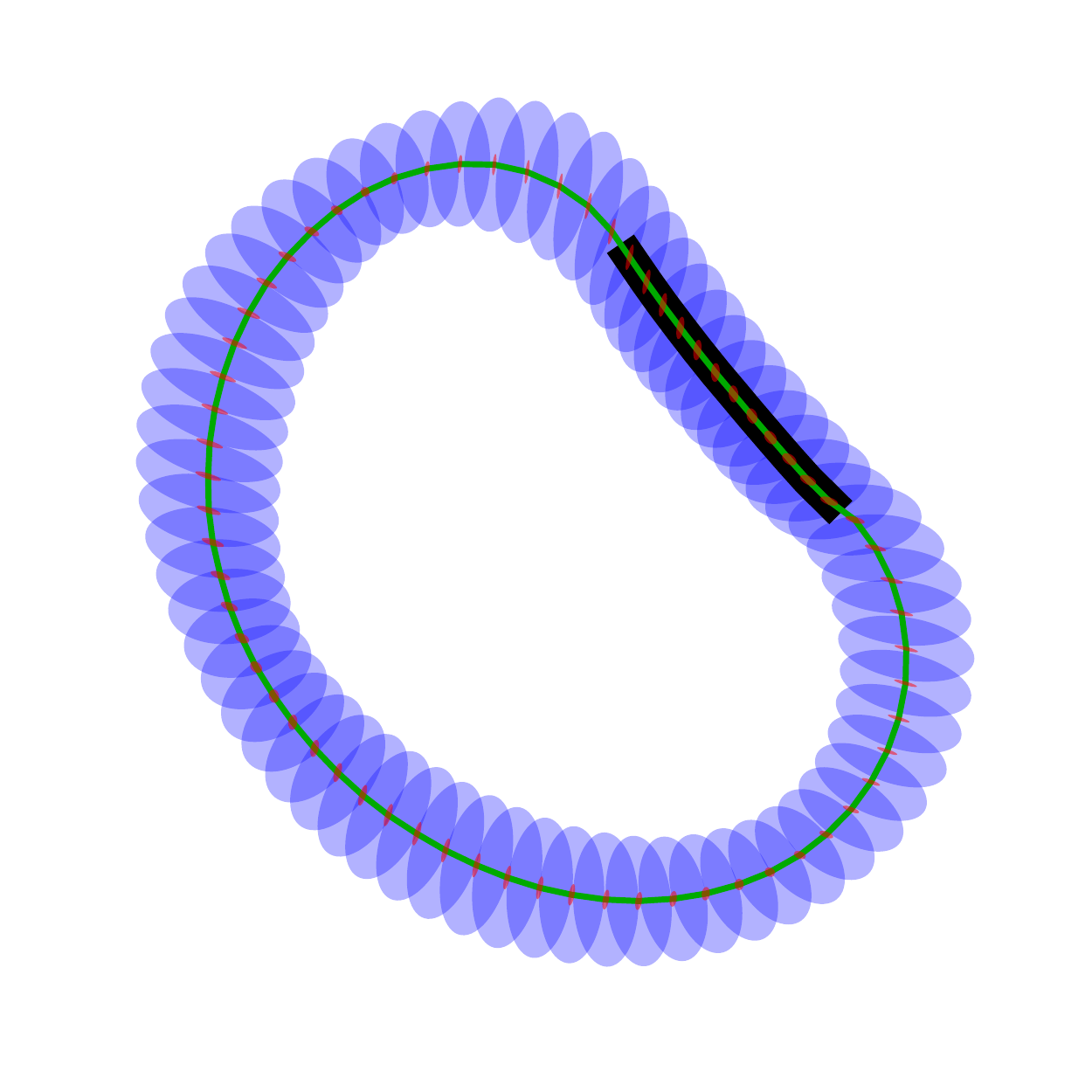}}
\\
\cline{1-7}
$\nch$ \T& \small 186\% &\small 200\% &\small 214\% &\small 286\% &\small 357\%& \\
&
\raisebox{-.5\height}{\includegraphics[width=\figureSize \linewidth]{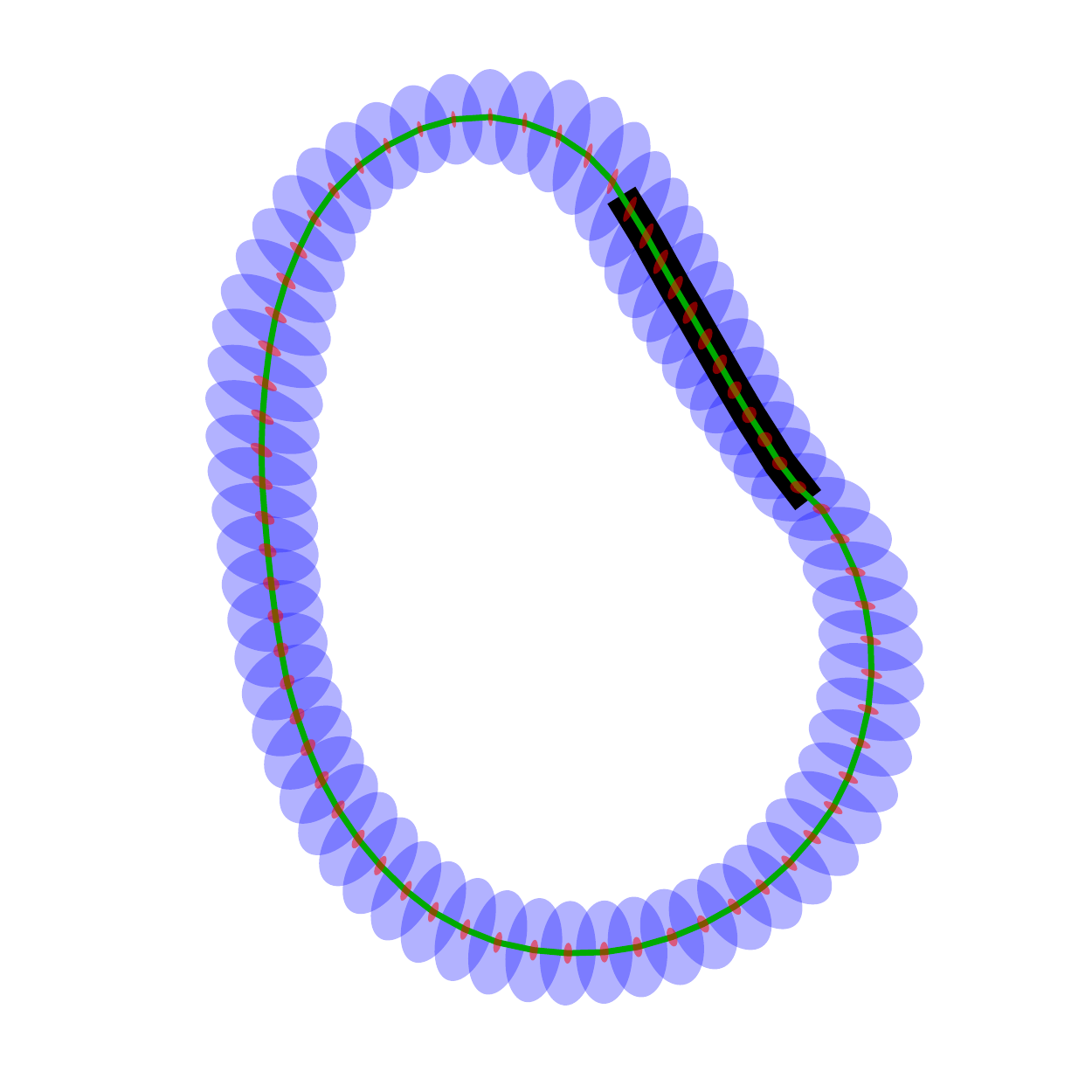}}
&
\raisebox{-.5\height}{\includegraphics[width=\figureSize \linewidth]{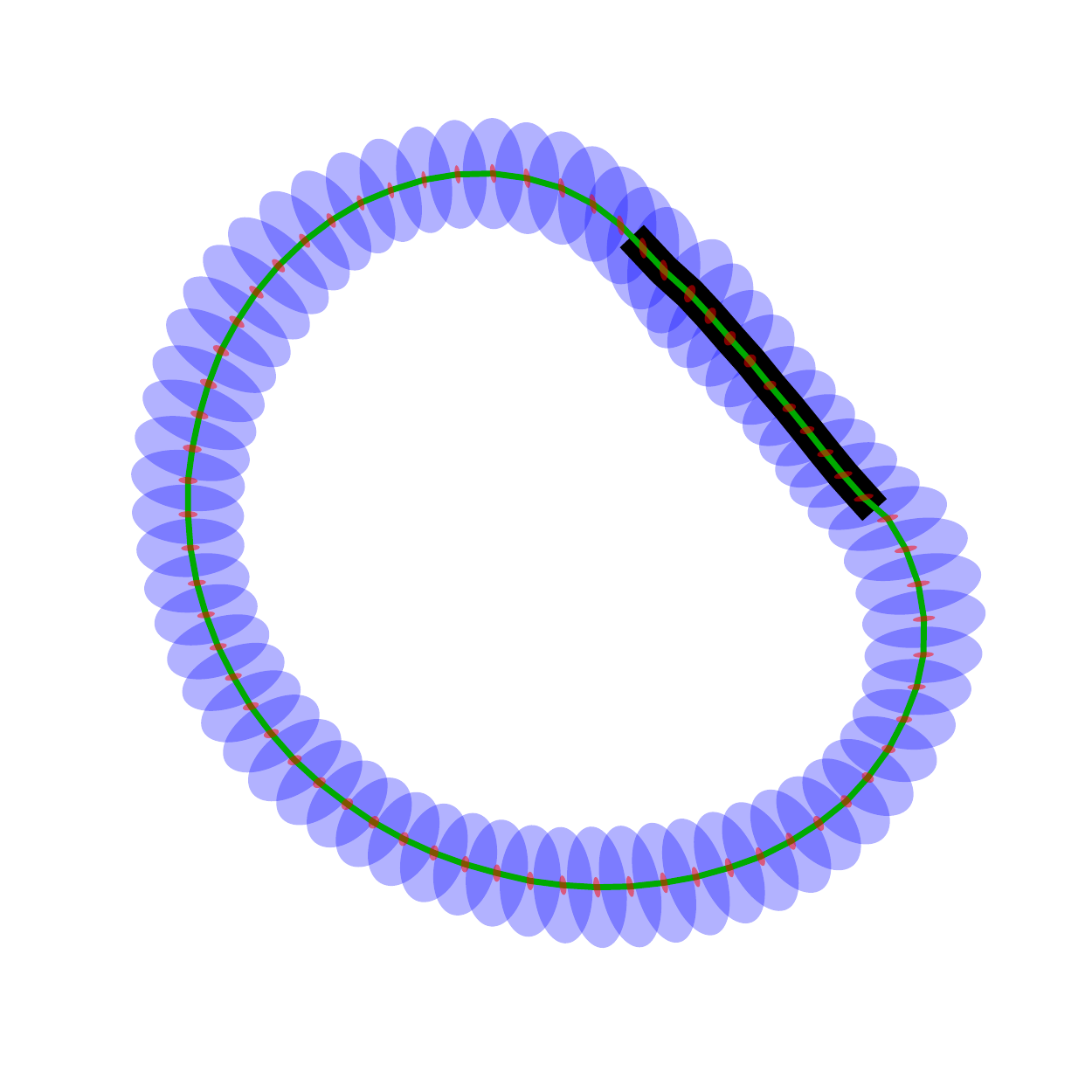}}
&
\raisebox{-.5\height}{\includegraphics[width=\figureSize \linewidth]{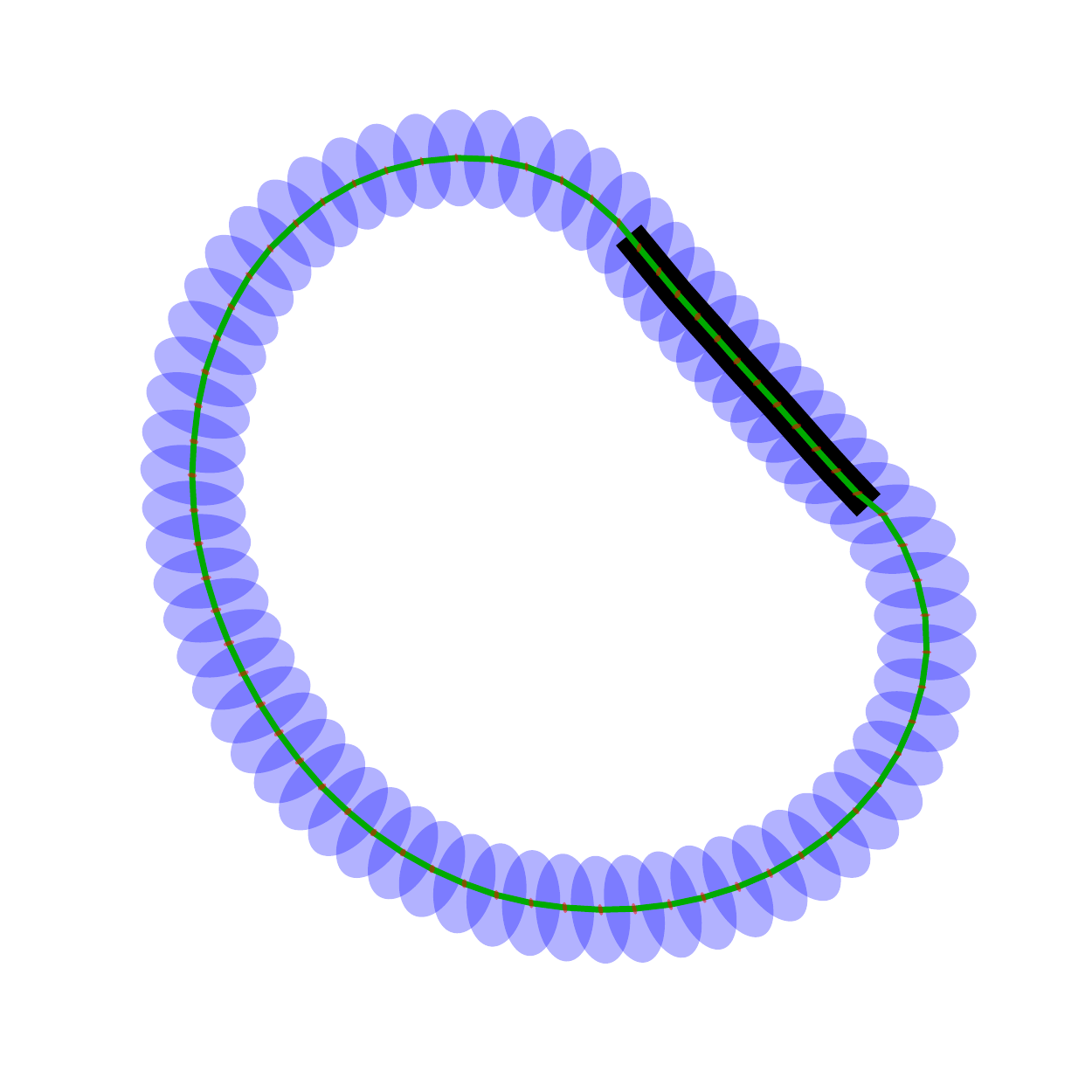}}
&
\raisebox{-.5\height}{\includegraphics[width=\figureSize \linewidth]{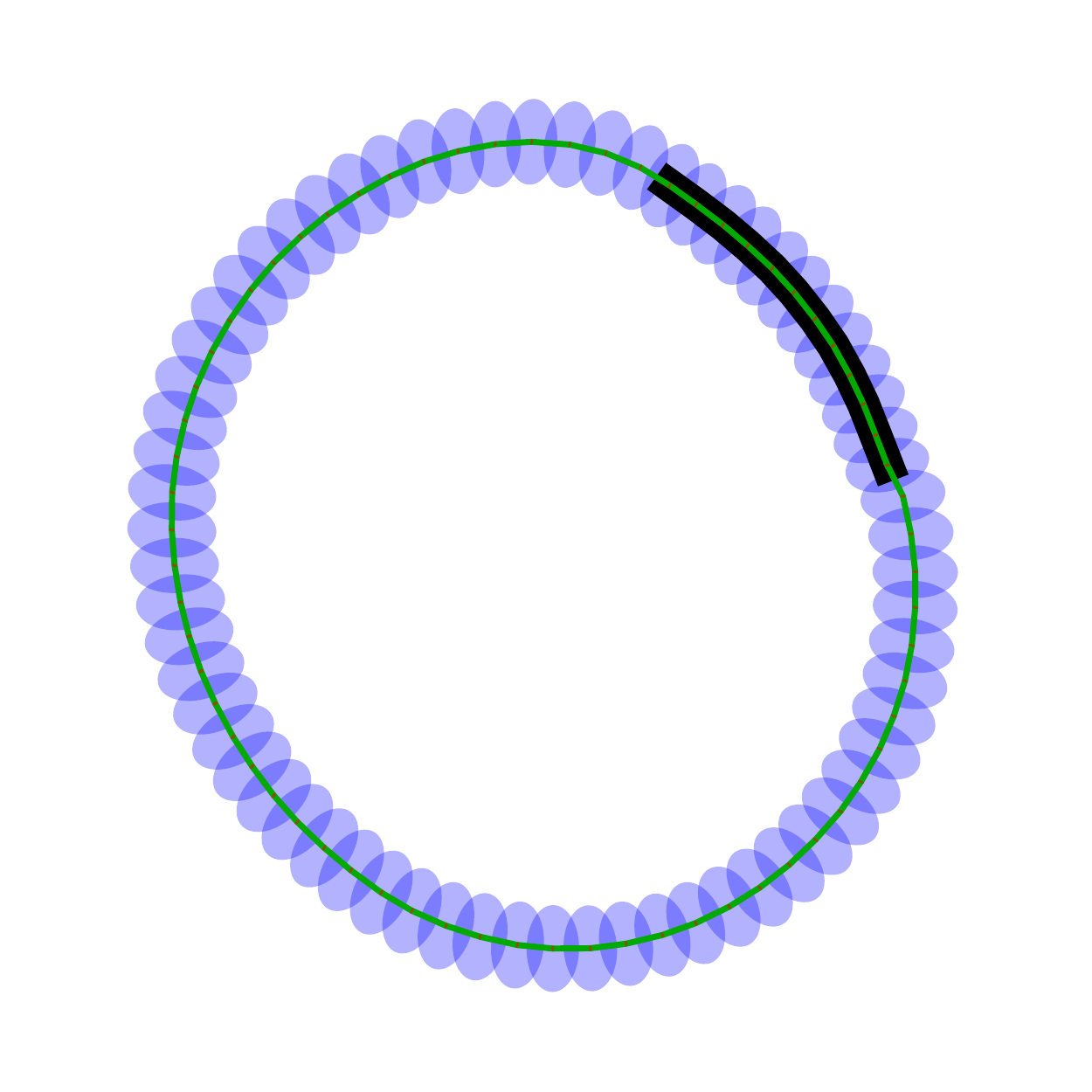}}
&
\raisebox{-.5\height}{\includegraphics[width=\figureSize \linewidth]{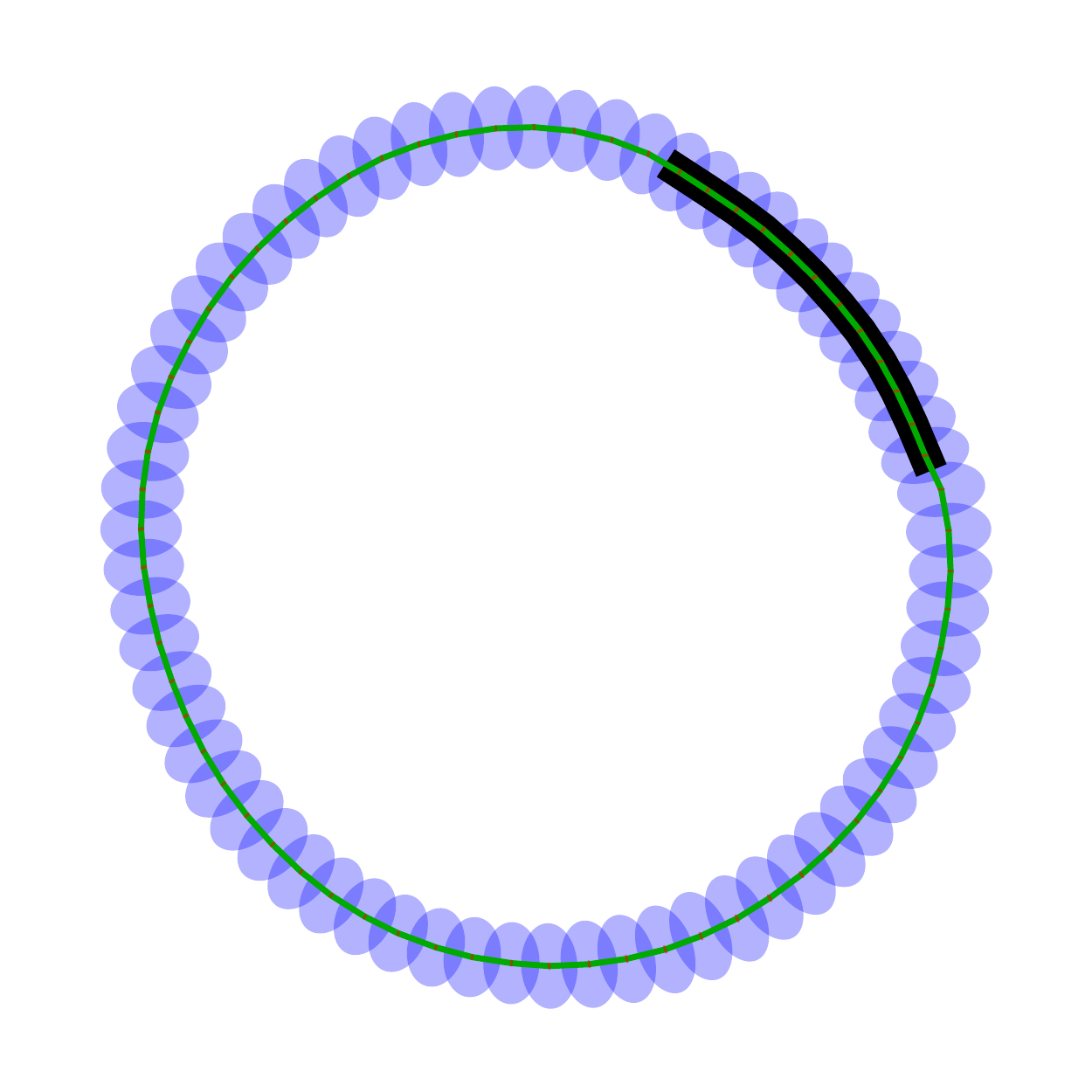}}
&
\\
\cline{1-7}
\end{tabular*}
\caption{Mean micelle shapes, thermal fluctuations, and errors in mean micelle shapes (illustrated in the manner of \cref{fig:baseCaseAverage}) as a function of the number of core beads $\nc$. 
As the number of core beads increases, the shapes become more circular, and, as illustrated by the size decrease of the blue ellipses and the fluctuations decrease.
}
\label{tab:coreShapes}
\end{table*}
\Cref{tab:coreShapes} shows the average shapes, fluctuations, and uncertainties in the average shapes resulting from varying the number of core beads $\nc$.
It is apparent from these results that the effect of increasing $\nc$ is to make the shapes more circular and decrease their fluctuations.
The character of these trends can be studied more precisely by plotting the shape features $\CR$ and $\delta$ against $\nc$, which we do in \cref{fig:corePlots}.
In this figure, it is clear that the shape features generally follow the trend apparent from \cref{tab:coreShapes}, and, excluding the largest values of $\nc$, the dependence of the shape features on $\nc$ is roughly linear.
The largest two values of $\nc$ indicate smaller slopes than the other data, consistent with the expectation that for large $\nc$, the curvature ratio should approach unity and the normalized fluctuations should go to zero.
Thus the data resulting from varying $\nc$ show a moderately sized domain of linearity.

\begin{figure*}
\subfloat[][]{%
\label{fig:nPhilCorePlots}%
\centering%
\includegraphics[width=.48\linewidth]{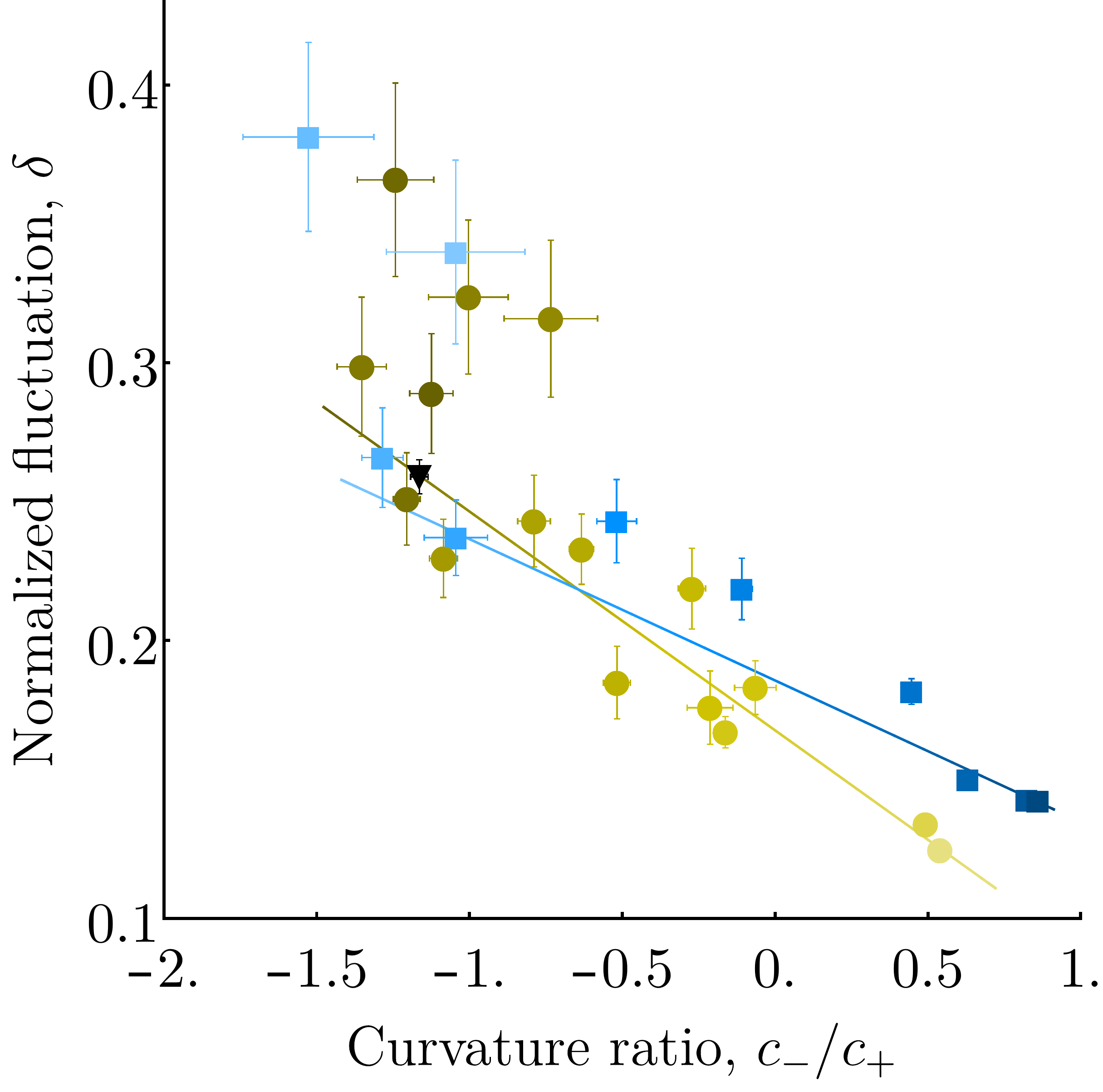}%
}\hfill%
\subfloat[][]{%
\label{fig:rPhilnPhoberPhobePlots}%
\centering%
\includegraphics[width=.48\linewidth]{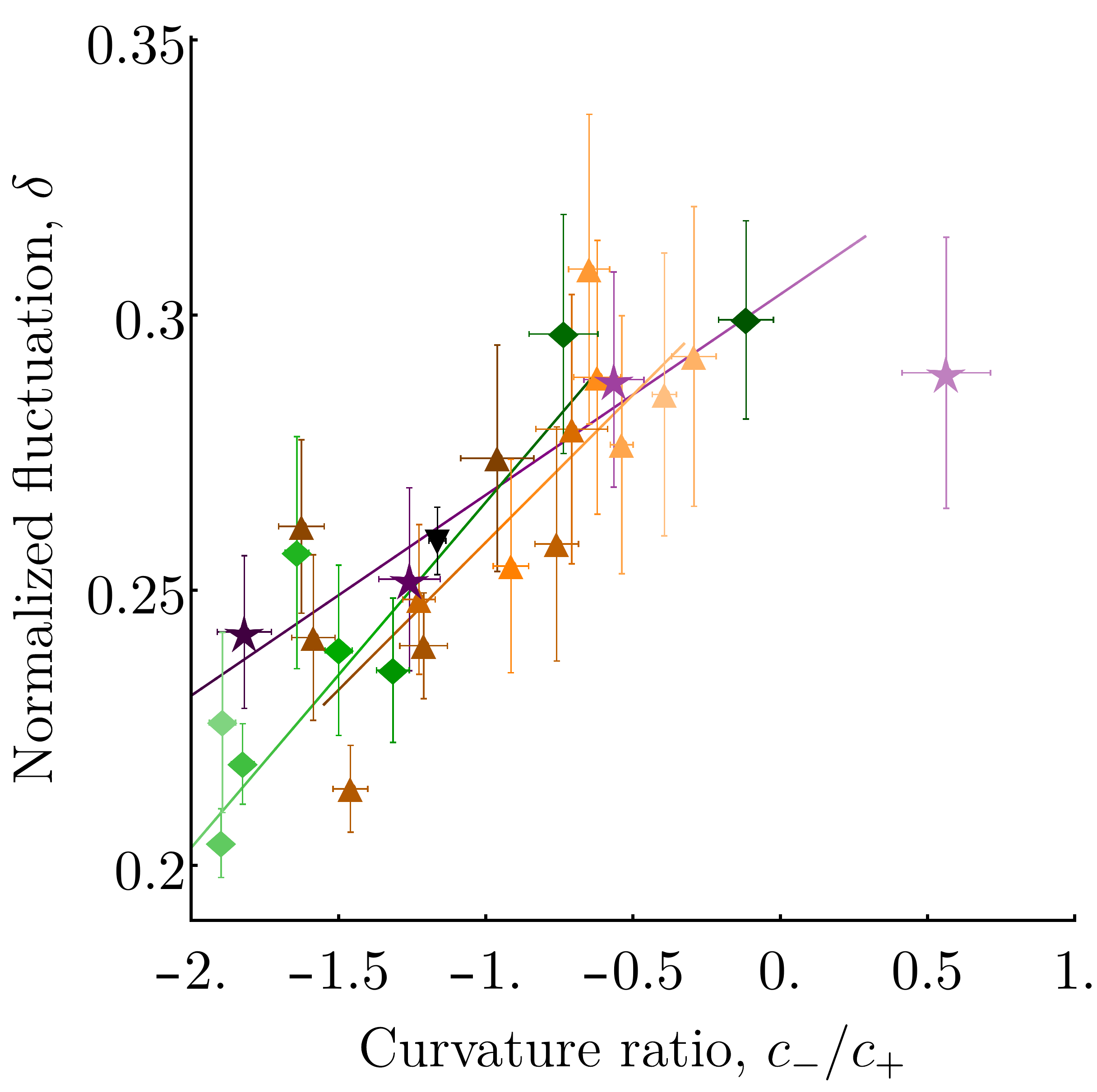}%
}

	\caption{
	Scatter plots showing curvature ratios and normalized fluctuations exhibited by micelles of several compositions.
	Each micelle composition has been obtained by changing one of the following parameters from the reference micelle value (the black downward-pointing triangle near $\CR=-1.2$, $\delta=0.26$): \protect\subref{fig:nPhilCorePlots} $\nc$ (yellow circles) or $\ni$ (blue squares); or \protect\subref{fig:rPhilnPhoberPhobePlots} $\no$ (orange upward-pointing triangles), $\ro$ (purple stars), or $\ri$ (green diamonds).
	The lightness of the data point represents the value of the parameter, with lighter points indicating larger values of the parameter in the manner of \cref{fig:corePlots}.
	To guide the eye and to assess linearity, we plot linear fits of the shape feature dependence on each of the five micelle composition parameters.
	To reduce crowding of the data, the five micelle compositions are partitioned into two plots according to the slope of the data resulting from varying the parameter.
	}
	\label{fig:combinedResults}
\end{figure*}

Next we present in \cref{fig:combinedResults} the dependences of the shape features on each of the five micelle composition parameters.
There are two types of trends that result from varying a micelle composition parameter: the first type of trend, shown in \cref{fig:nPhilCorePlots}, is where the fluctuation increases as the dimple becomes more pronounced (i.e., curvature ratio becomes more negative); the second type of trend, shown in \cref{fig:rPhilnPhoberPhobePlots}, involves the opposite relationship between the shape features, with the fluctuation instead decreasing as the dimple becomes more pronounced.
The first type of trend results from varying $\nc$ and $\ni$, while the second type of trend results from varying $\ro$, $\ri$, and $\no$.

We now propose an explanation for why varying $\nc$ and $\ni$ (the shapes resulting from varying $\ni$ are shown in \cref{tab:nPhilShapes}) both cause the normalized fluctuation and the strength of the dimple to respond in the same direction.
Since increasing $\nc$ tends to increase the size and therefore the perimeter of the micelle, and decreasing $\ni$ decreases the number of diblocks on the micelle perimeter, either of these changes tends to decrease the density of diblocks on the micelle surface.
As the surface density of the diblocks is decreased, we expect their surfactant-like effect to be reduced so that the surface tension of the micelle would increase.
This surface tension increase should have two effects.
The first effect is to reduce fluctuations in the micelle shape, and the second effect is to make the micelle shape more circular, reducing the strength of the dimple.
Thus we expect that increasing $\nc$ or decreasing $\ni$ both decreases the fluctuations (i.e., decreases $\delta$) and decreases the strength of the dimple (i.e., makes $\CR$ more positive).
By this reasoning, changing either $\nc$ or $\ni$ would cause $\CR$ and $\delta$ to change in opposite directions, consistent with the negative slope in \cref{fig:corePlots}.
\begin{table}
	\newcommand\T{\rule{0pt}{2.6ex}}
	\begin{tabular}{|c|cccccc|}
		\hline 
		$\nih$ \T& 45\% &55\% &64\% &73\% &82\% &91\%\\
		&
		\raisebox{-.5\height}{\includegraphics[width=\figureSize \linewidth]{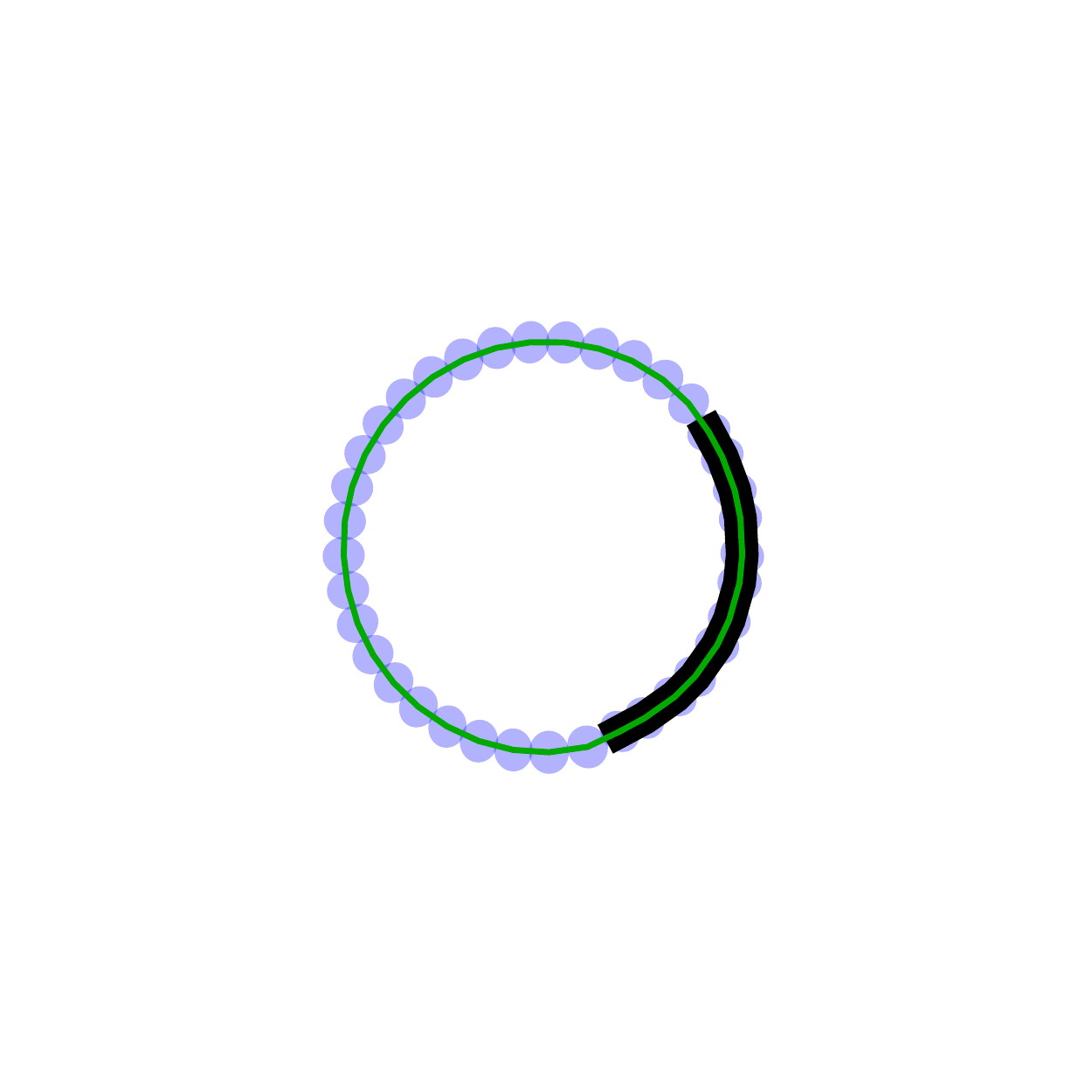}}
		&
		\raisebox{-.5\height}{\includegraphics[width=\figureSize \linewidth]{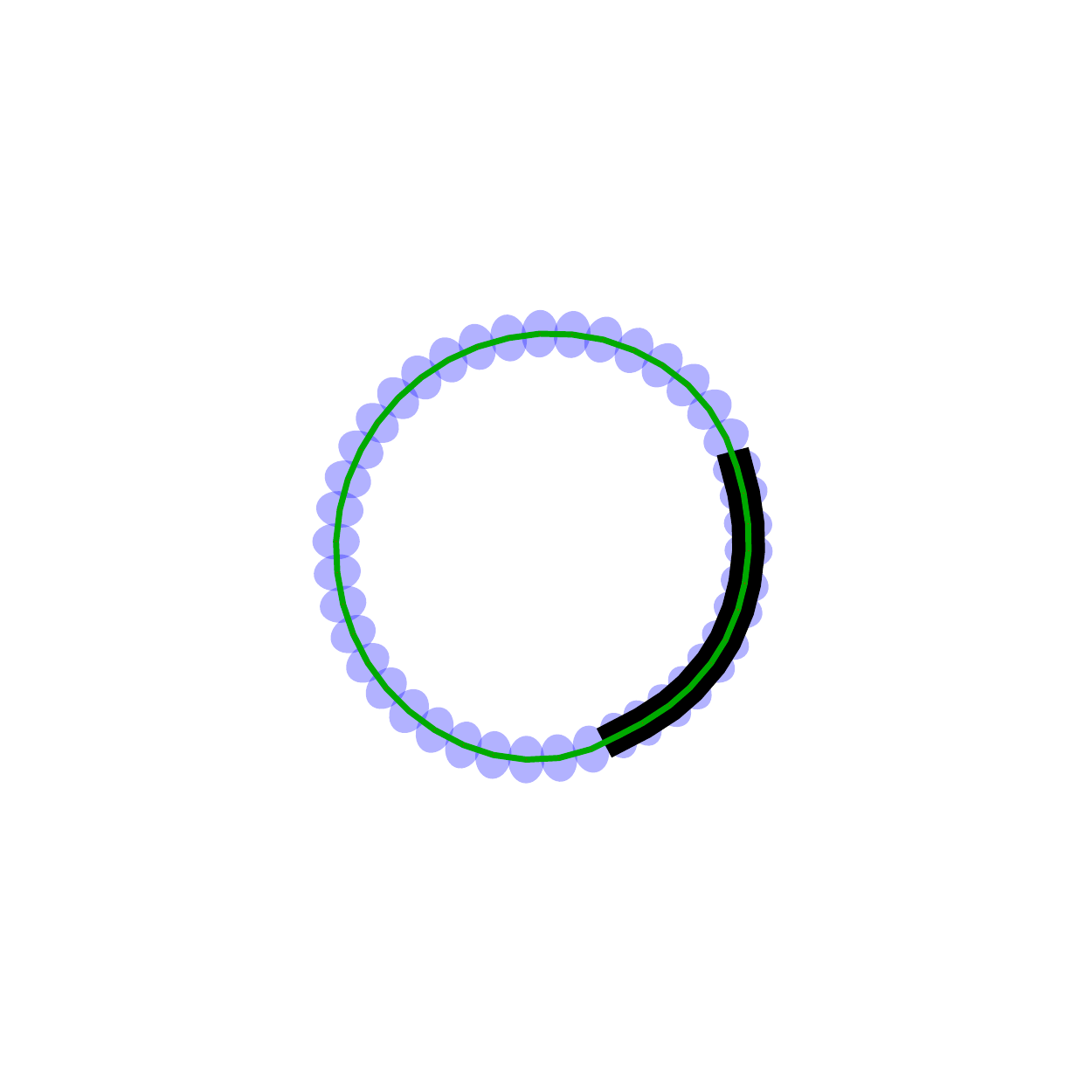}}
		&
		\raisebox{-.5\height}{\includegraphics[width=\figureSize \linewidth]{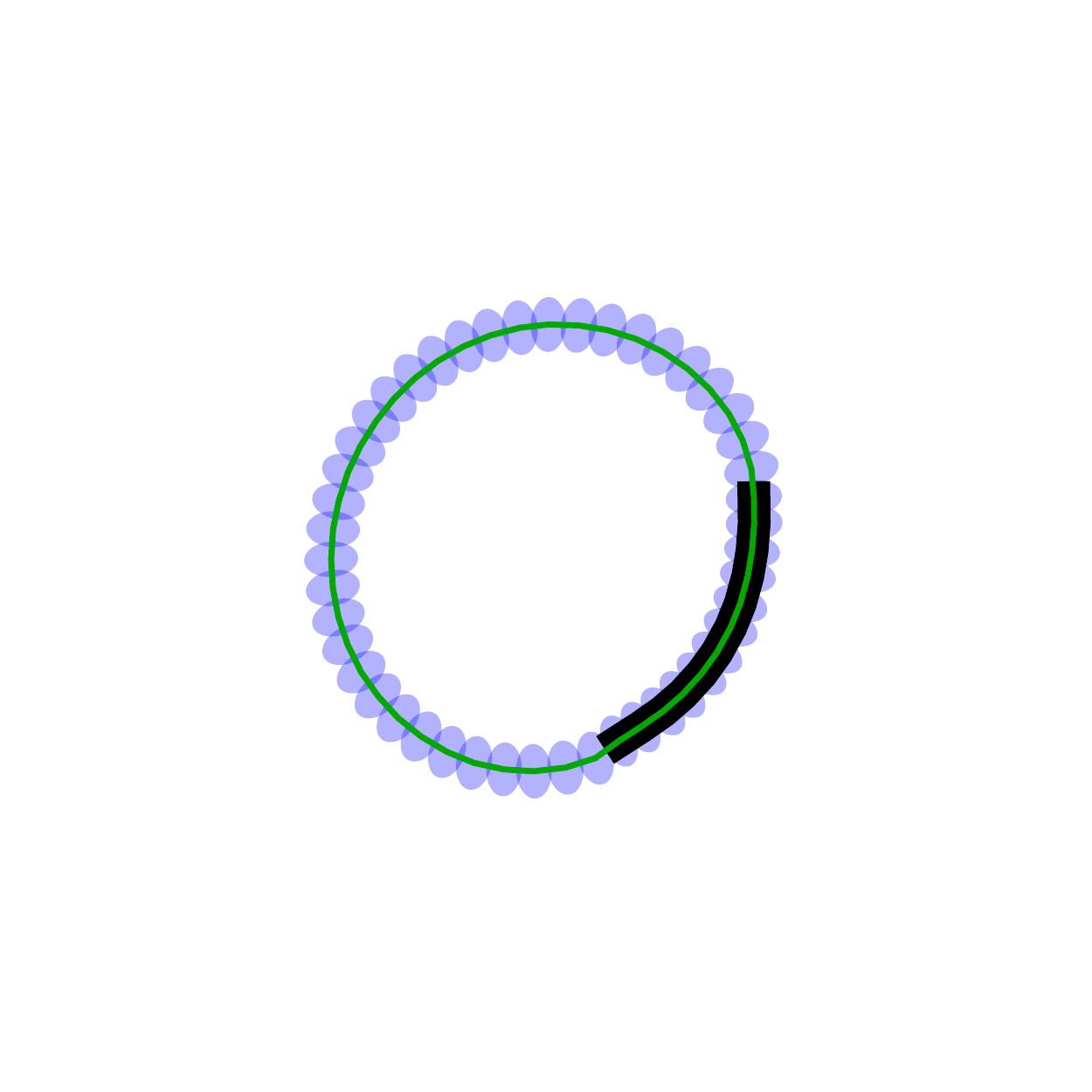}}
		&
		\raisebox{-.5\height}{\includegraphics[width=\figureSize \linewidth]{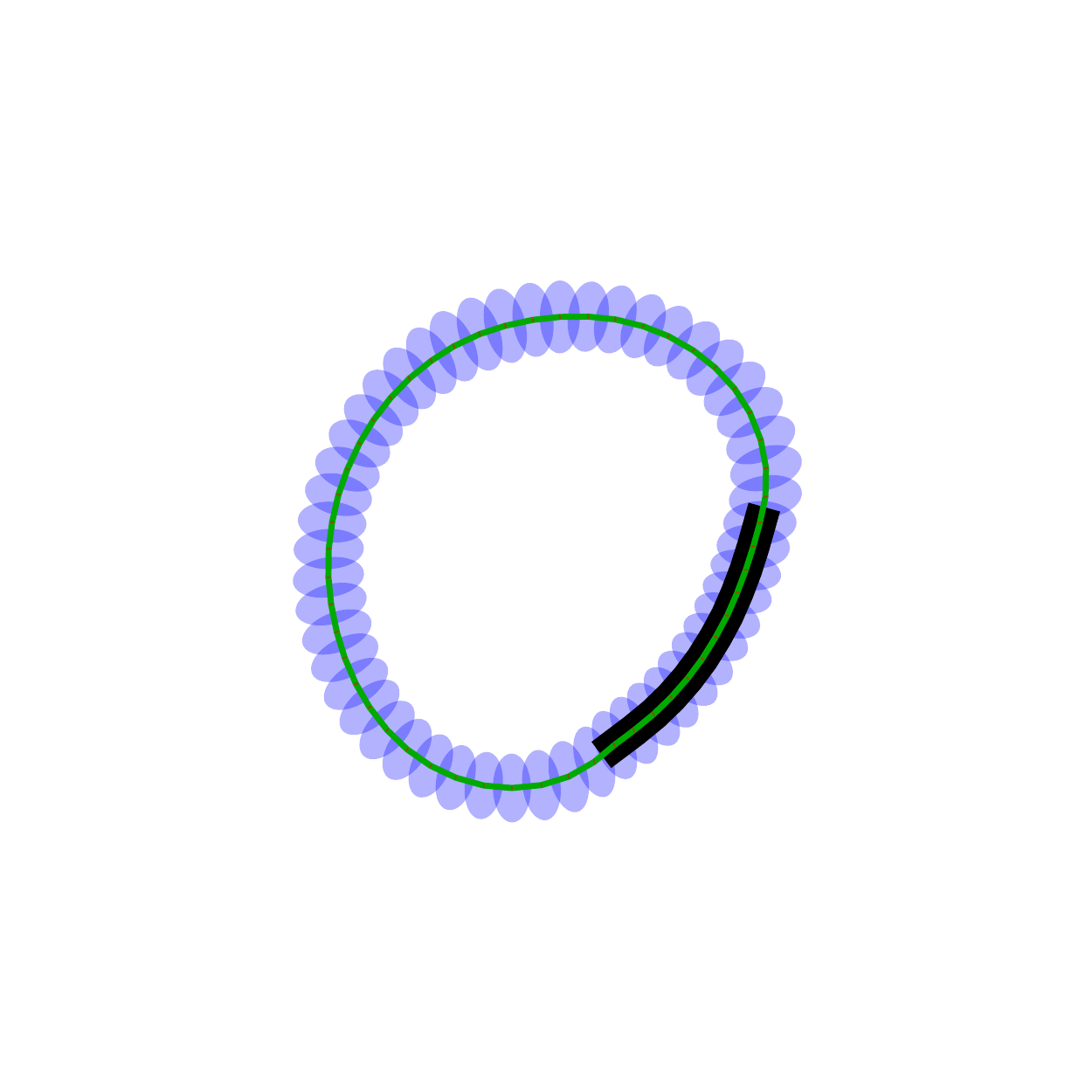}}
		&
		\raisebox{-.5\height}{\includegraphics[width=\figureSize \linewidth]{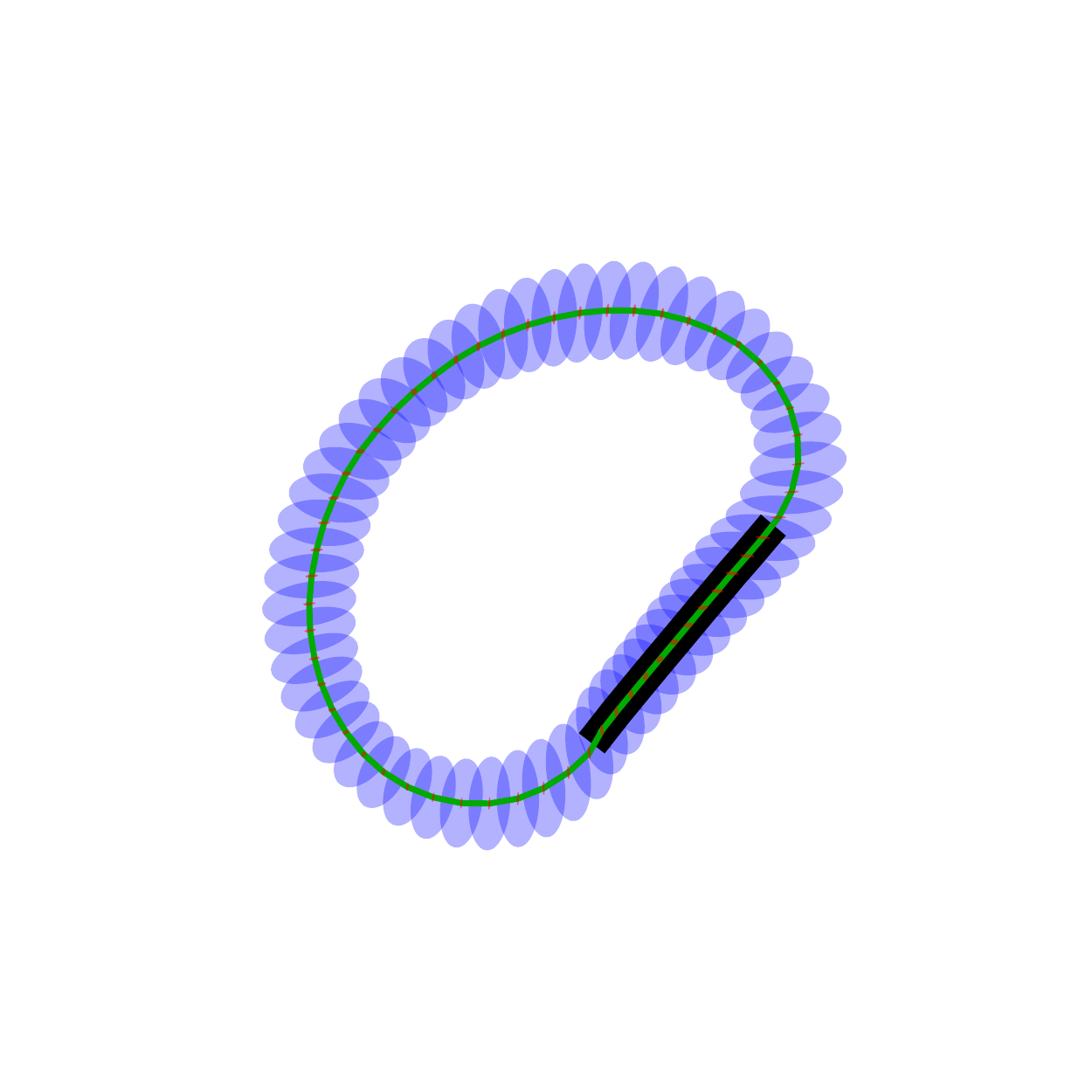}}
		&
		\raisebox{-.5\height}{\includegraphics[width=\figureSize \linewidth]{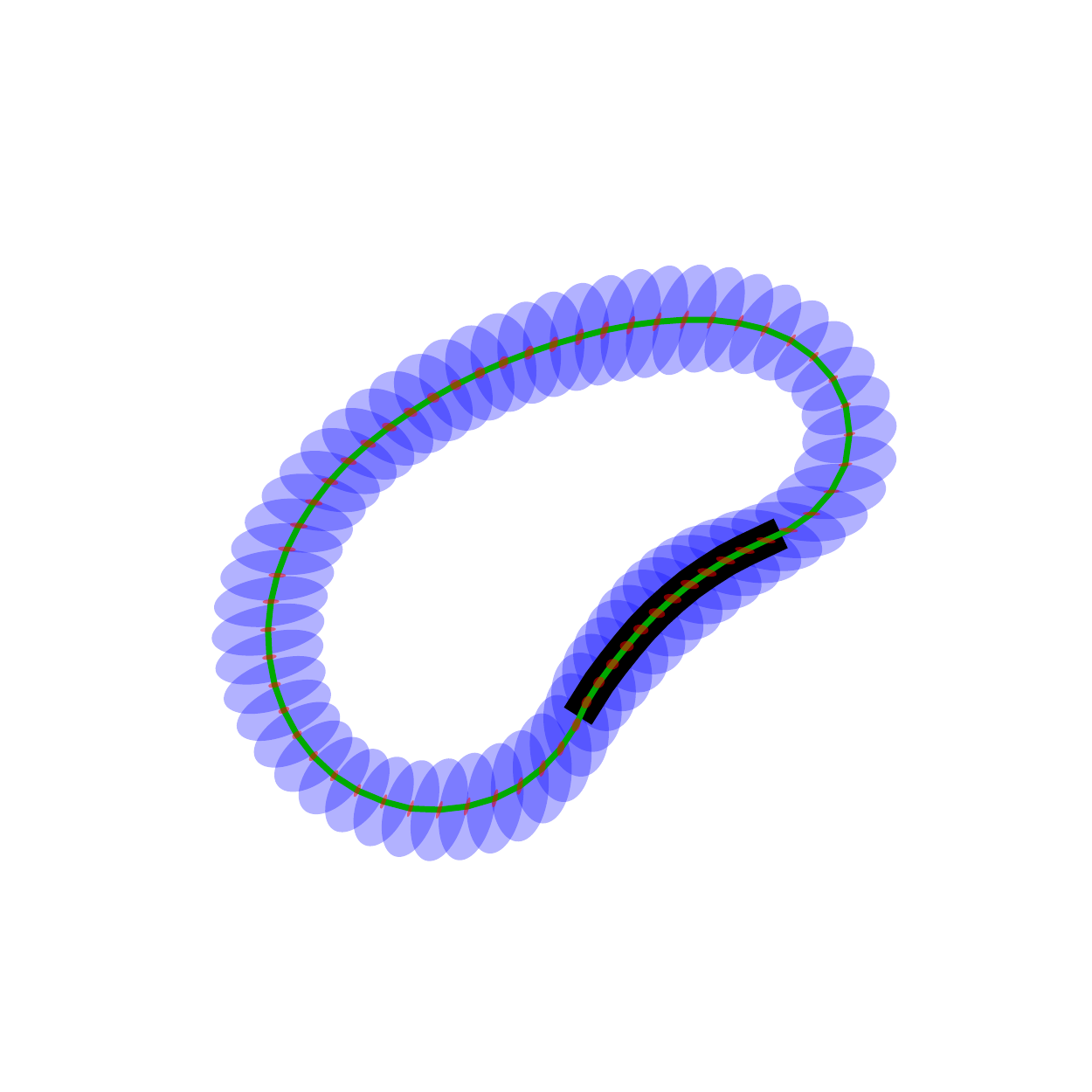}}\\
		\hline
		$\nih$ \T& 100\% &109\% &118\% &127\%&136\%& \\
		&
		\raisebox{-.5\height}{\includegraphics[width=\figureSize \linewidth]{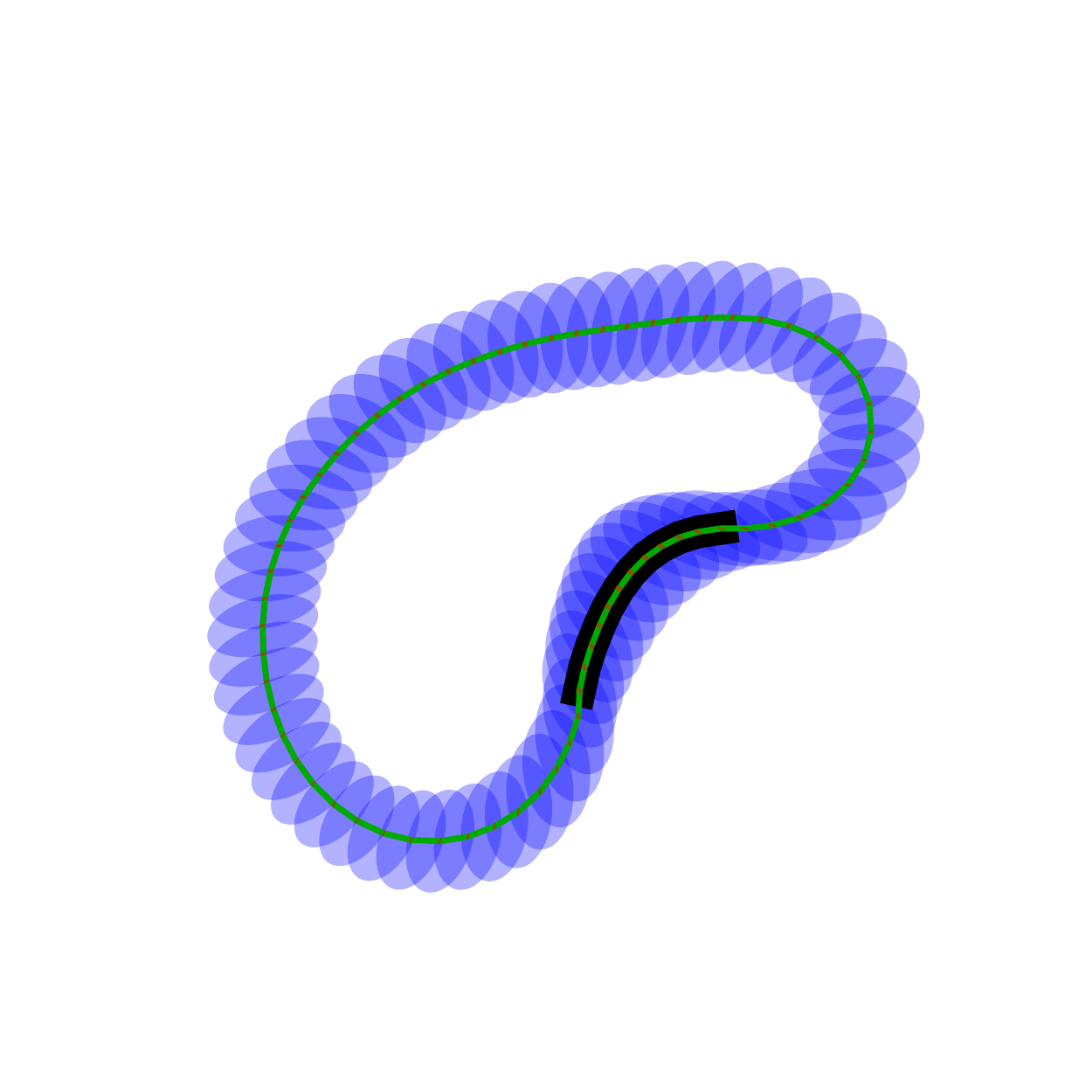}}
		&
		\raisebox{-.5\height}{\includegraphics[width=\figureSize \linewidth]{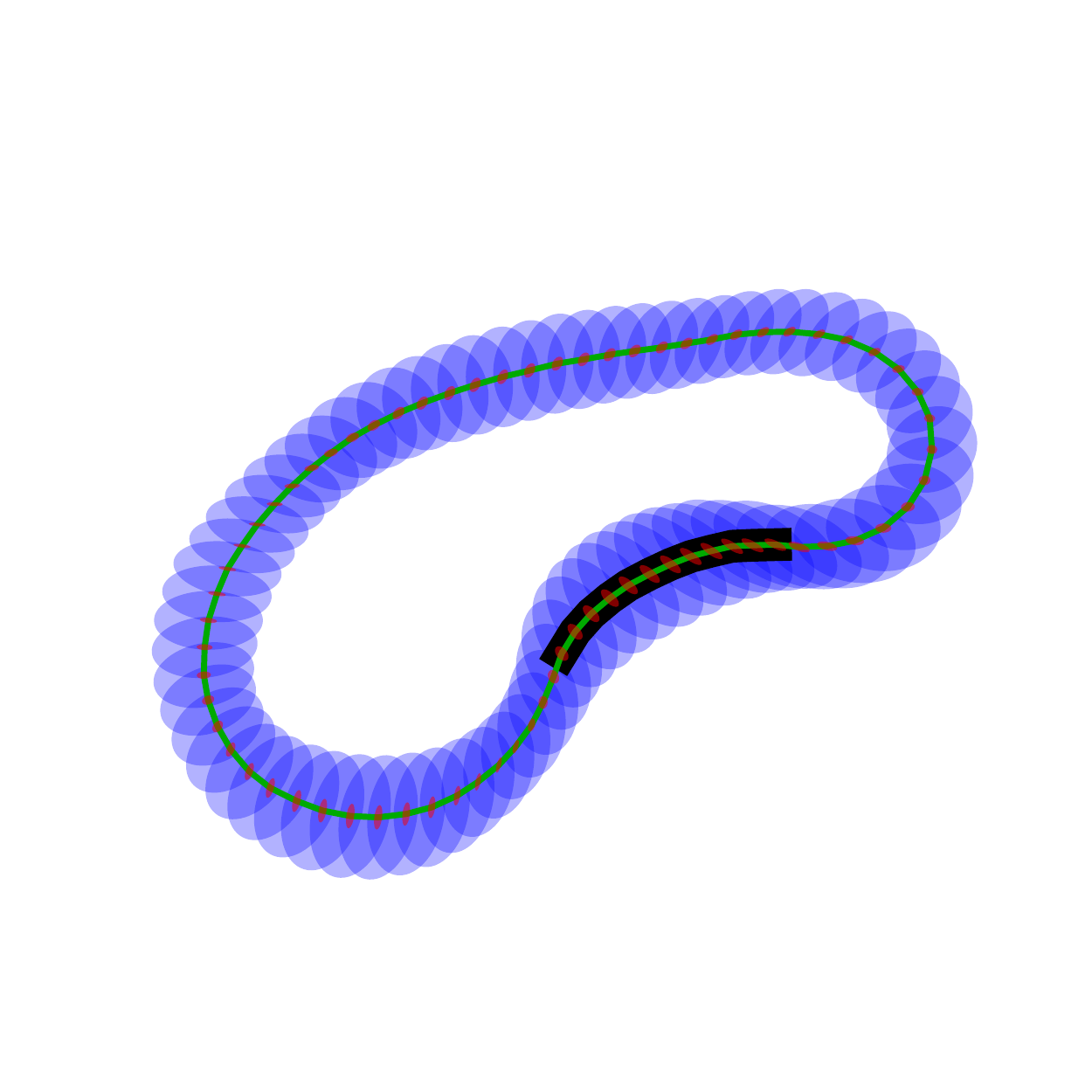}}
		&
		\raisebox{-.5\height}{\includegraphics[width=\figureSize \linewidth]{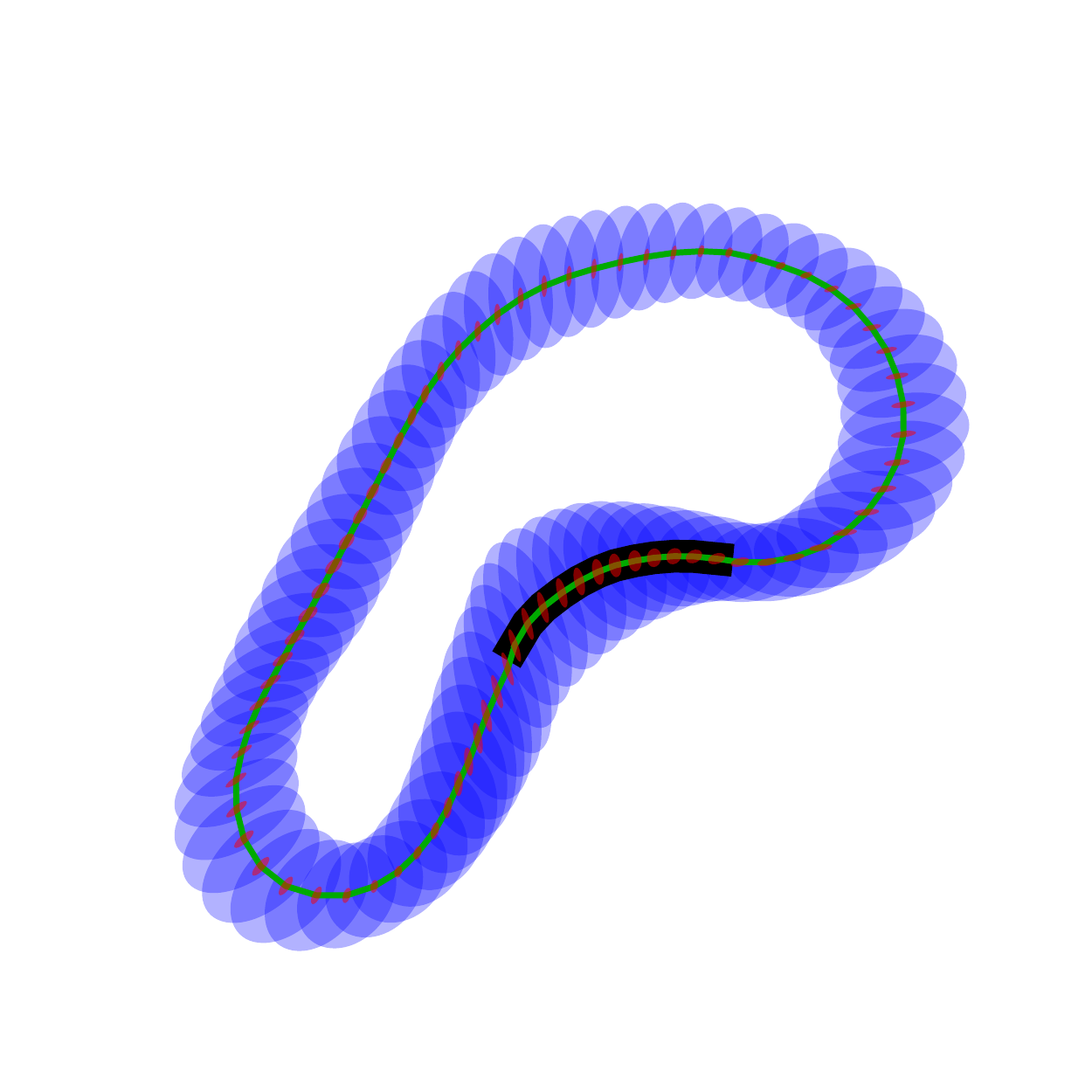}}
		&
		\raisebox{-.5\height}{\includegraphics[width=\figureSize \linewidth]{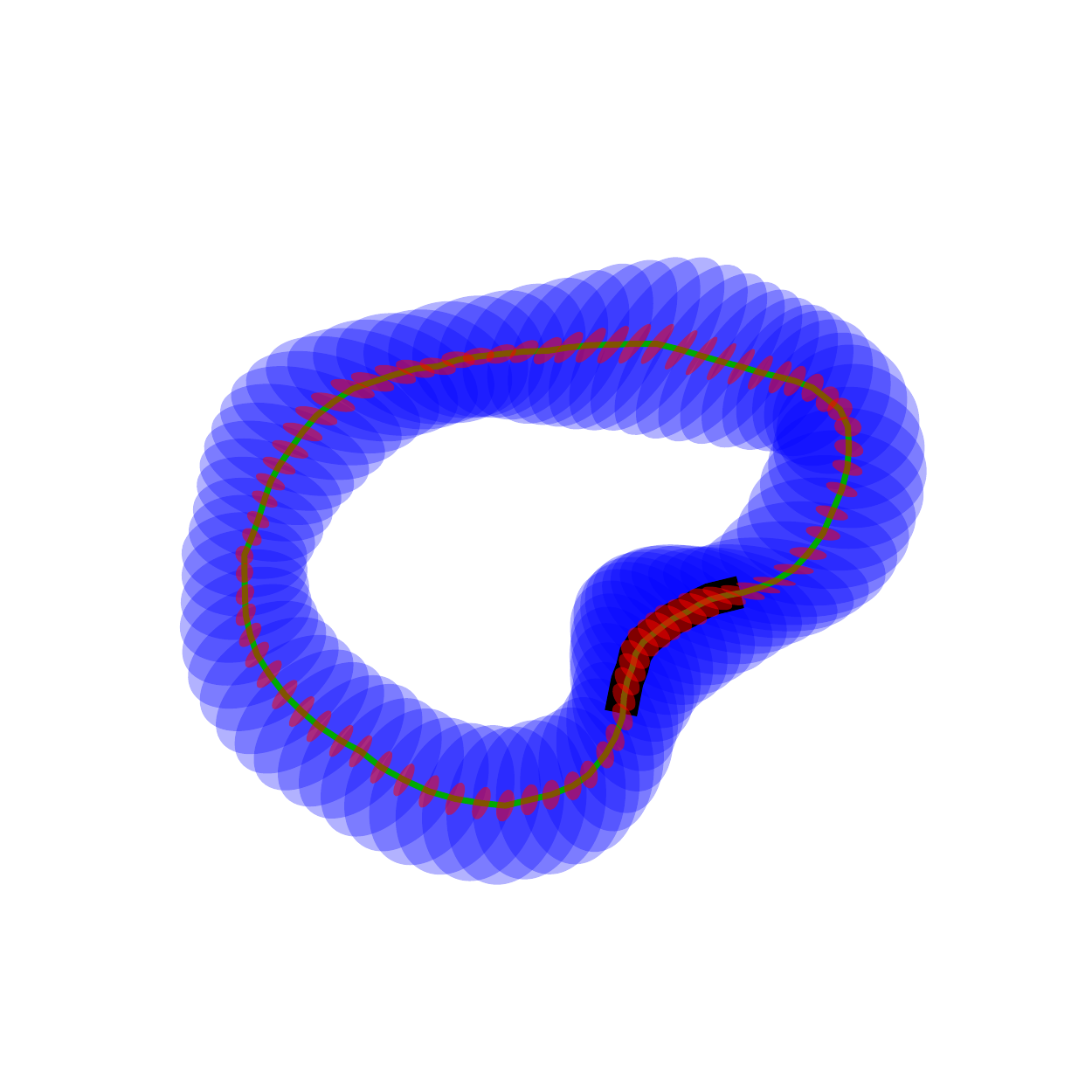}}
		&
		\raisebox{-.5\height}{\includegraphics[width=\figureSize \linewidth]{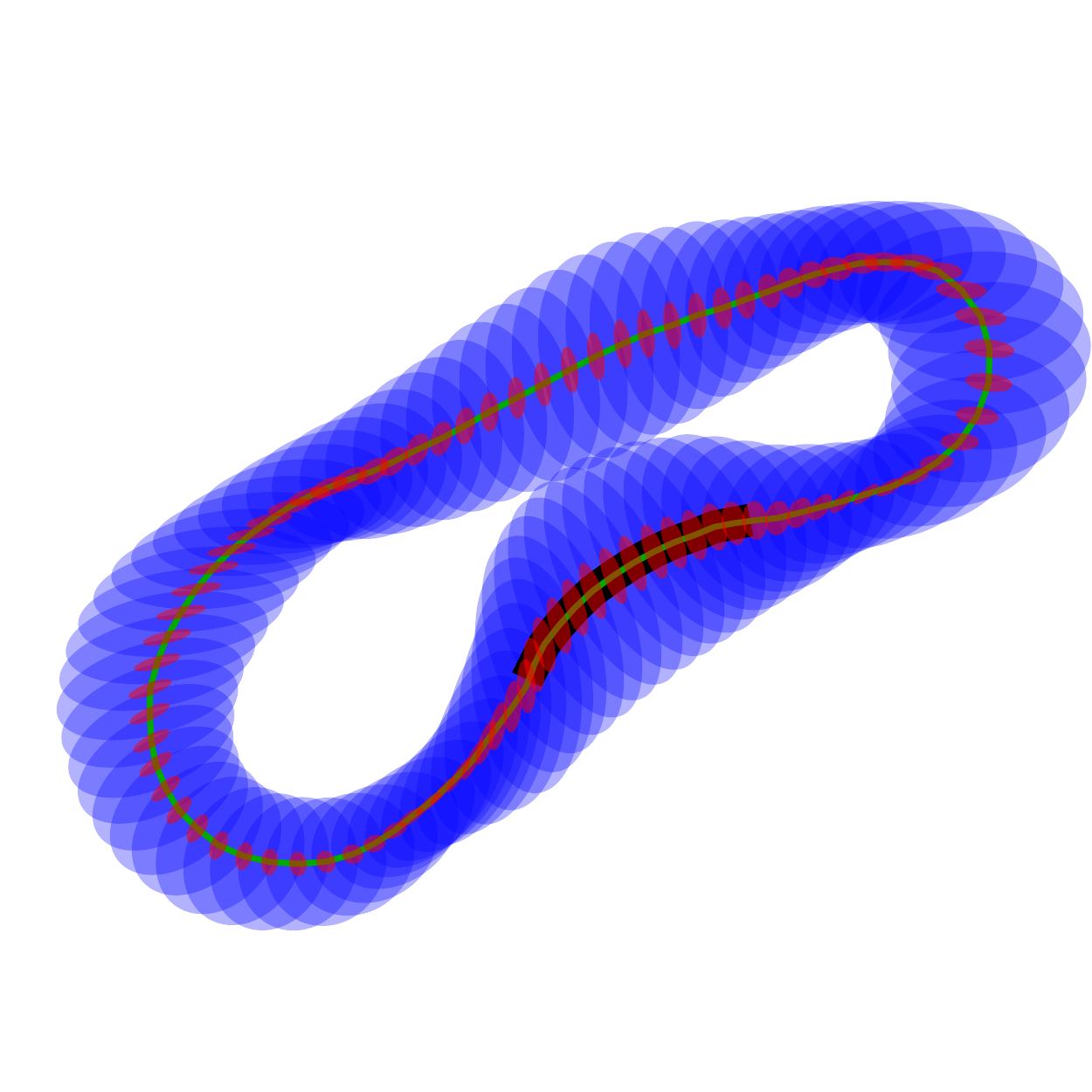}} &\\
		\hline
	\end{tabular} 
	\caption{Mean micelle shapes, thermal fluctuations, and errors in mean micelle shapes (illustrated in the manner of \cref{fig:baseCaseAverage}) as a function of $\nih$, the number of solvophilic-rich diblocks expressed as a percentage of the reference micelle value. 
	As the number of solvophilic-rich diblocks increases, the shapes become less circular and the fluctuations increase.}
	\label{tab:nPhilShapes}
\end{table}

Having discussed the two micelle composition parameters which affect the normalized fluctuation and the dimple strength in the same way, we now discuss the remaining three composition parameters, where the responses of the shape features are opposite to each other as shown in \cref{fig:rPhilnPhoberPhobePlots}.
First we propose explanations for the results of varying asymmetry ratio $\ro$ of the solvophobic-rich diblocks, those designed to sit at the micelle's dimple.
The resulting shapes are shown in \cref{tab:rPhobeShapes}.
\begin{table}
	\begin{tabular}{|c|cccccc|}
		\hline 
		$\roh$ & $74\%$ & $83\%$ & $ 91\% $ & $ 100\% $ & $ 109\% $ & $ 117\% $\\
		&
		\raisebox{-.5\height}{\includegraphics[width=\figureSize \linewidth]{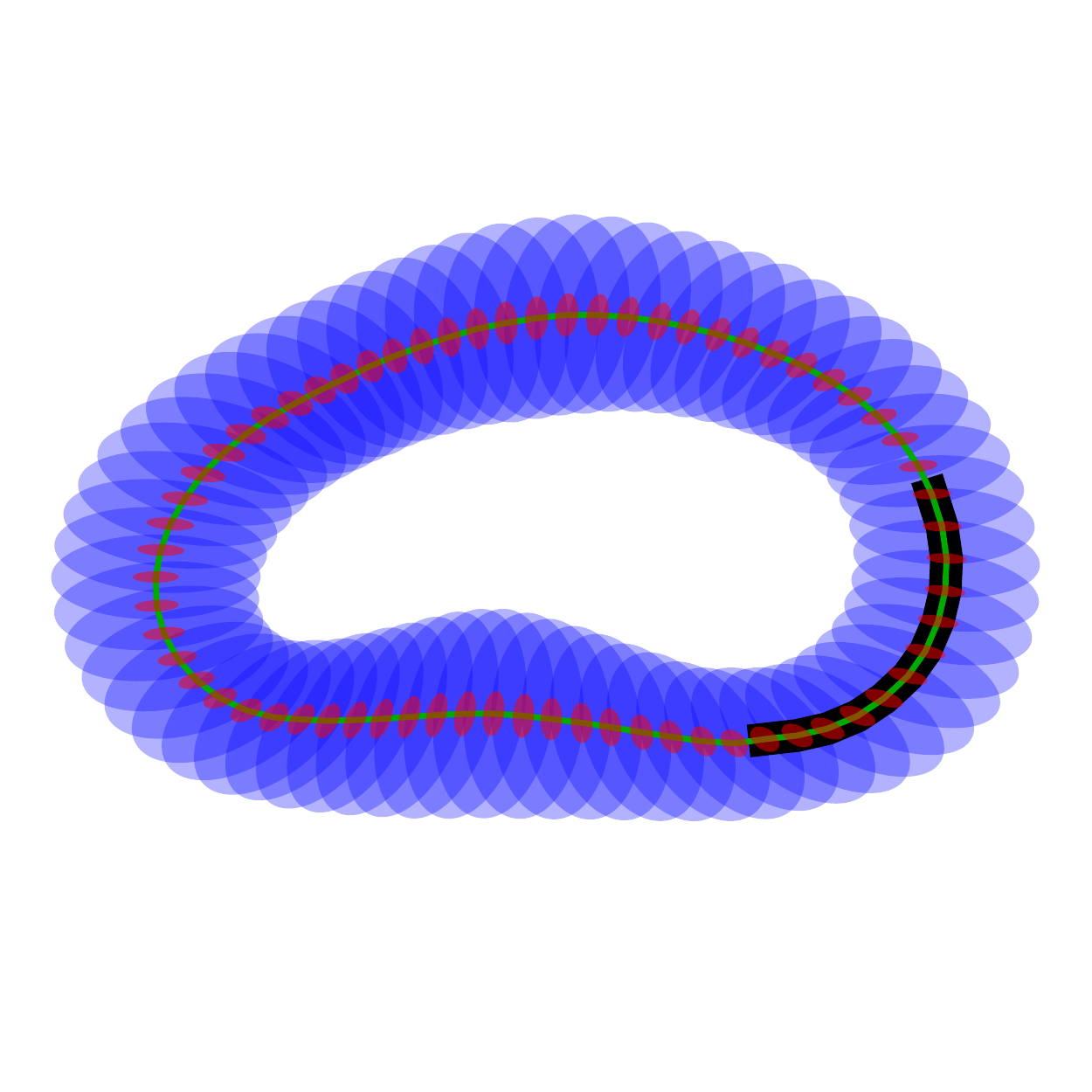}}
		&
		\raisebox{-.5\height}{\includegraphics[width=\figureSize \linewidth]{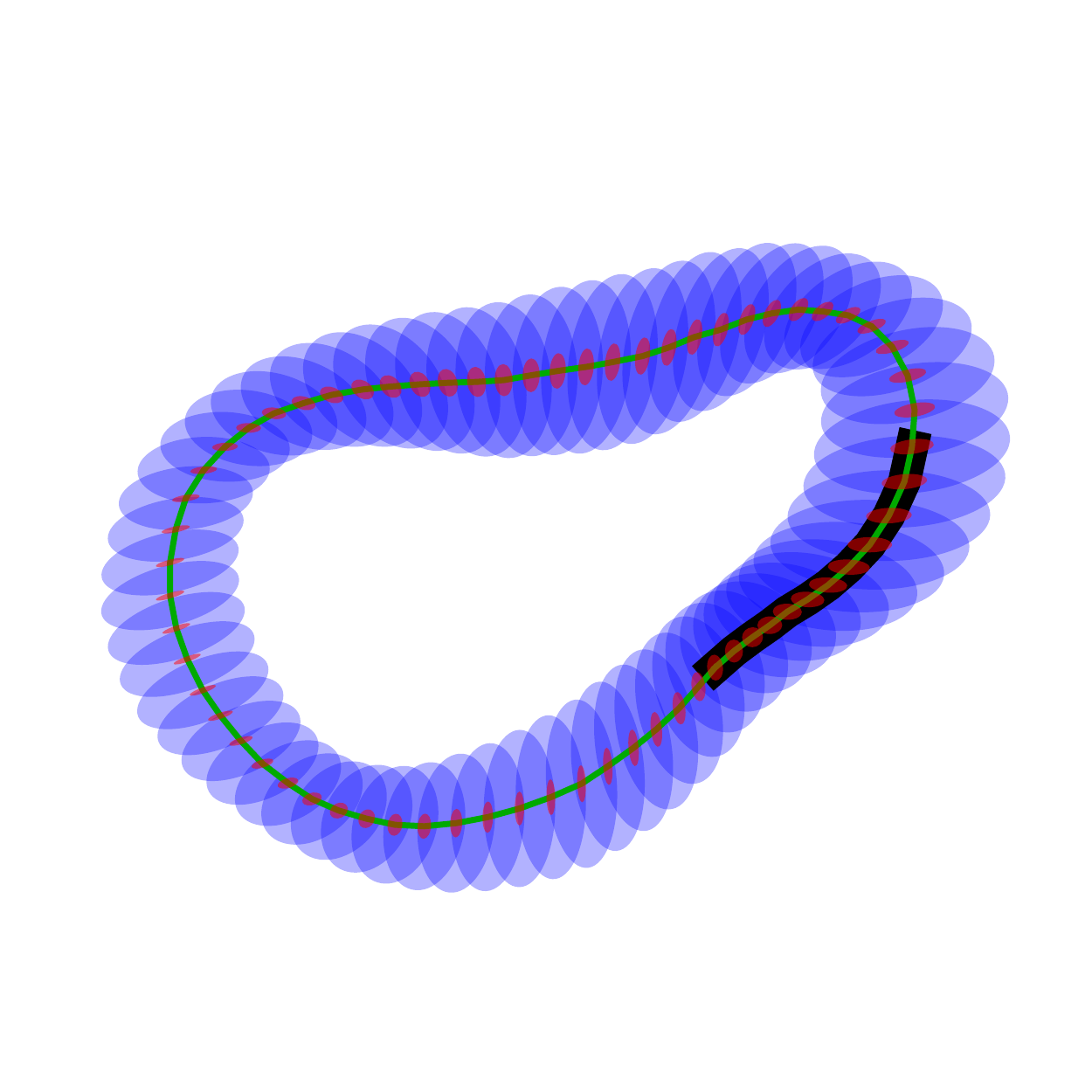}}
		&
		\raisebox{-.5\height}{\includegraphics[width=\figureSize \linewidth]{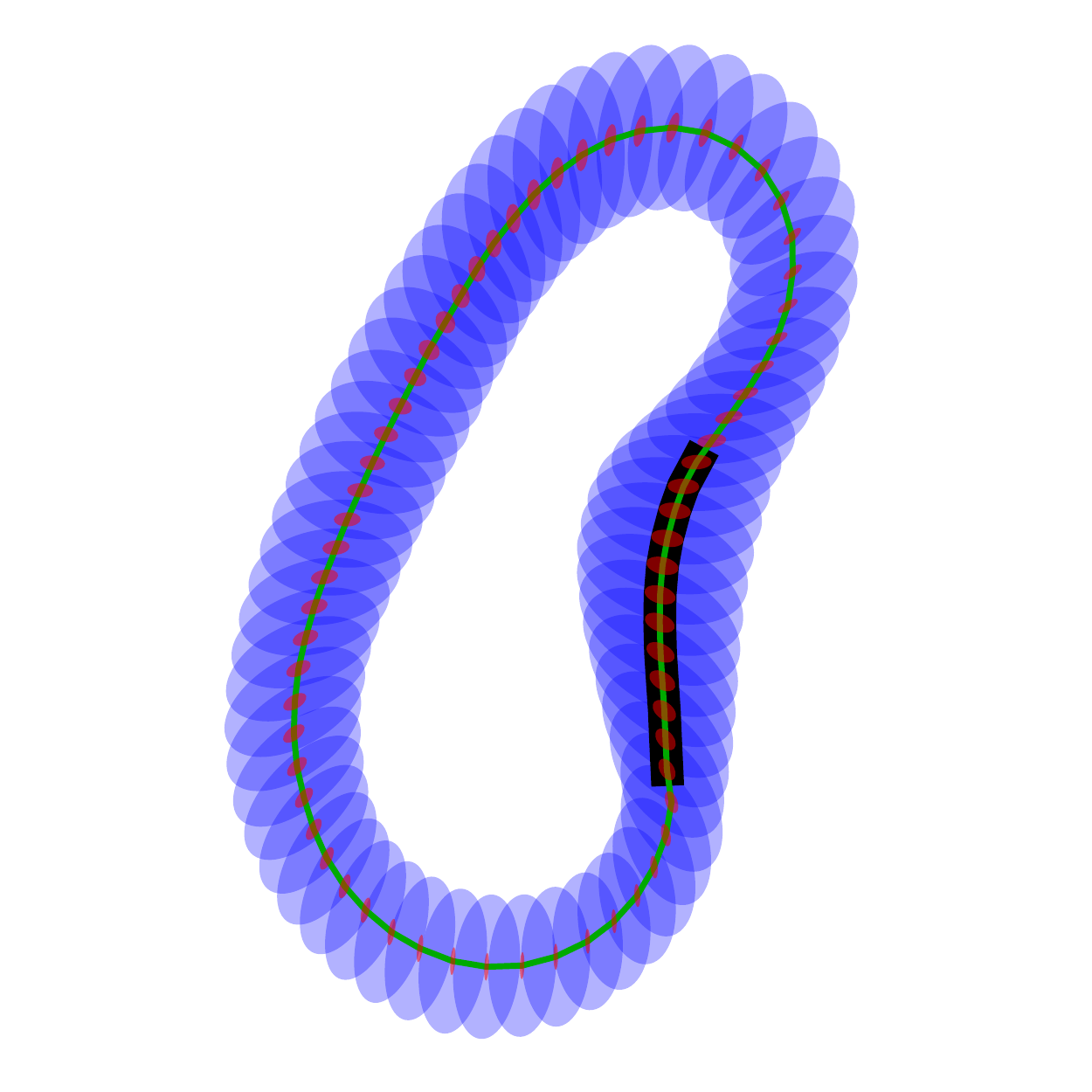}}
		&
		\raisebox{-.5\height}{\includegraphics[width=\figureSize \linewidth]{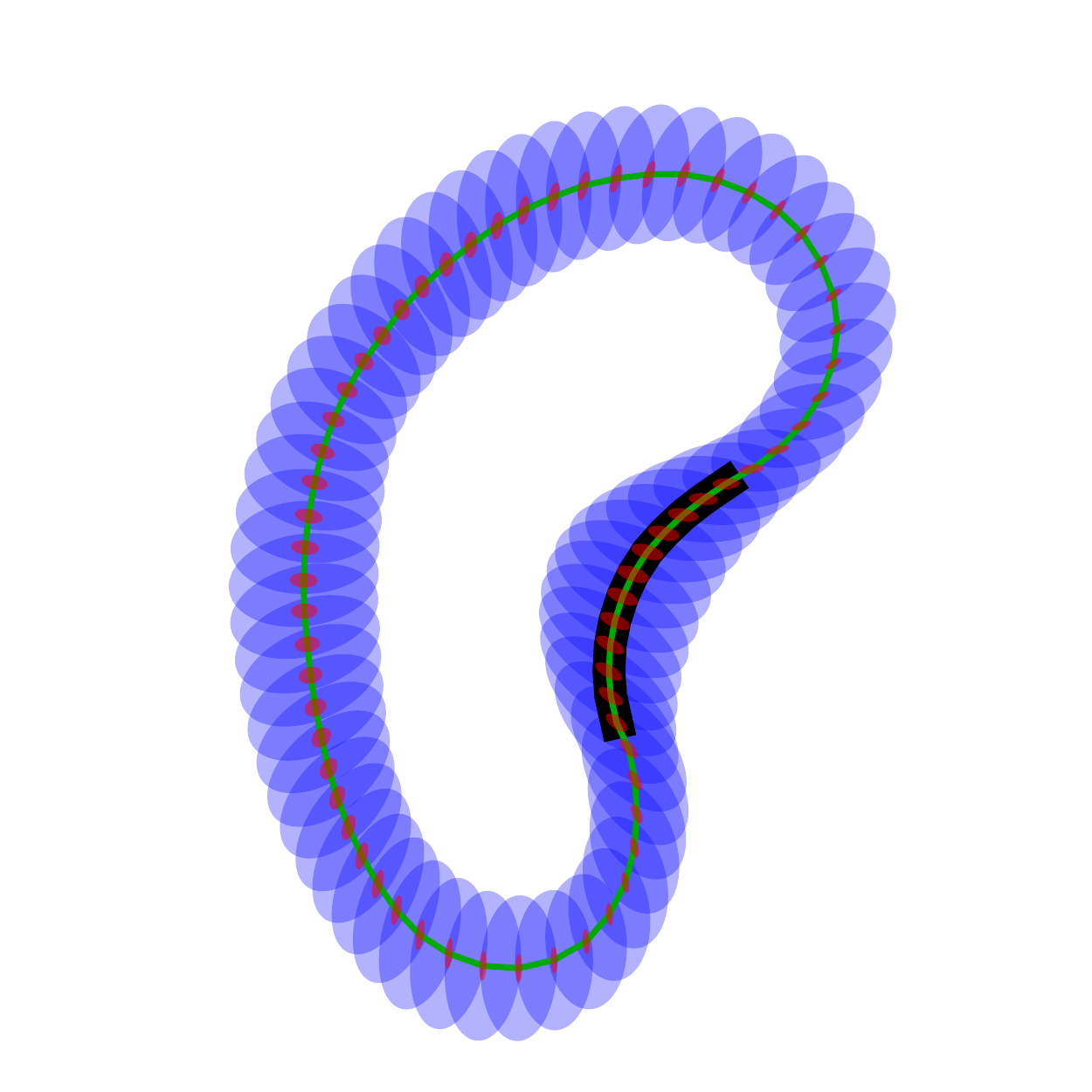}}
		&
		\raisebox{-.5\height}{\includegraphics[width=\figureSize \linewidth]{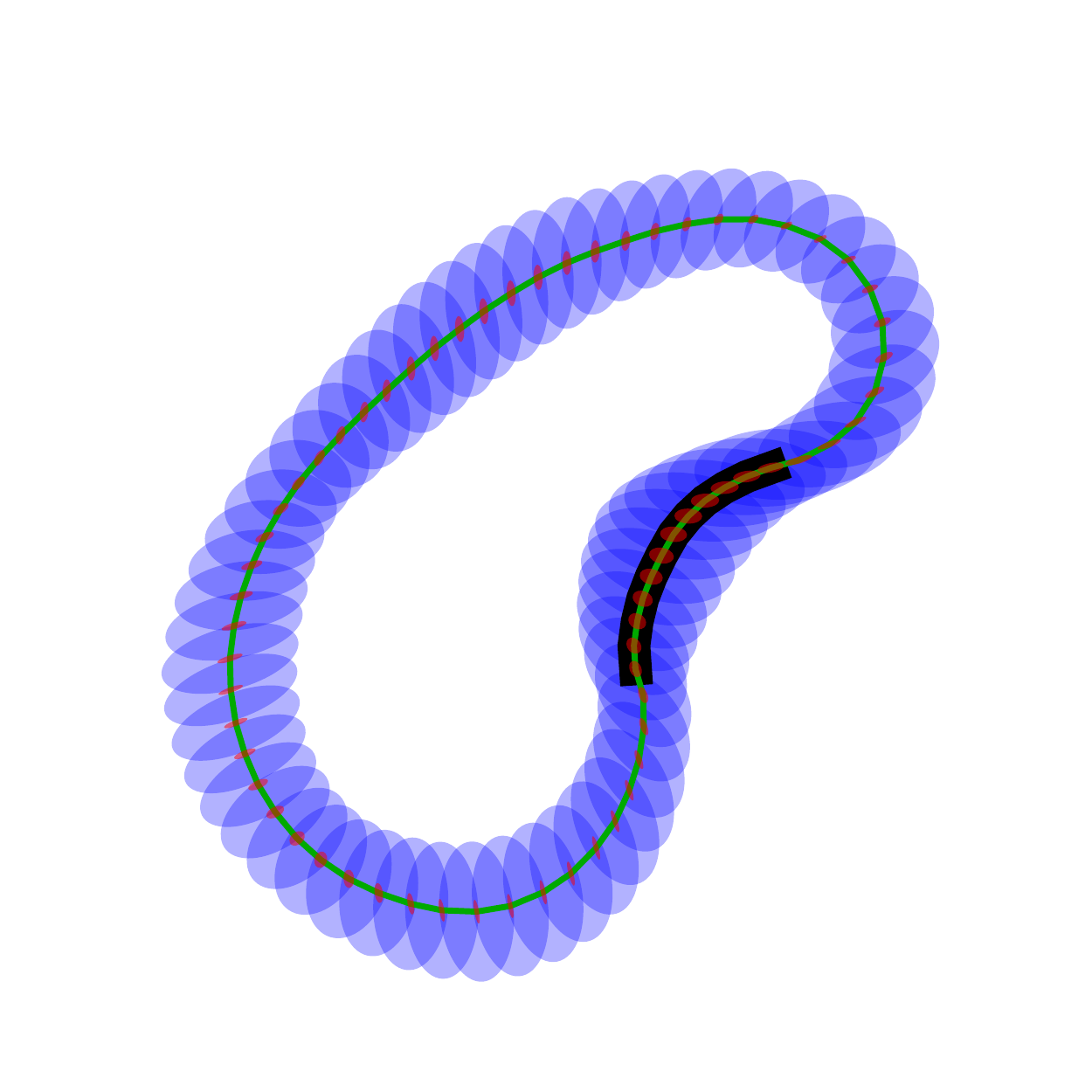}}
		&
		\raisebox{-.5\height}{\includegraphics[width=\figureSize \linewidth]{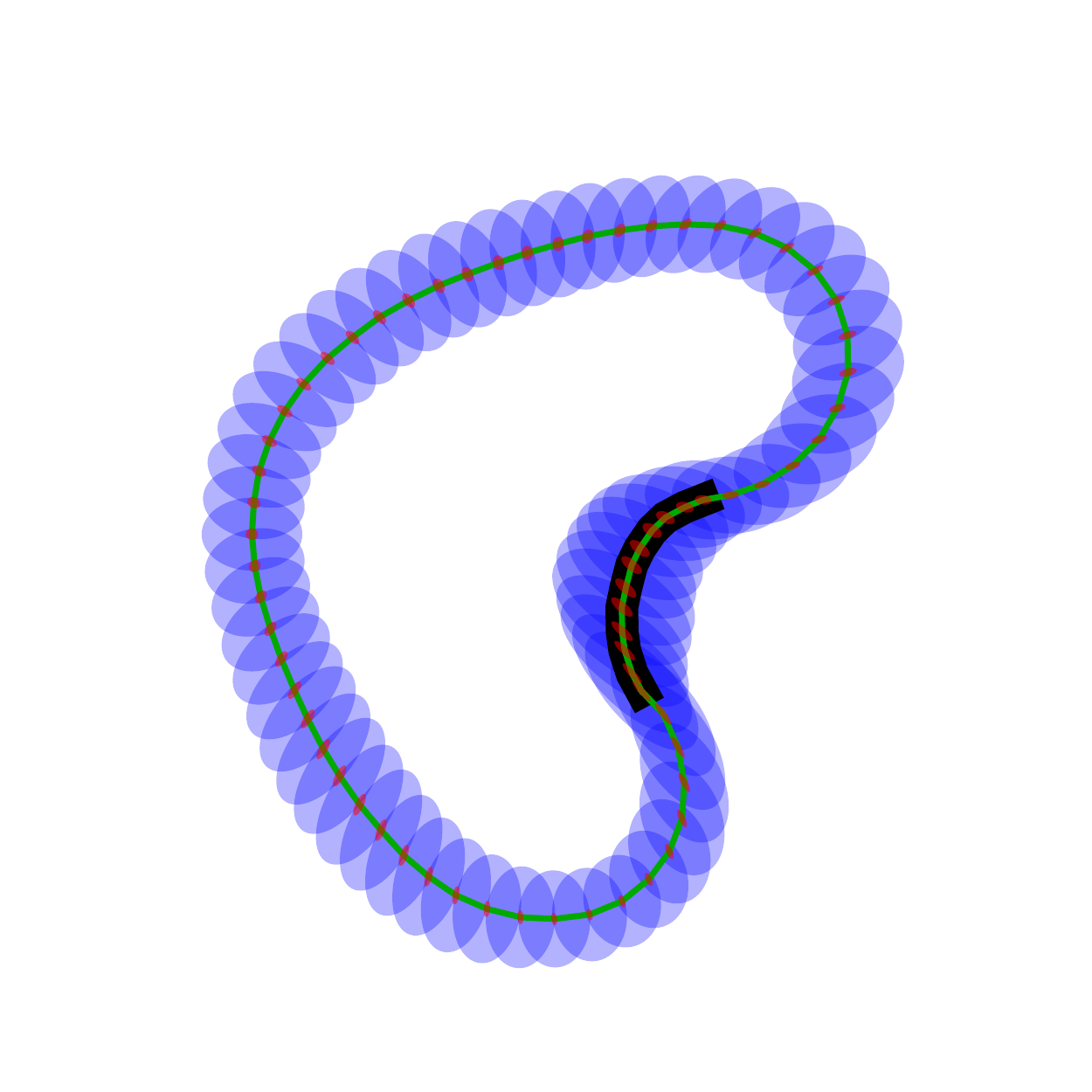}}
		\\
		\hline
	\end{tabular} 
	\caption{Mean micelle shapes, thermal fluctuations, and errors in mean micelle shapes (illustrated in the manner of \cref{fig:baseCaseAverage}) as a function of $\roh$, the asymmetry ratio of the solvophobic-rich diblocks expressed as a percentage of the reference micelle value. 
	As the solvophobic-rich diblocks become more asymmetric, making a sharper contrast with the solvophilic-rich diblocks, the shapes become less circular, the fluctuations decrease.}
	\label{tab:rPhobeShapes}
\end{table}
For values of $\ro$ closer to zero (i.e., more symmetric diblocks), the solvophobic-rich diblocks are very similar in composition to the solvophilic-rich diblocks, and so their preferred curvatures are similar, which we expect to result in a weak dimple (i.e., $\CR$ should become less negative).
If the dimple is weak, then the shape should be nearly circular, so that less perimeter is required to enclose the same amount of volume, and indeed we expect that the volume enclosed by the micelle depends only weakly on the diblock composition so that micelle perimeter does decrease.
A decrease in perimeter causes a higher density of diblocks and since we expect the diblock composition only weakly affects the preferred density of diblocks, we therefore expect a lower surface tension, leading to greater shape fluctuations (i.e., an increase in $\delta$).
We conclude that $\ro$ should change $\CR$ and $\delta$ in the same direction, as observed.

The key point to the above argument was that the difference between the asymmetries of the micelle's two species of diblock determines how circular the micelle shape is.
In the case we discussed, this asymmetry contrast was controlled by changing $\ro$, but it could just as well been controlled by changing  $\ri$ (results shown in \cref{tab:rPhilShapes}).
\begin{table}
	\begin{tabular}{|c|ccccc|}
		\hline 
		$\rih$  & $135\%$ & $124\%$ & $ 112\% $ & $ 100\% $ & $ 88\%$ \\
		&
		\raisebox{-.5\height}{\includegraphics[width=\figureSize \linewidth]{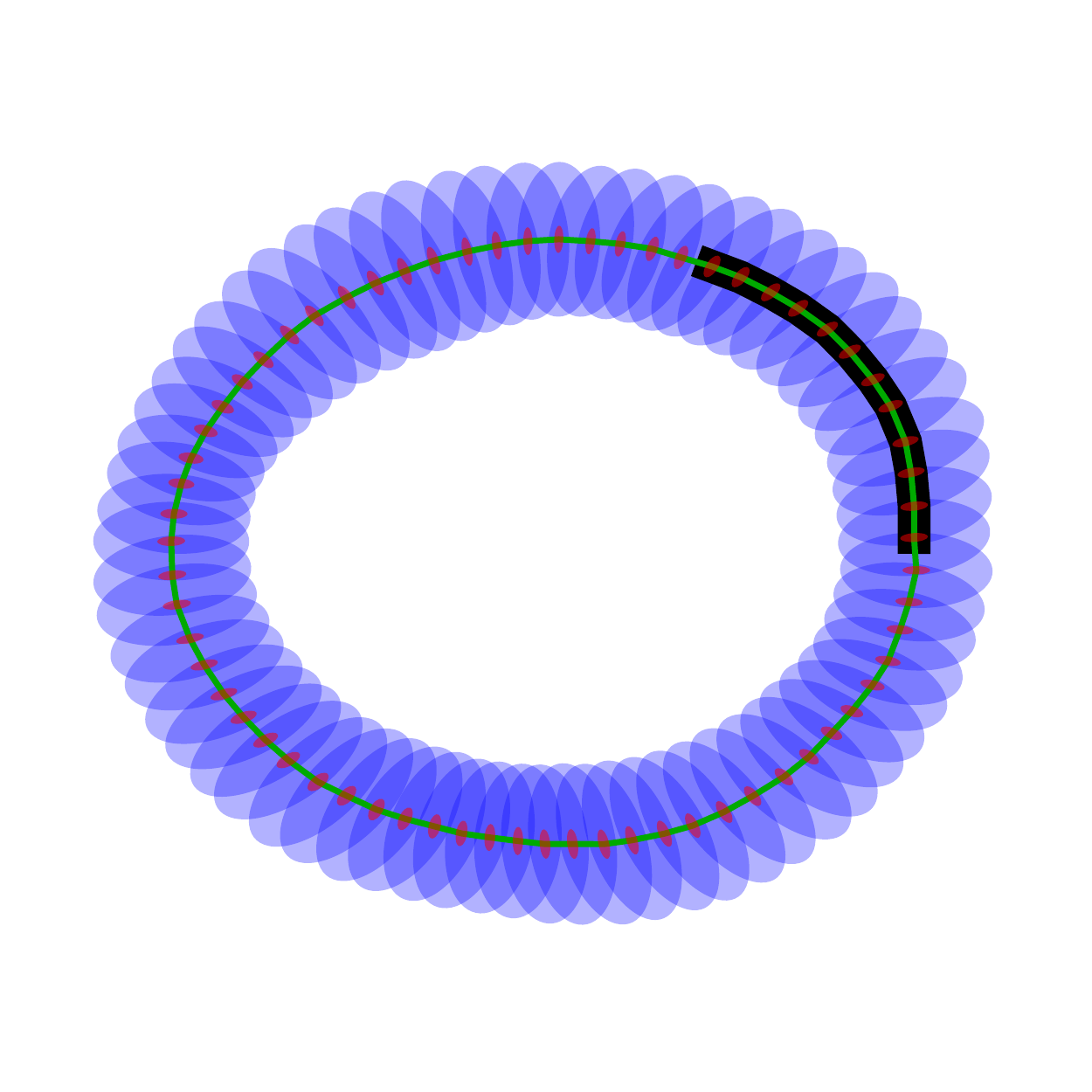}}
		&
		\raisebox{-.5\height}{\includegraphics[width=\figureSize \linewidth]{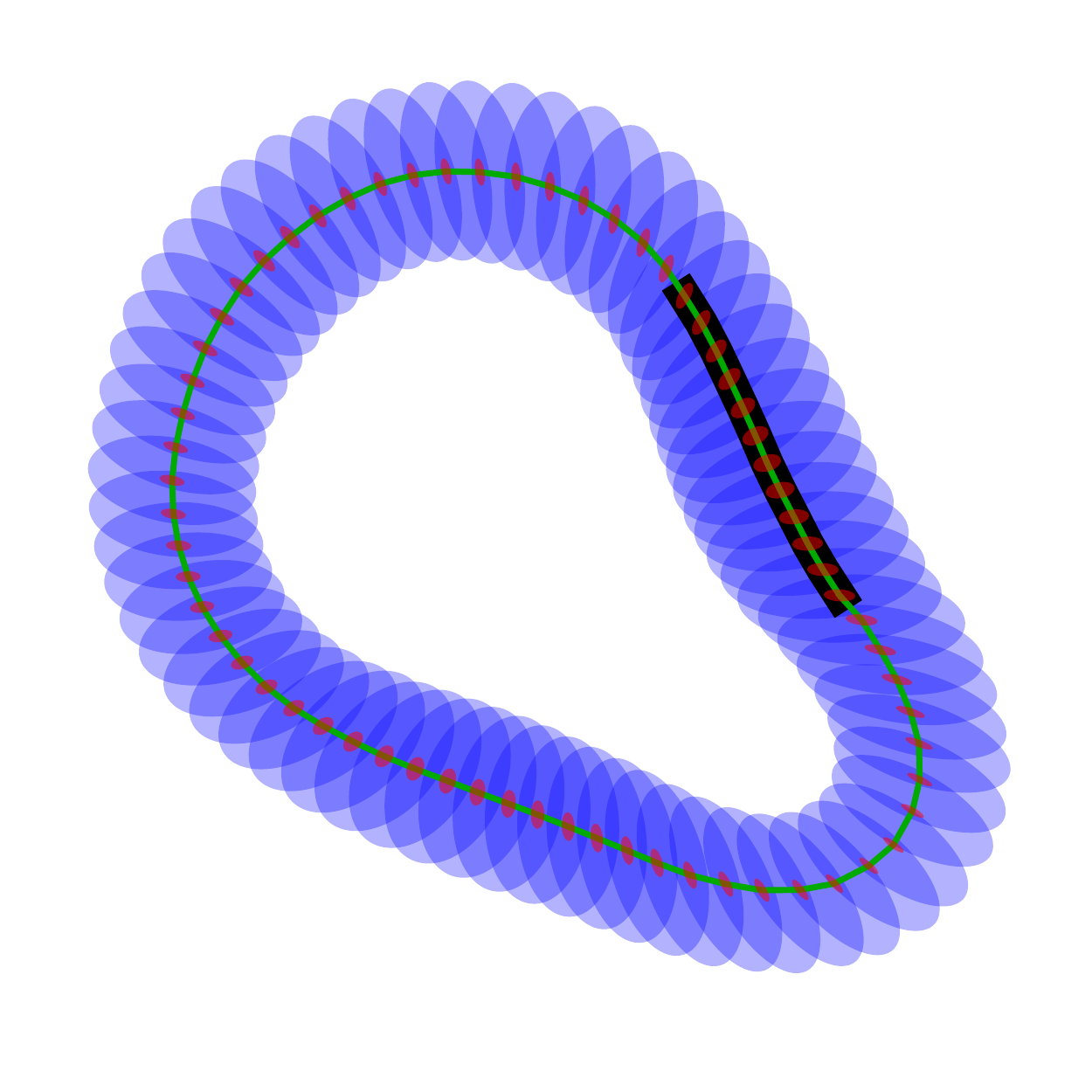}}
		&
		\raisebox{-.5\height}{\includegraphics[width=\figureSize \linewidth]{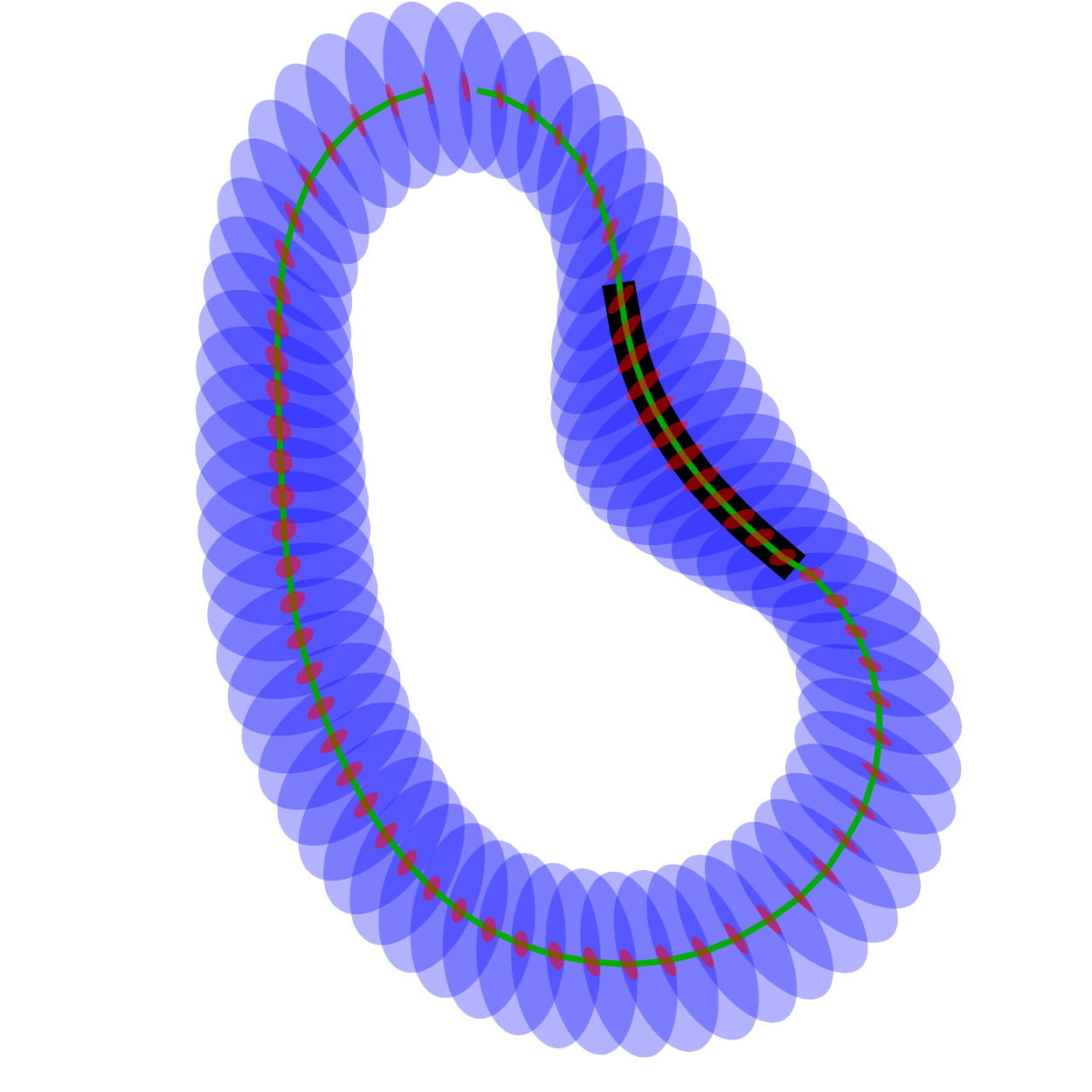}}
		&
		\raisebox{-.5\height}{\includegraphics[width=\figureSize \linewidth]{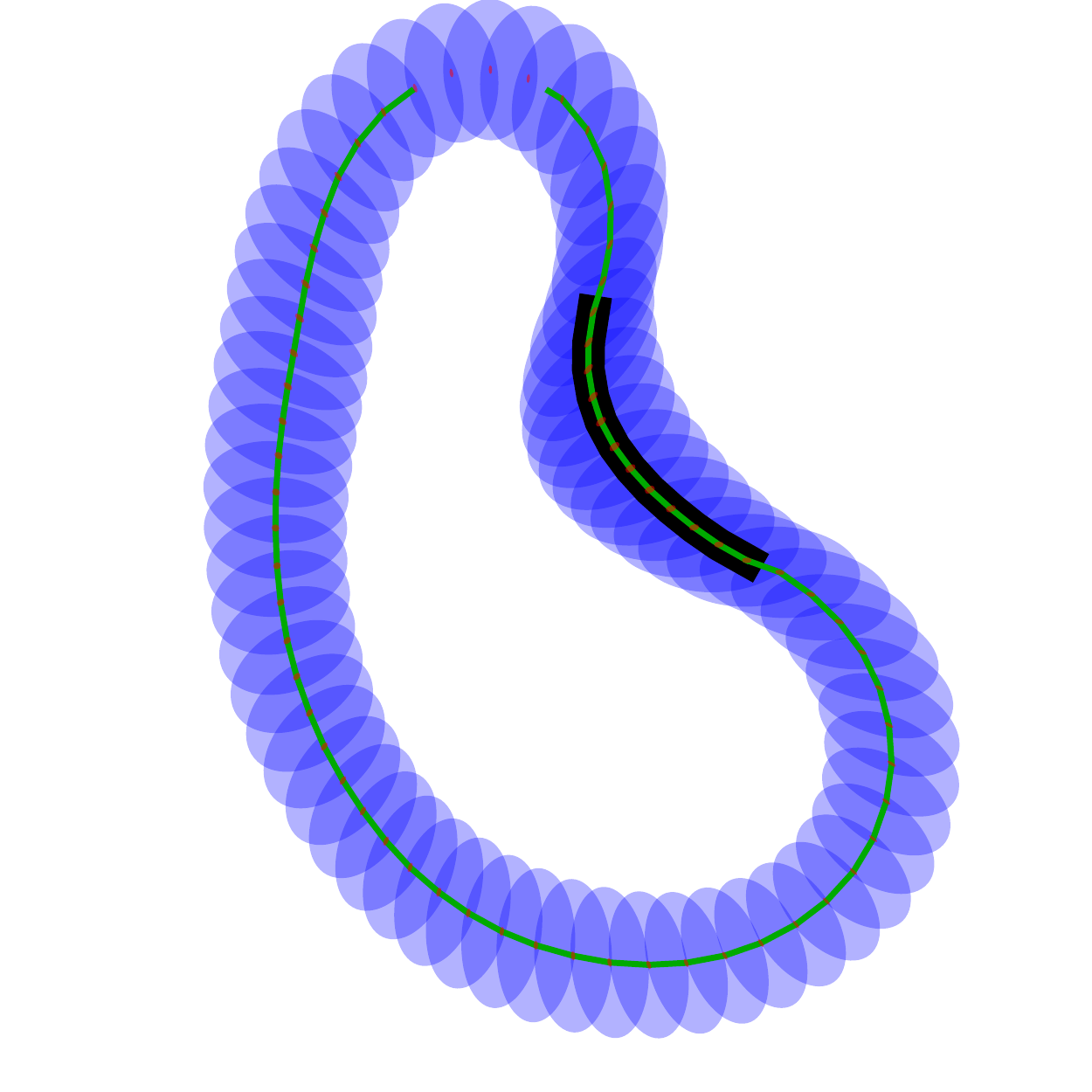}}
		&
		\raisebox{-.5\height}{\includegraphics[width=\figureSize \linewidth]{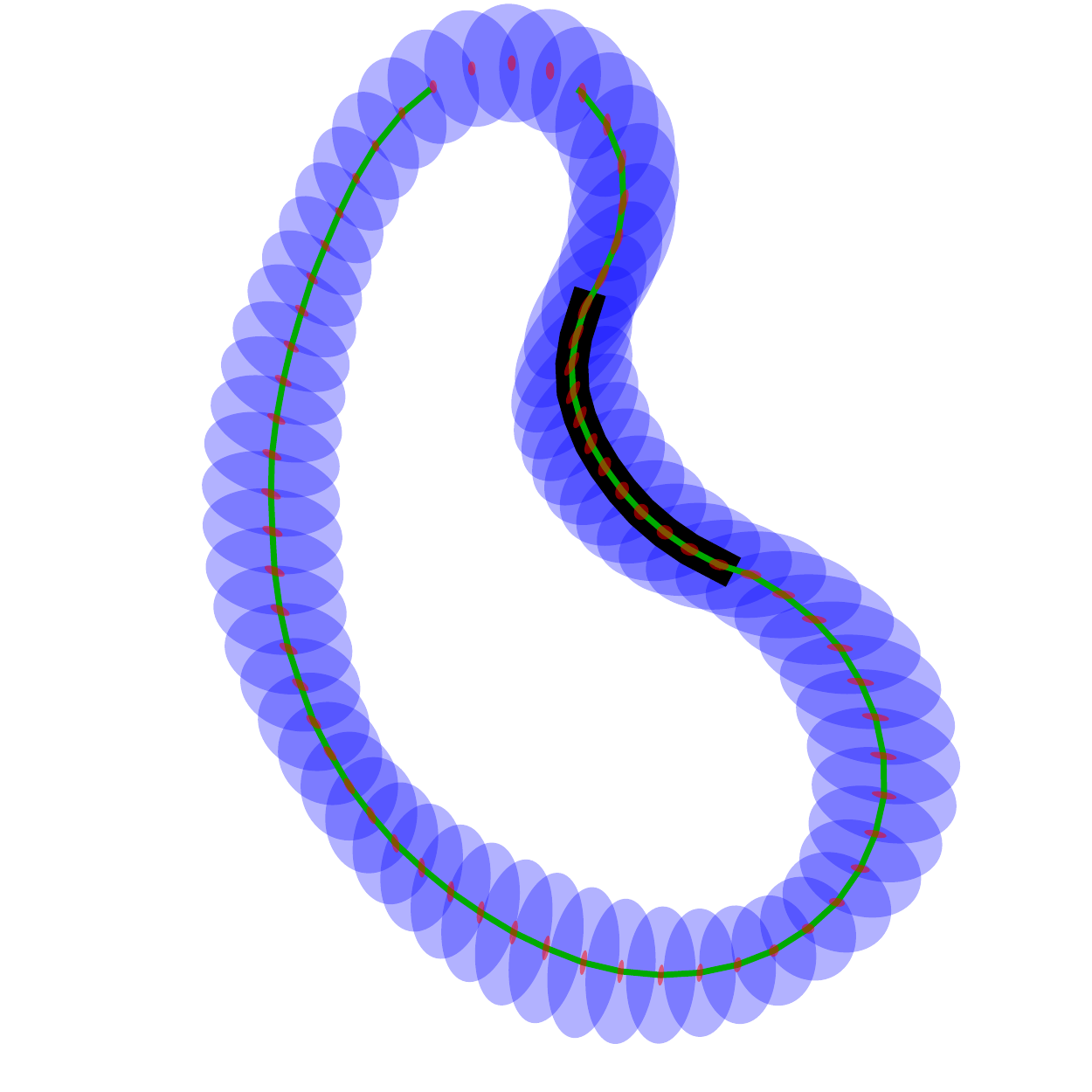}}\\
		\hline
		$\rih$   & $ 76\% $& $ 65\% $ & $ 53\% $ & $ 41\% $ & $ 29\% $\\
		&
		\raisebox{-.5\height}{\includegraphics[width=\figureSize \linewidth]{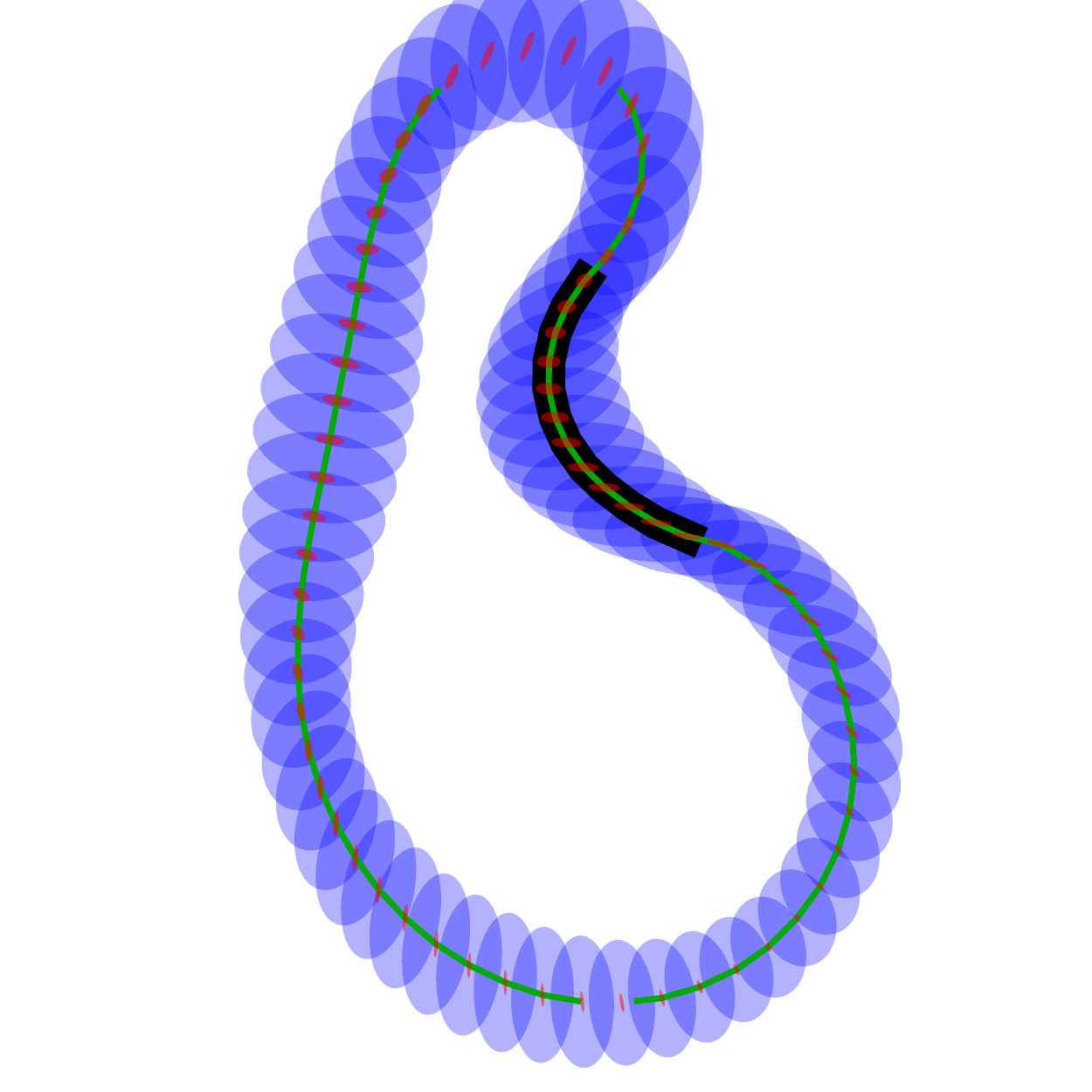}}
		&
		\raisebox{-.5\height}{\includegraphics[width=\figureSize \linewidth]{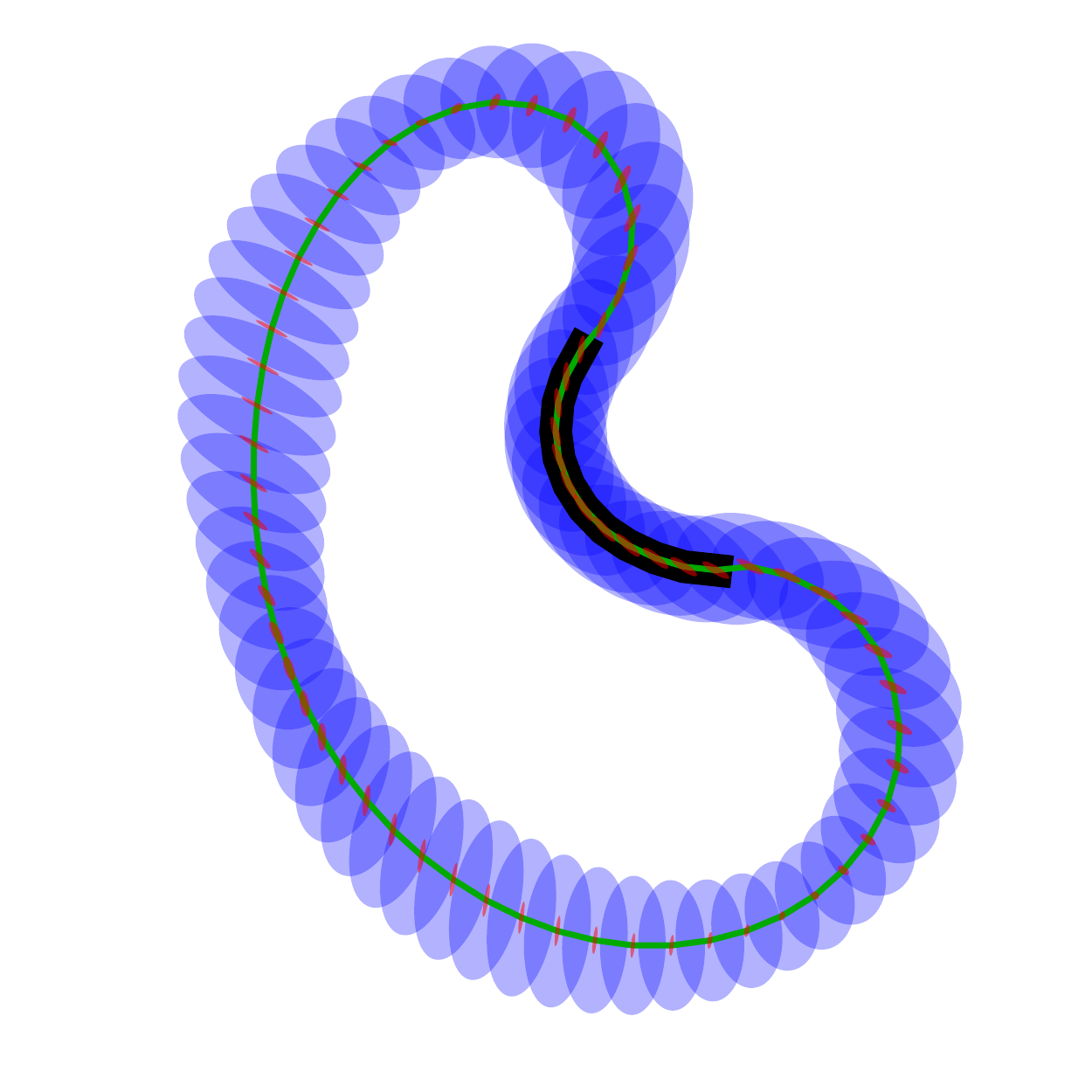}}
		&
		\raisebox{-.5\height}{\includegraphics[width=\figureSize \linewidth]{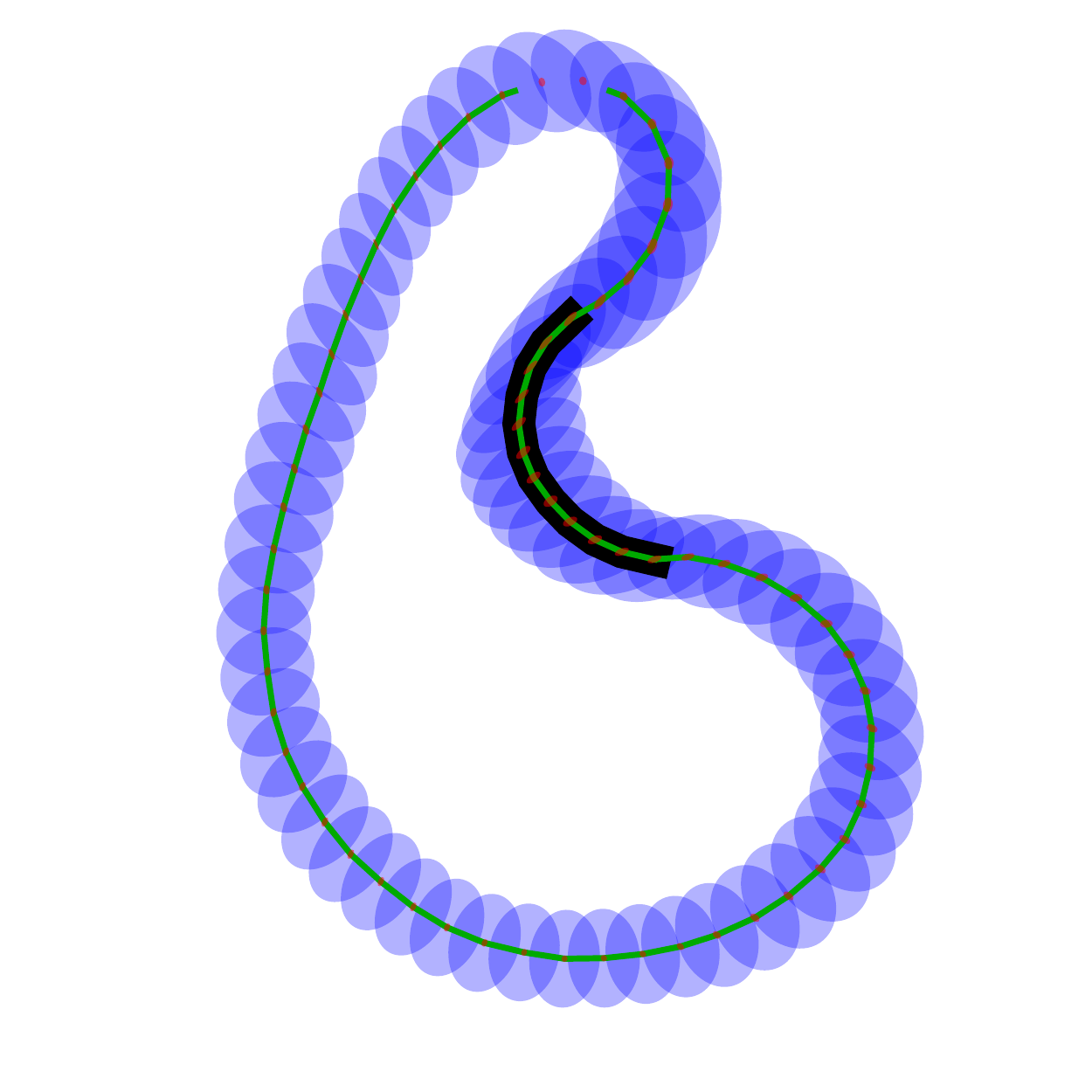}}
		&
		\raisebox{-.5\height}{\includegraphics[width=\figureSize \linewidth]{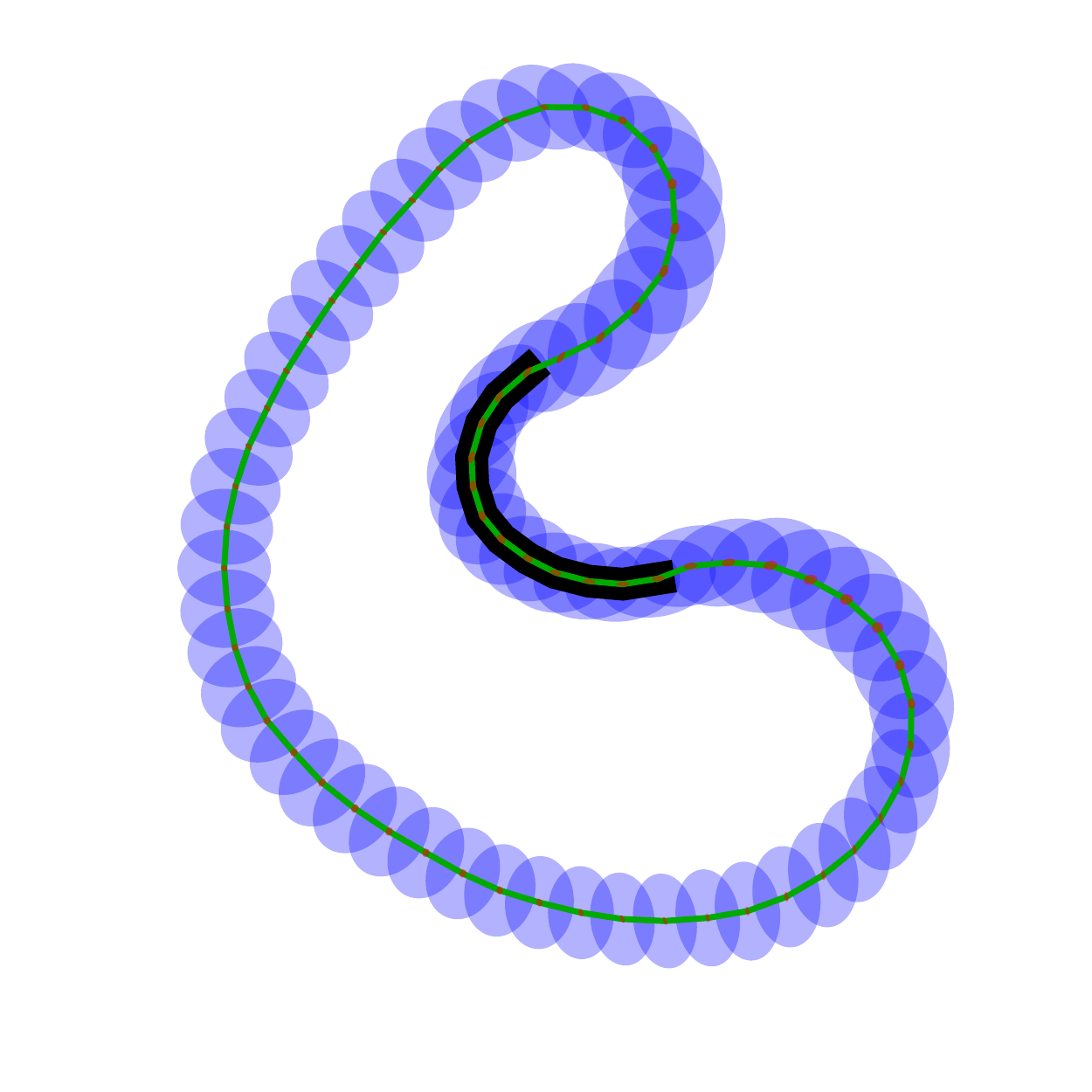}}
		&
		\raisebox{-.5\height}{\includegraphics[width=\figureSize \linewidth]{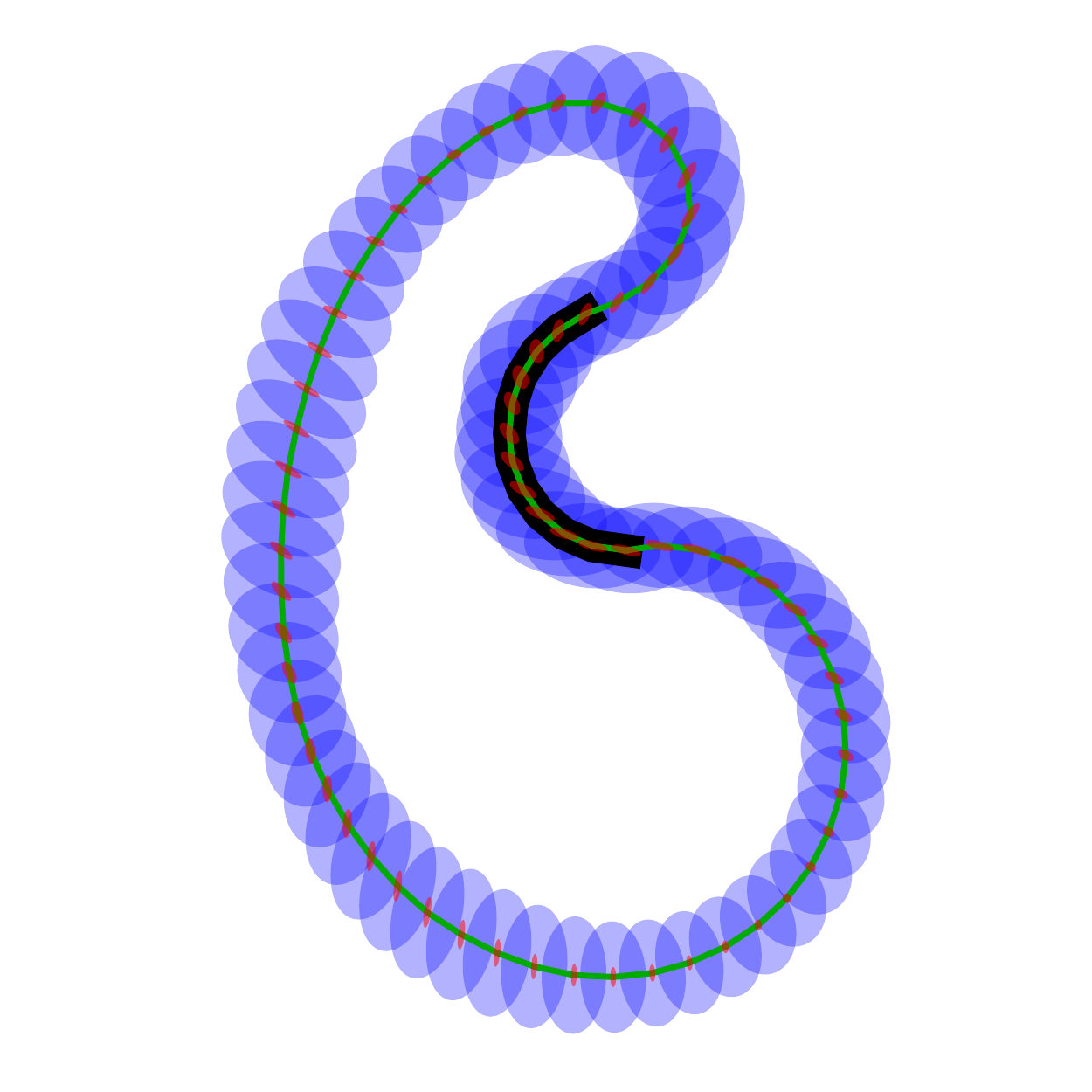}}
		\\
		\hline
	\end{tabular} 
	\caption{Mean micelle shapes, thermal fluctuations, and errors in mean micelle shapes (illustrated in the manner of \cref{fig:baseCaseAverage}) as a function of $\rih$, the asymmetry ratio of the solvophilic-rich diblocks expressed as a percentage of the reference micelle value. 
	As the solvophilic-rich diblocks become less asymmetric, making a sharper contrast with the solvophobic-rich diblocks, the shapes become less circular and the fluctuations decrease.}
	\label{tab:rPhilShapes}
\end{table}
Therefore, as the solvophilic-rich diblocks are made more asymmetric, they become more similar to the solvophobic-rich diblocks, and so we expect the shapes to become more circular and to fluctuate more.
Thus, like $\ro$, $\ri$ should affect $\CR$ and $\delta$ in the same direction.

The results of varying the number of solvophobic-rich diblocks $\no$ (see \cref{tab:nPhobeShapes}) mostly follow the same trend as the results of varying the diblock asymmetries $\ri$ and $\ro$, but a different explanation is required.
In this case we expect decreasing $\no$ to decrease the density of diblocks on the micelle surface, thereby increasing the surface tension and decreasing the fluctuations as measured by $\delta$.
However, decreasing $\no$ also decreases the length of micelle perimeter that has to deform in order to achieve its preferred curvature.
Therefore, we expect micelles with small $\no$ may have a more strongly curved dimple, so that $\CR$ becomes more negative.
By this reasoning $\no$, like $\ri$ and $\ro$, would affect $\CR$ and $\delta$ in the same direction.

\begin{table}
	\newcommand\T{\rule{0pt}{2.6ex}}
	\begin{tabular}{|c|ccccc|}
		\hline
		$\noh$ \T & 33\% & 42\% &50\% &58\%& 67\%   \\ 
		&
		\raisebox{-.5\height}{\includegraphics[width=\figureSize \linewidth]{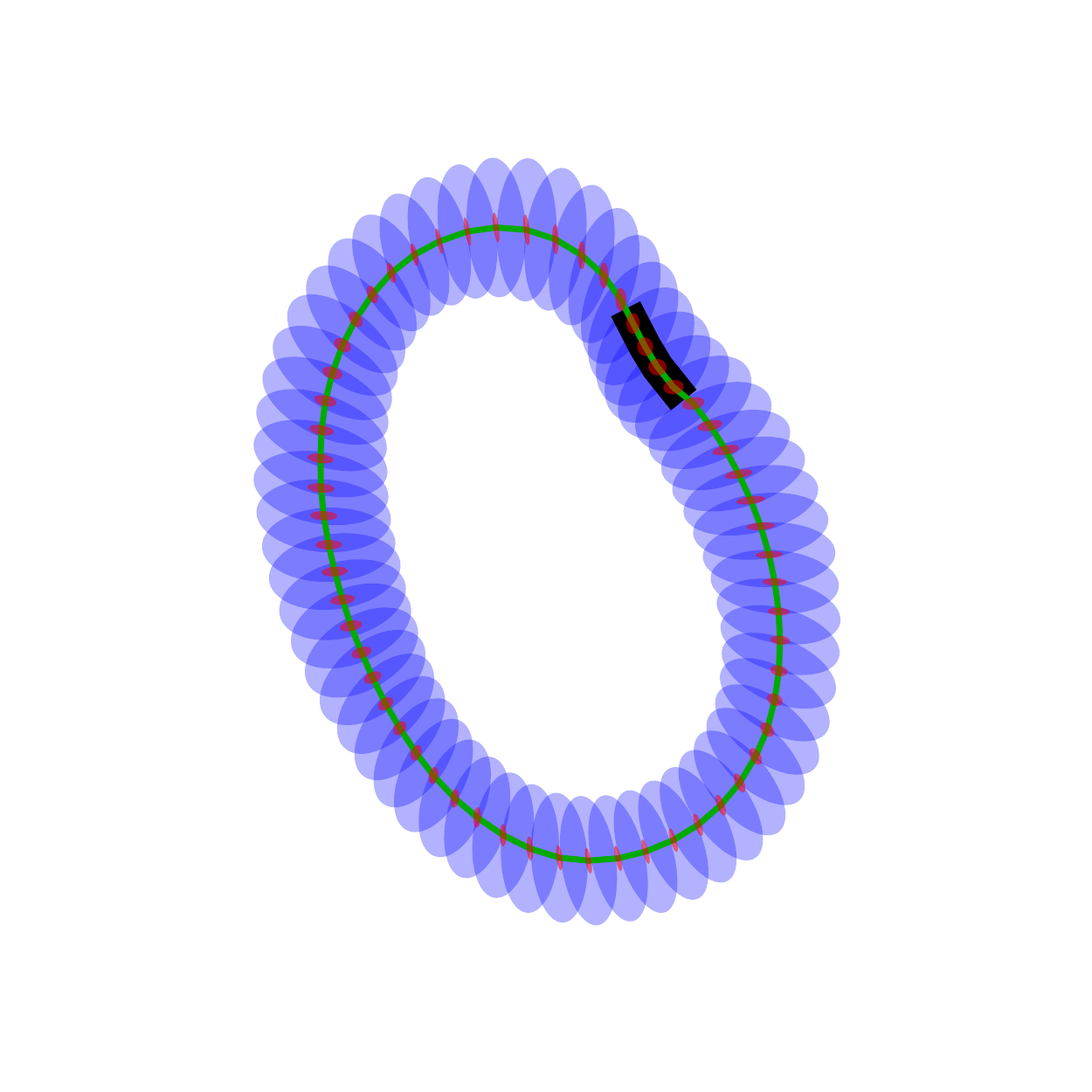}}
		&
		\raisebox{-.5\height}{\includegraphics[width=\figureSize \linewidth]{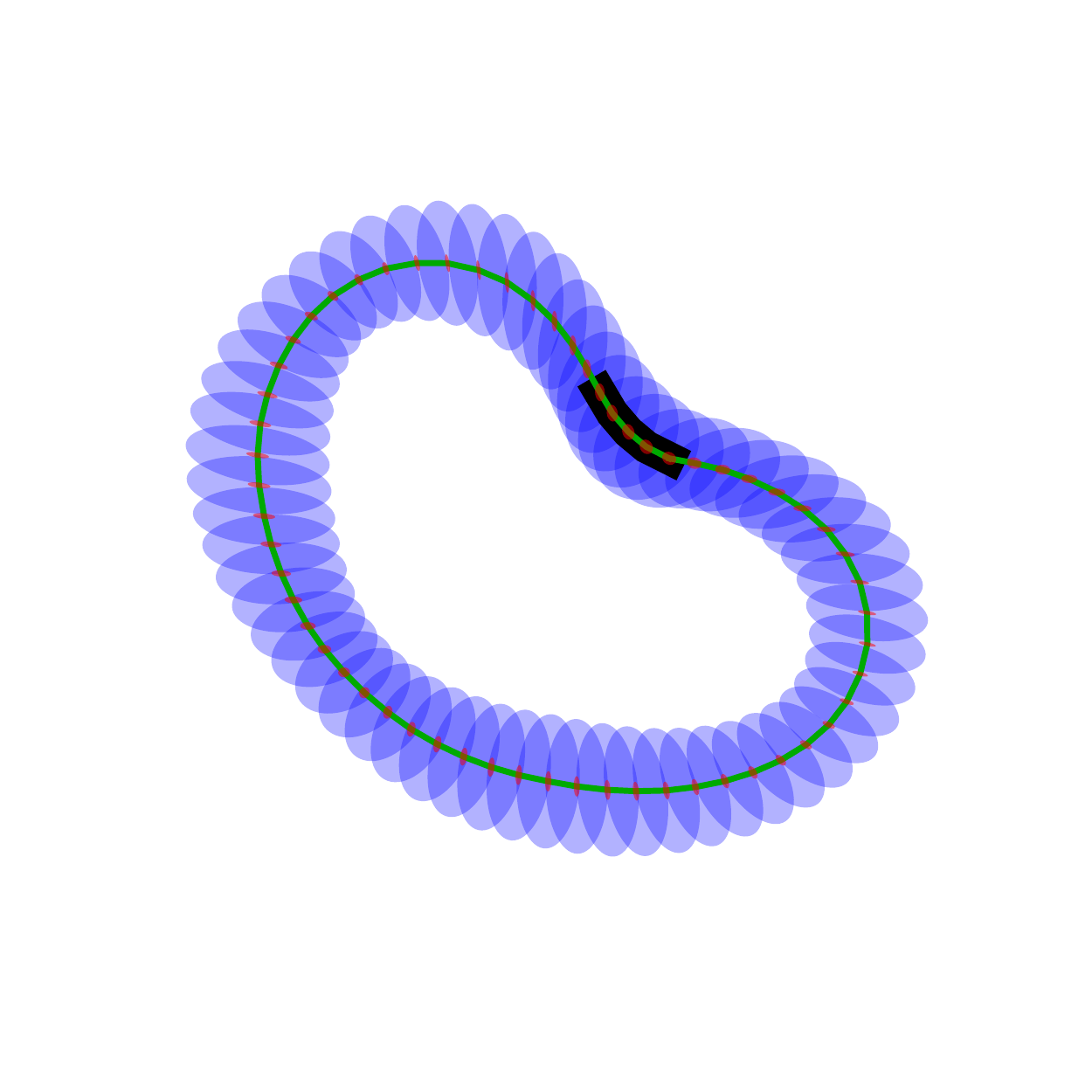}}
		&
		\raisebox{-.5\height}{\includegraphics[width=\figureSize \linewidth]{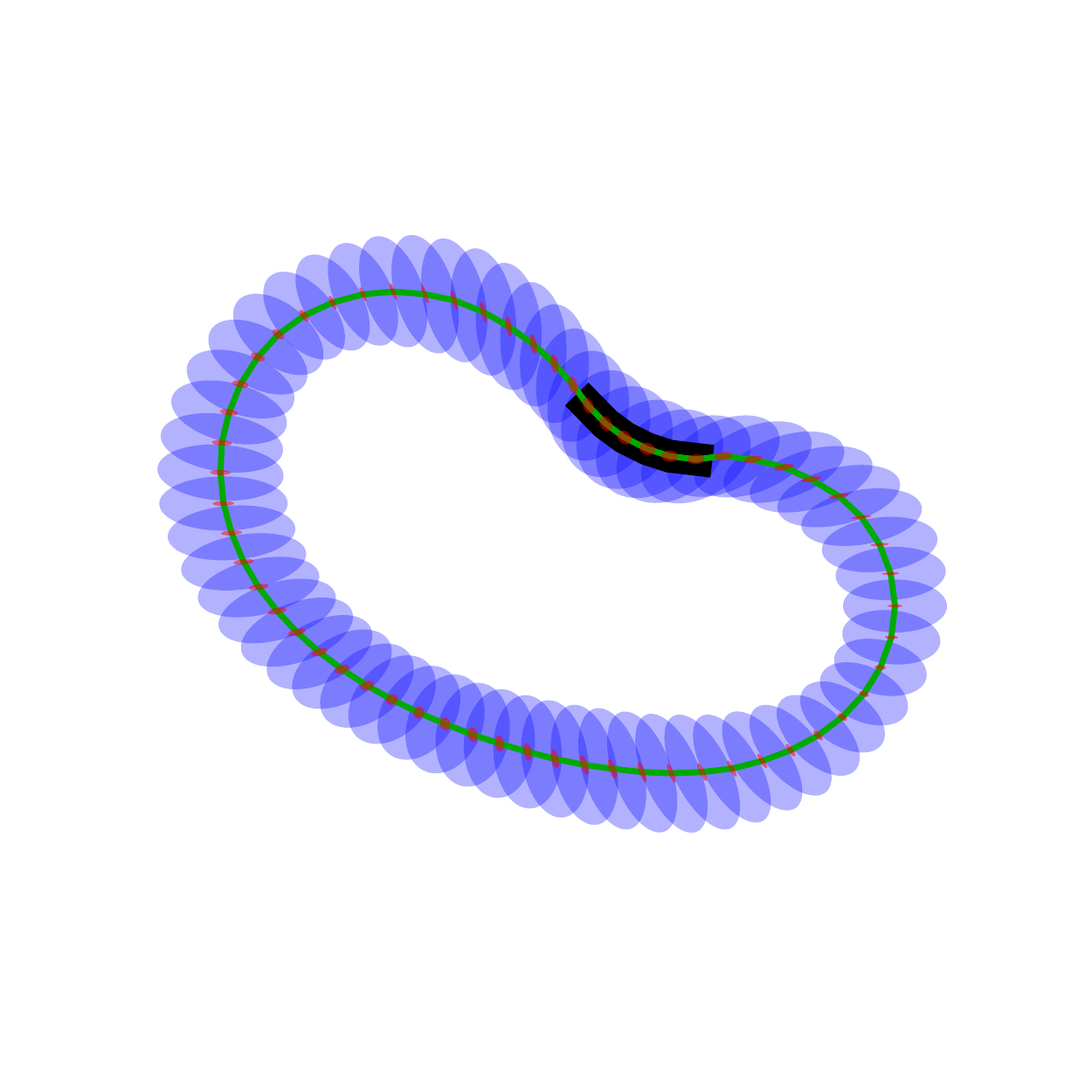}}
		&
		\raisebox{-.5\height}{\includegraphics[width=\figureSize \linewidth]{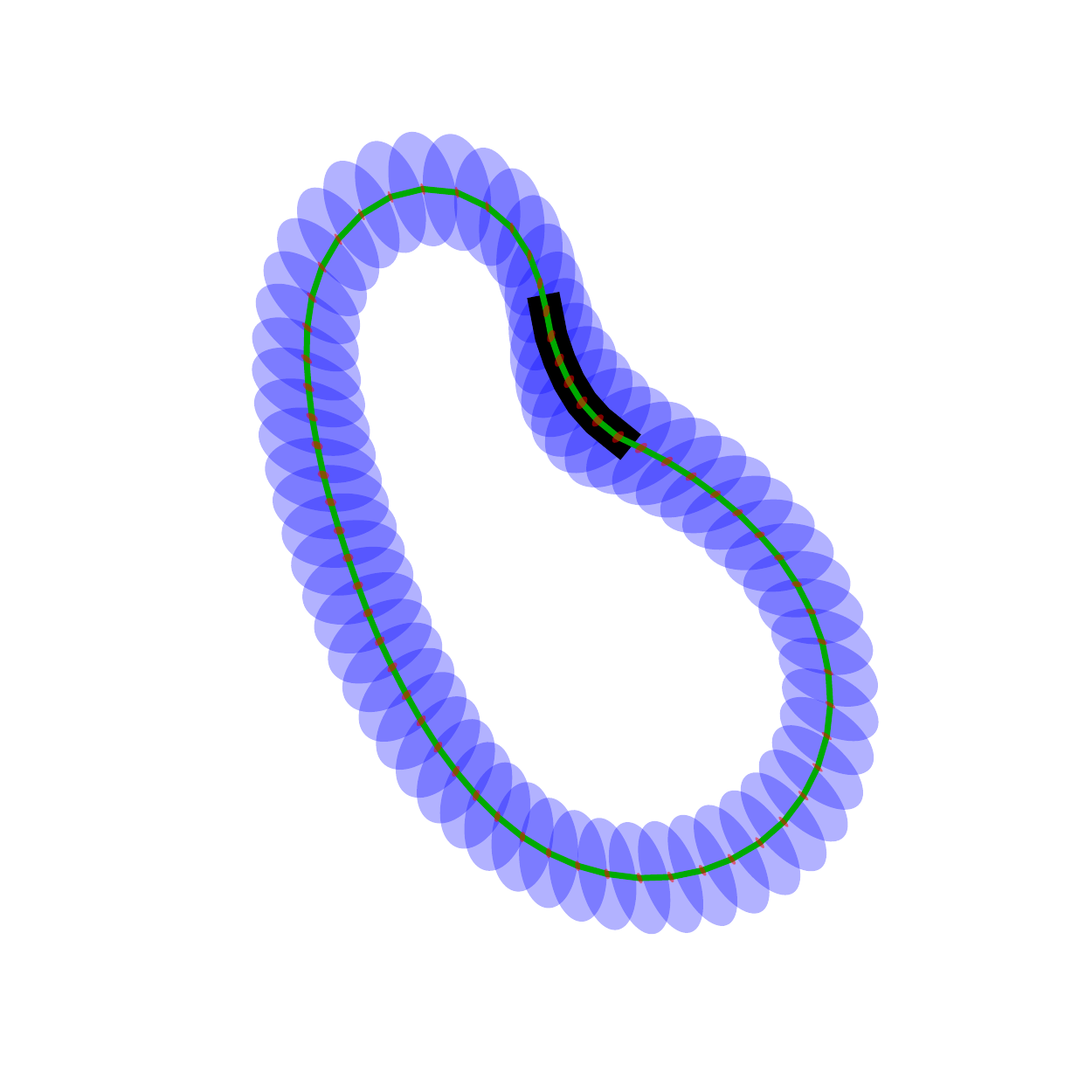}}
		&
		\raisebox{-.5\height}{\includegraphics[width=\figureSize \linewidth]{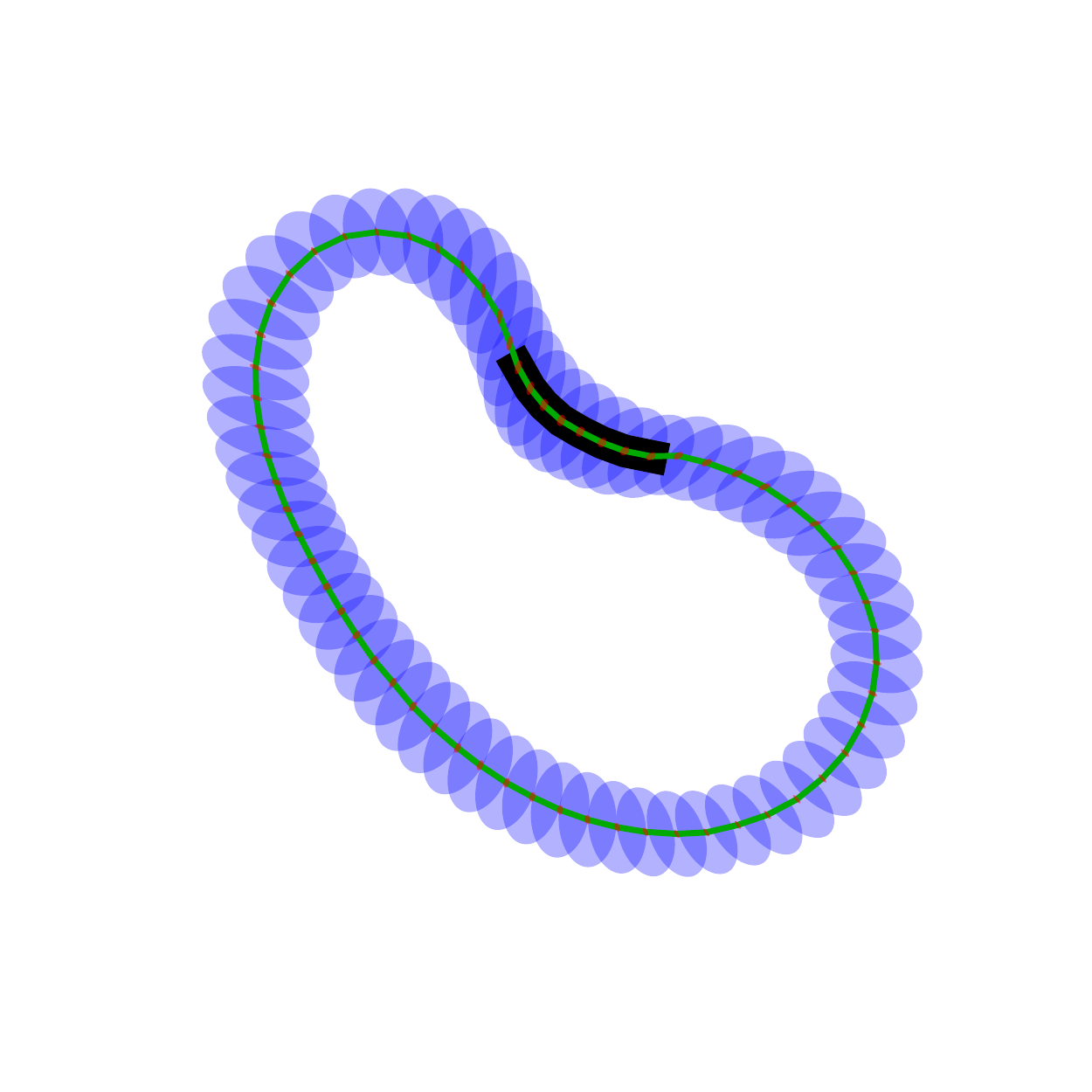}}\\
				\hline
		$\noh$ \T & 75\% & 83\% & 92\% & 100\%& 117\% \\
		&
		\raisebox{-.5\height}{\includegraphics[width=\figureSize \linewidth]{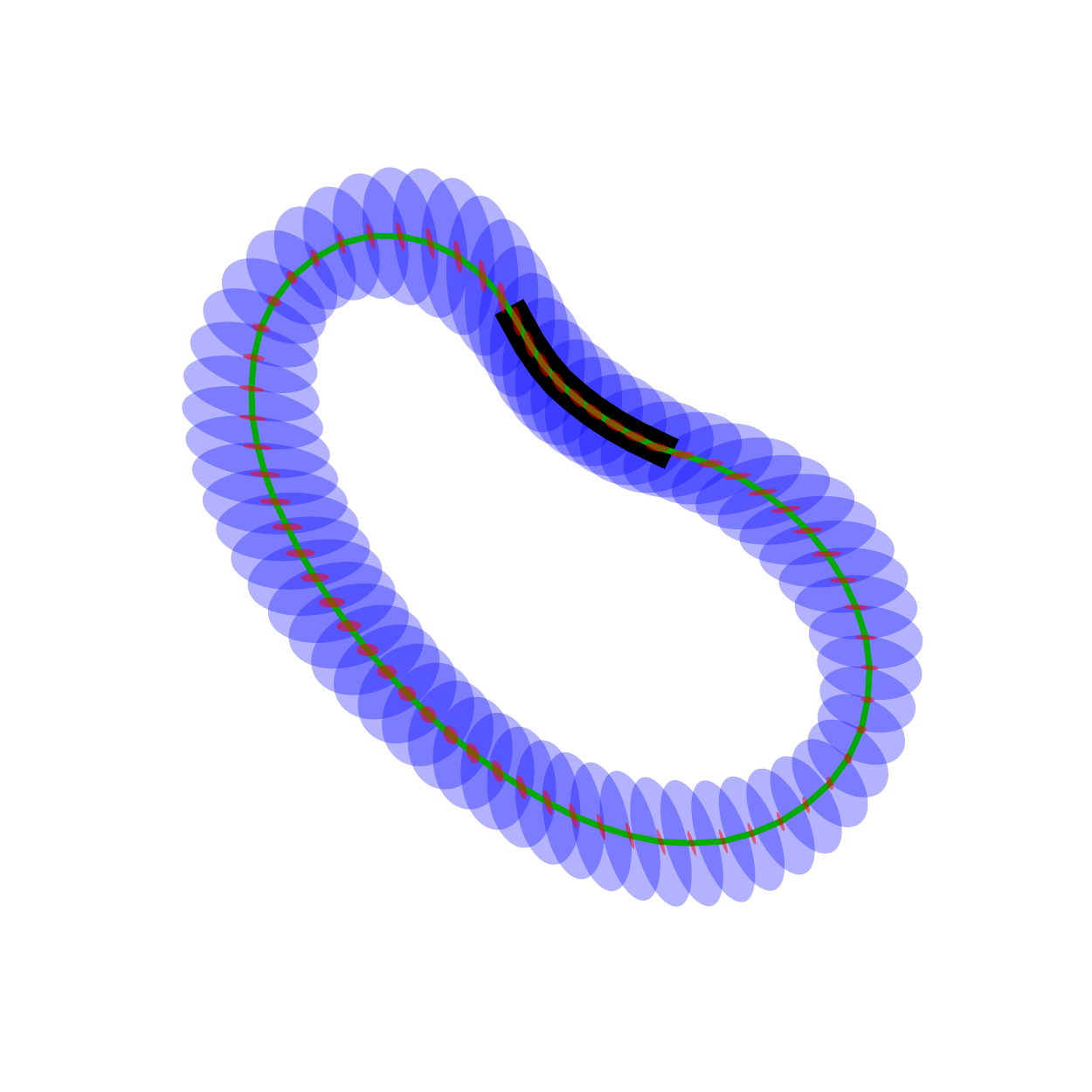}}
		&
		\raisebox{-.5\height}{\includegraphics[width=\figureSize \linewidth]{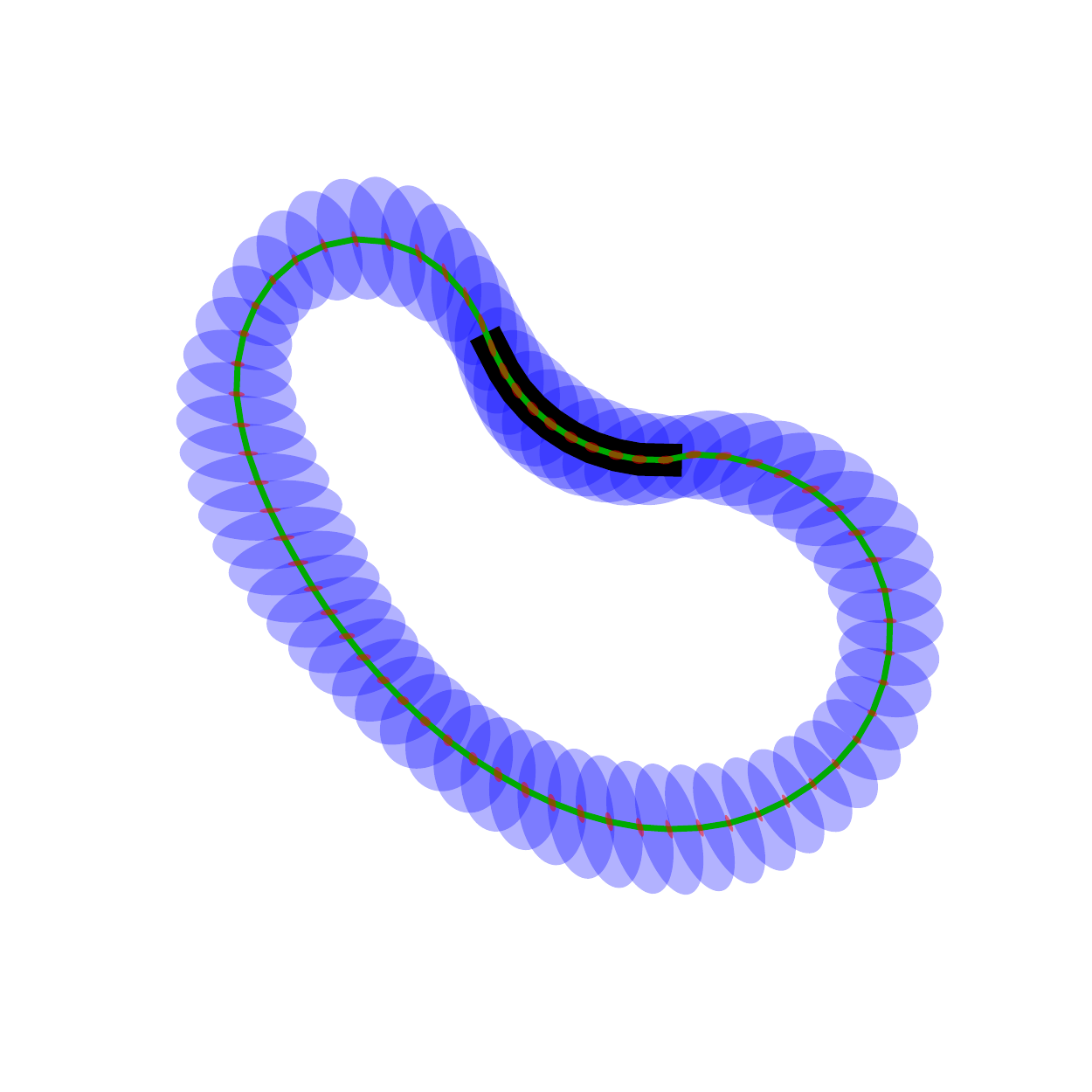}}
		&
		\raisebox{-.5\height}{\includegraphics[width=\figureSize \linewidth]{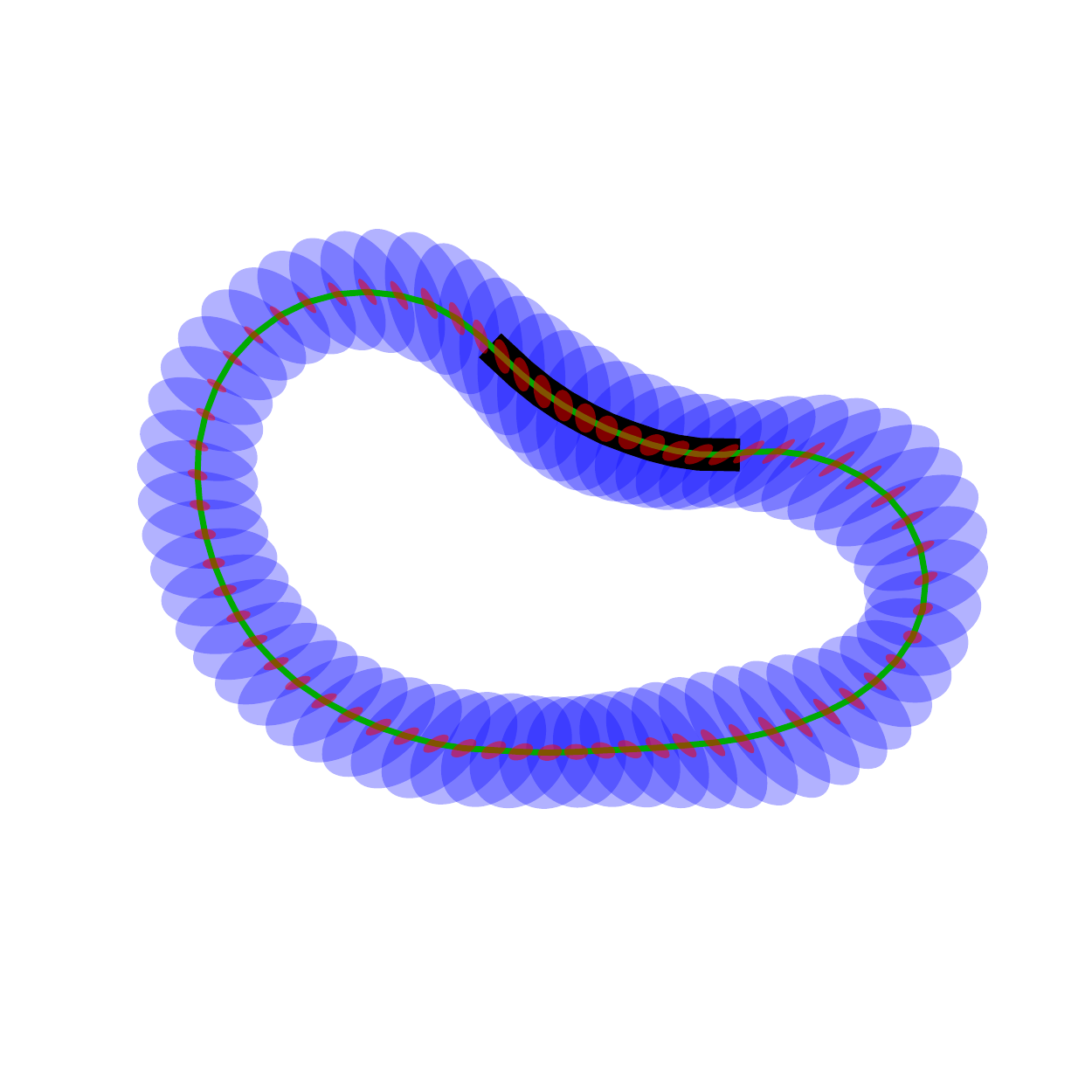}}
		&
		\raisebox{-.5\height}{\includegraphics[width=\figureSize \linewidth]{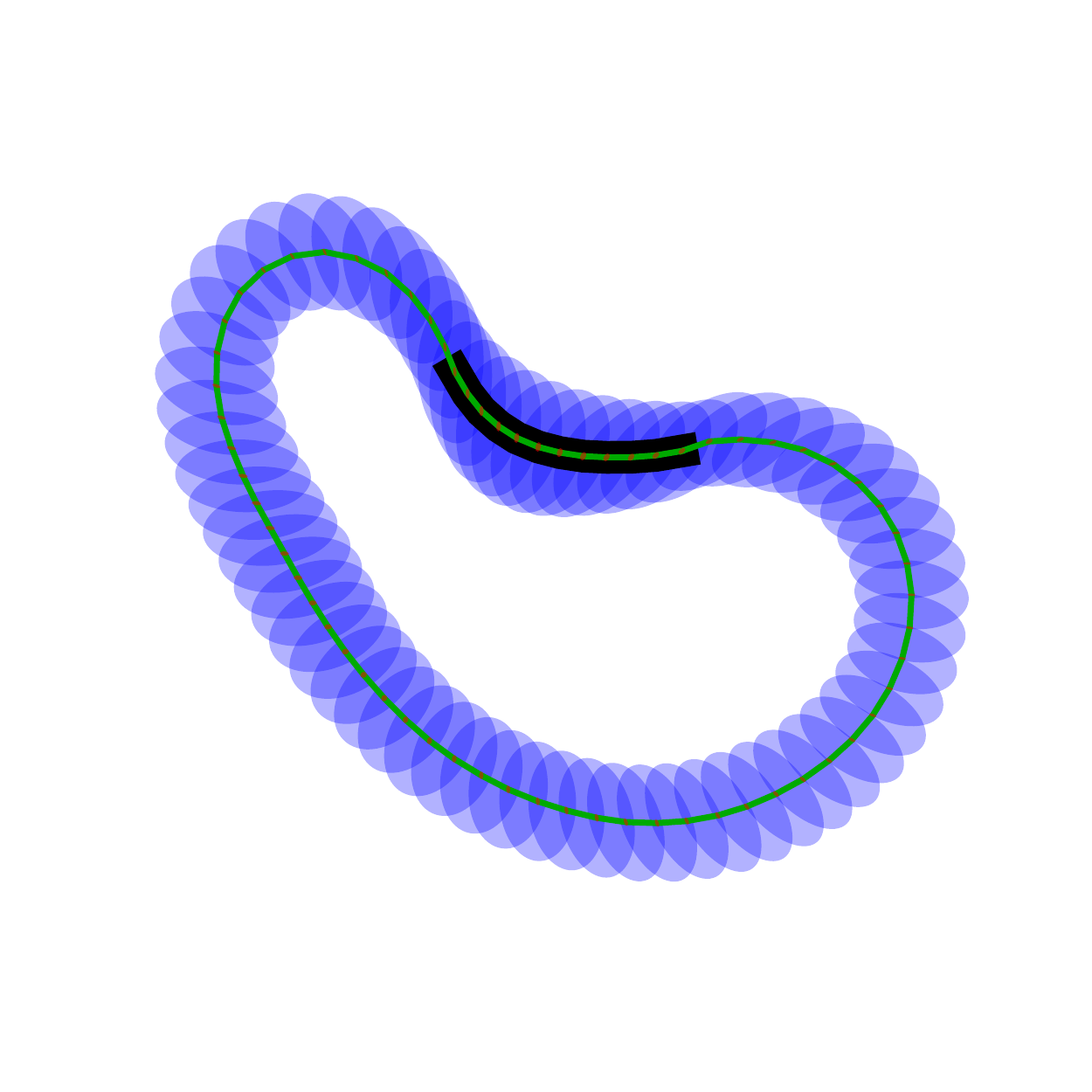}}
		&
		\raisebox{-.5\height}{\includegraphics[width=\figureSize \linewidth]{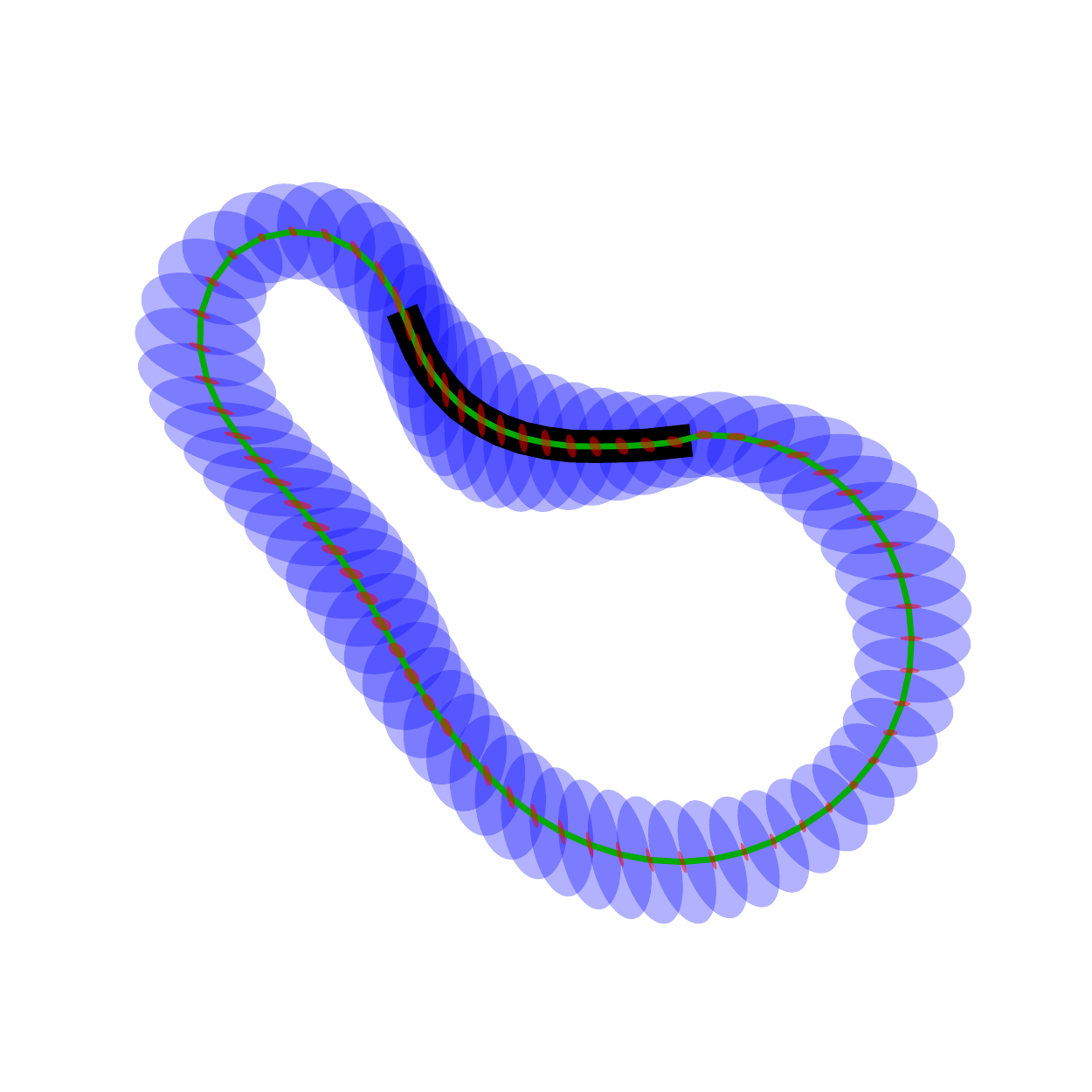}}
		\\
		\hline
		$\noh$ \T & 133\% &150\% &167\% &183\% &200\%  \\
		&
		\raisebox{-.5\height}{\includegraphics[width=\figureSize \linewidth]{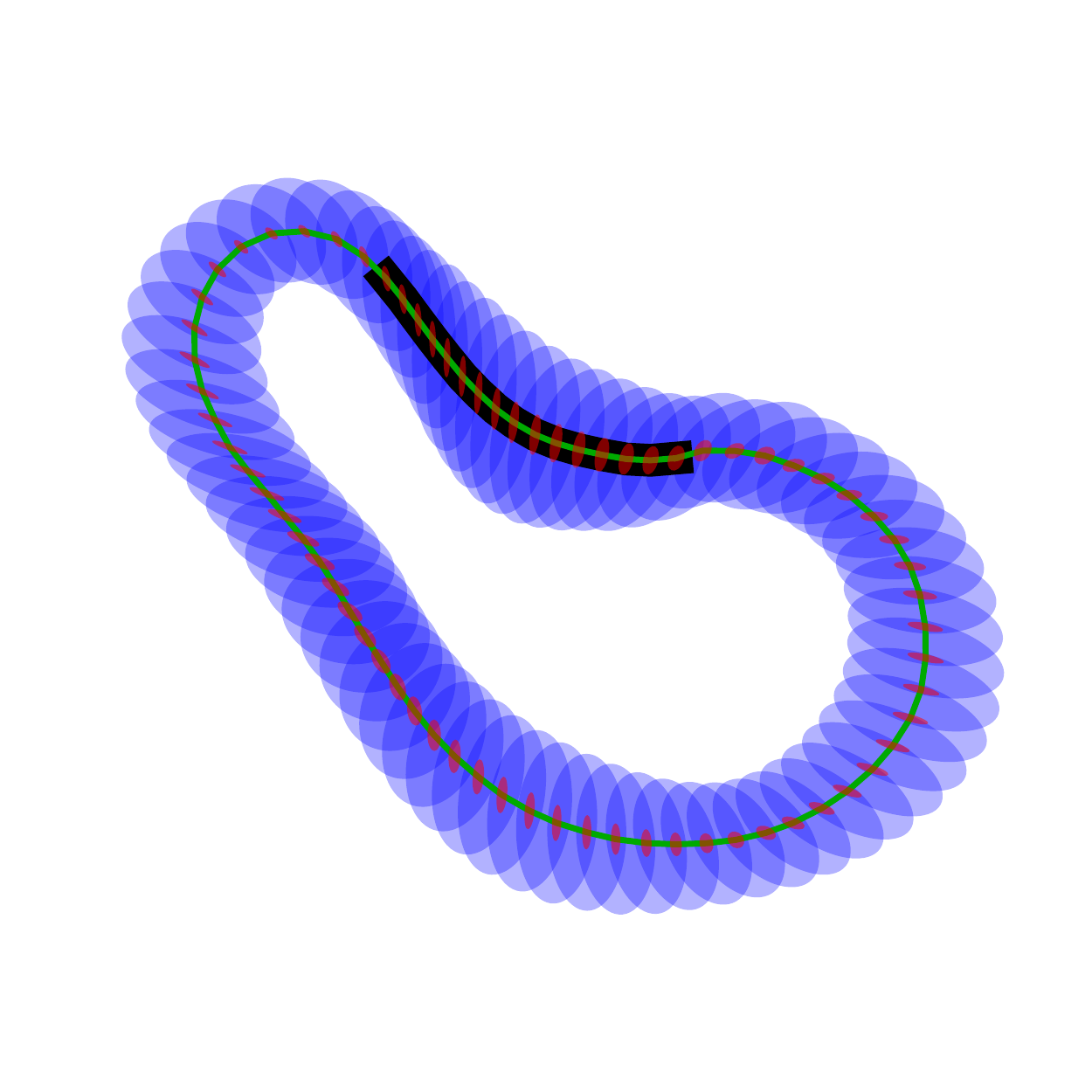}}
		&
		\raisebox{-.5\height}{\includegraphics[width=\figureSize \linewidth]{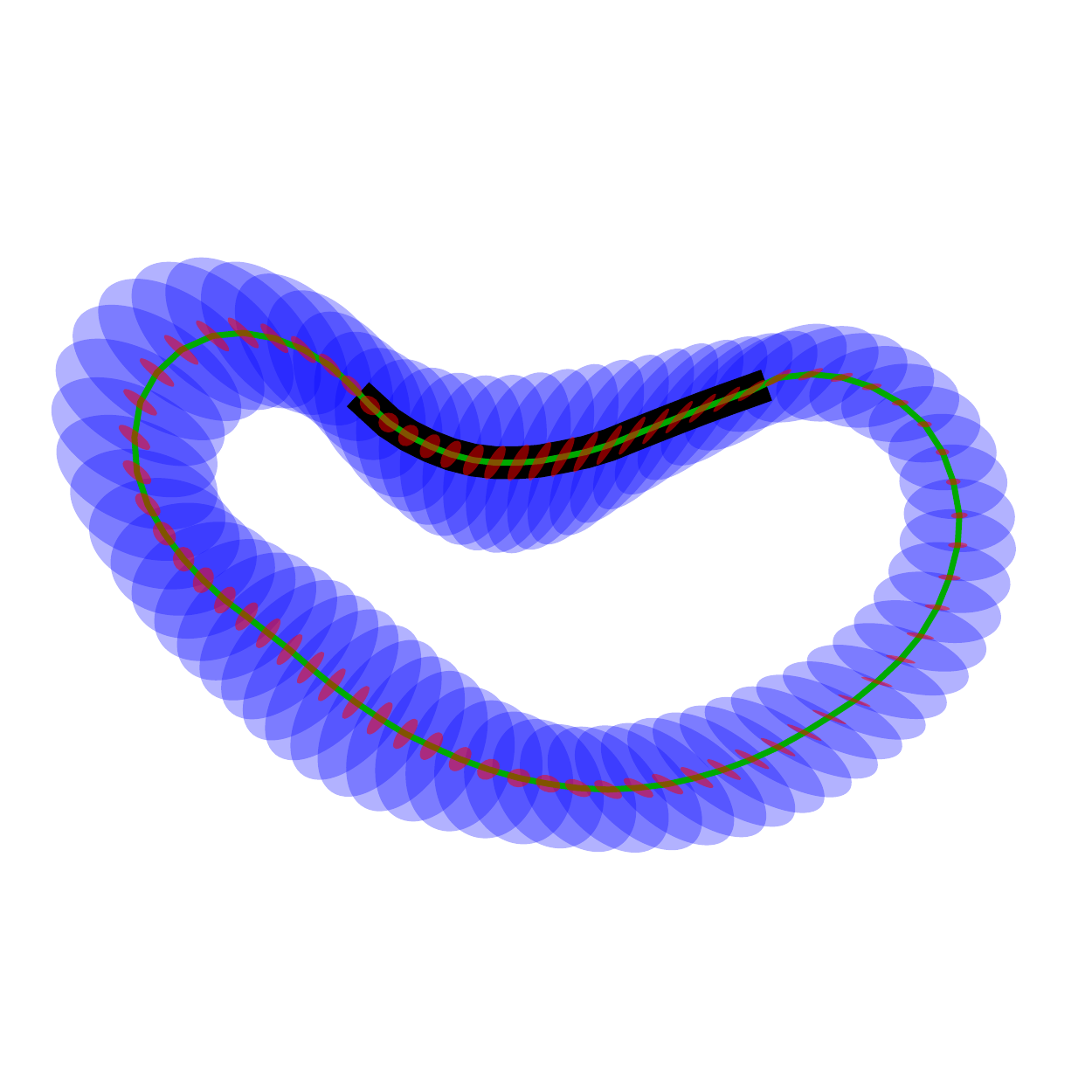}}
		&
		\raisebox{-.5\height}{\includegraphics[width=\figureSize \linewidth]{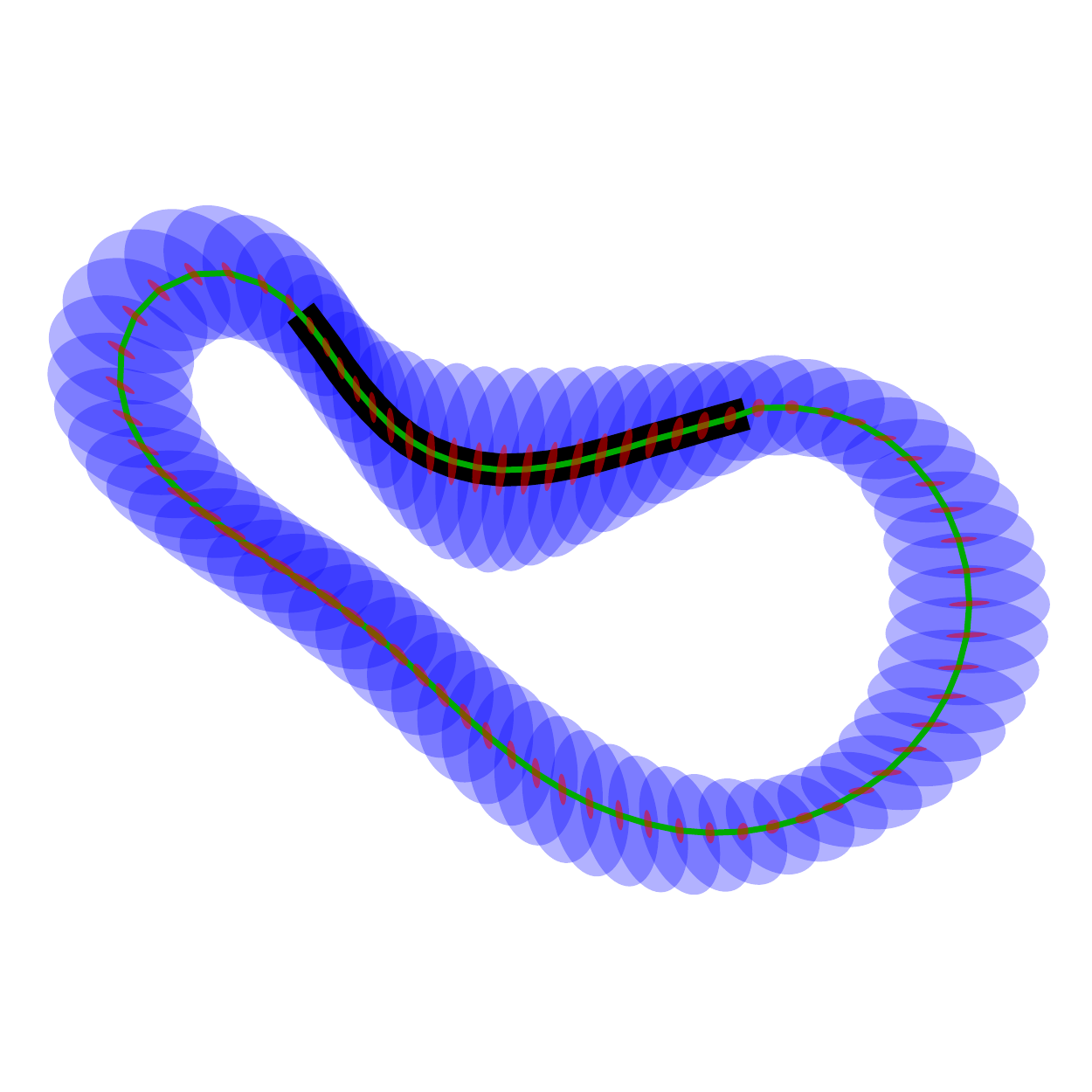}}
		&
		\raisebox{-.5\height}{\includegraphics[width=\figureSize \linewidth]{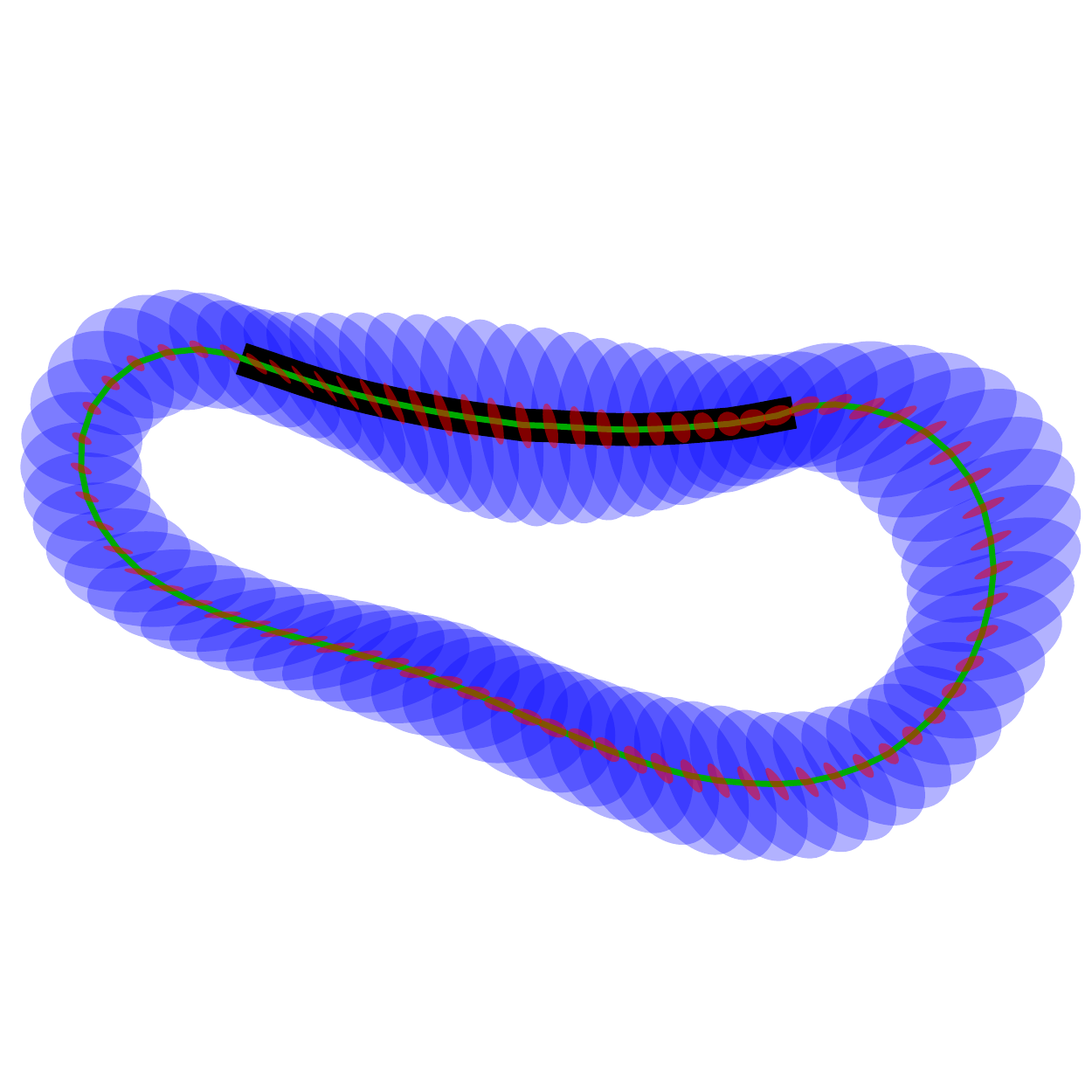}}
		&
		\raisebox{-.5\height}{\includegraphics[width=\figureSize \linewidth]{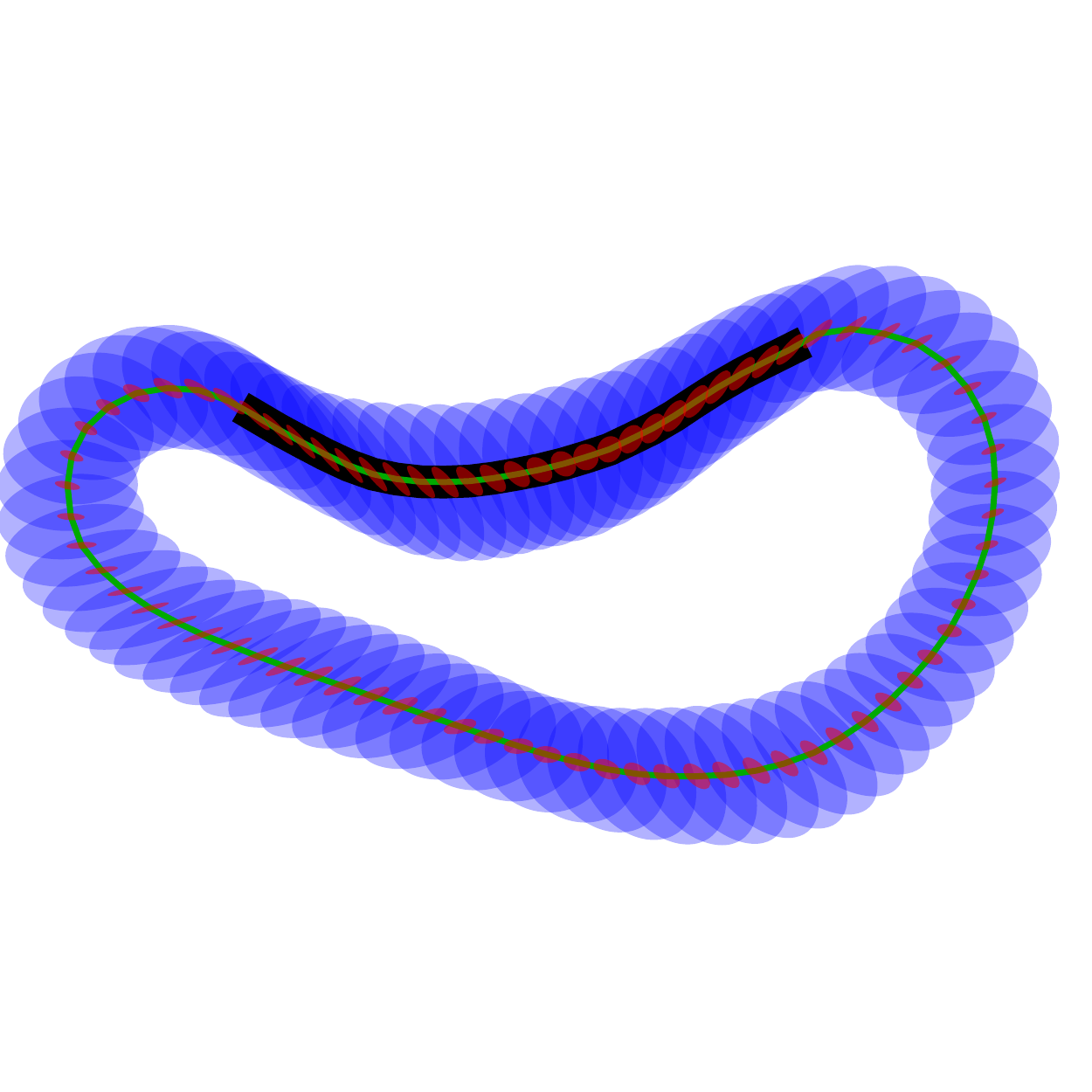}}
		\\ 
		\hline
	\end{tabular}
	\caption{Mean micelle shapes, thermal fluctuations, and errors in mean micelle shapes (illustrated in the manner of \cref{fig:baseCaseAverage}) as a function  of $\noh$, the number of solvophobic-rich diblocks expressed as a percentage of the reference micelle value.
	As the number of solvophobic-rich diblocks beads increases, the shapes become less circular and the fluctuations increase.
	}
	\label{tab:nPhobeShapes}
\end{table}

While the results of varying the number of solvophobic-rich diblocks $\no$ do mostly follow a smooth trend, we note one nonmonotonic feature of this data.
The simulated micelle with the smallest $\no$ (four solvophobic-rich diblocks), which according to the trend of the data should have the most extreme dimple, actually has a less pronounced dimple than even the base micelle.
One might hypothesize that there is a minimum number of solvophobic-rich diblocks needed to nucleate a dimple.
Whatever the case, this nonmonotonicity that a naive linear model explaining the micelle shape may not be sufficient for shape design.

In the preceding results, we have varied a single parameter of the micelle composition to observe the effect on two micelle shape features and have seen that the data lie on only two trend lines.
This contradicts our expectation that each composition parameter change the shape features in a unique direction in shape feature space.
Instead, we find that three shape composition parameters change the micelle shape in the same direction, meaning that at the level of a linear approximation, there are two independent combinations of these three parameters which have no effect on the micelle shape.
The other two composition parameters also change the micelle shape features in a common direction, so that there would be one combination of the parameters which have no effect on the micelle shape features.

To produce a micelle shape not falling on either of the two trends, it is necessary to change multiple micelle composition parameters at once.
For ease of shape design, we would hope that the effect of simultaneously changing two composition parameters could be naively inferred by linearly extrapolating from the individual effects of the parameters.
While the effects of varying individual micelle composition parameters were not independent as we expected, they were indeed often roughly linear.
If linearity of the shape dependence is assumed, then the effect of varying any combination of micelle composition parameters can be inferred from the data presented above.
One can then determine precisely how to change the composition parameters to produce a desired shape change (e.g., to reduce normalized fluctuations while holding the curvature ratio fixed).
Additionally, by determining which composition parameters have no effect on the shape, one has freedom in picking the composition parameters.
This freedom may be used to choose the most convenient parameters resulting in a desired shape.
To test if things are this simple in practice, we have performed simulations where two micelle composition parameters are varied simultaneously. %

In the first set of simulations, $\nc$ and $\no$ are varied to interpolate between the $\nch=214\%$ and $\noh=33\%$ data points of \cref{tab:coreShapes} and \cref{tab:nPhobeShapes}.
The simulated shapes are shown in \cref{tab:coreNPhobeShapes}.
\begin{table}
	\newcommand\T{\rule{0pt}{2.6ex}}
	\begin{tabular}{|c|ccccc|}
		\hline 
		$\nch$ \T & \small 100\% & \small 114\% &\small 129\% &\small 143\% & 157\% \\
		$\noh$ & 33\% & 42\% & 50\% &58\% & 67\%   \\
		&
		\raisebox{-.5\height}{\includegraphics[width=\figureSize \linewidth]{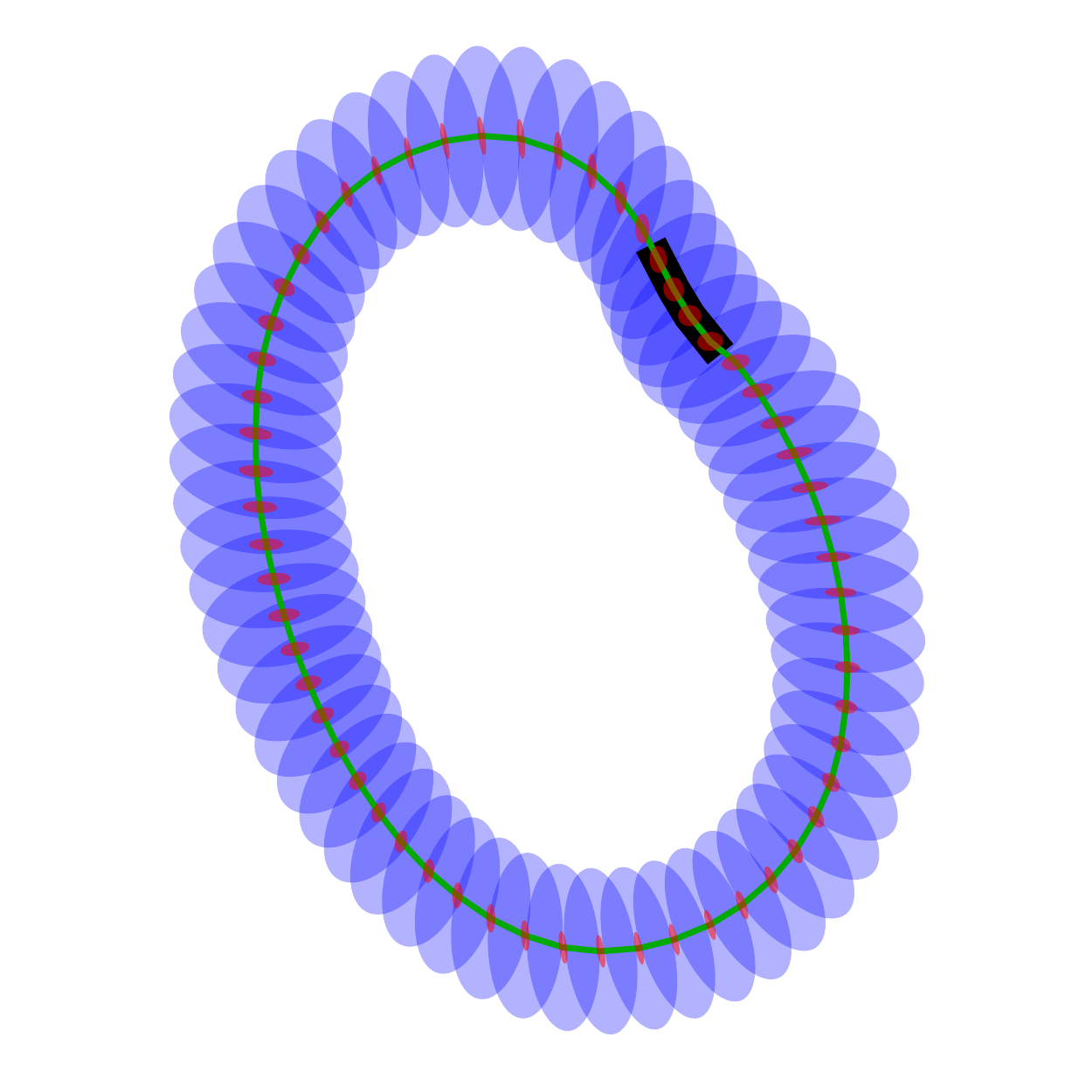}}
		&
		\raisebox{-.5\height}{\includegraphics[width=\figureSize \linewidth]{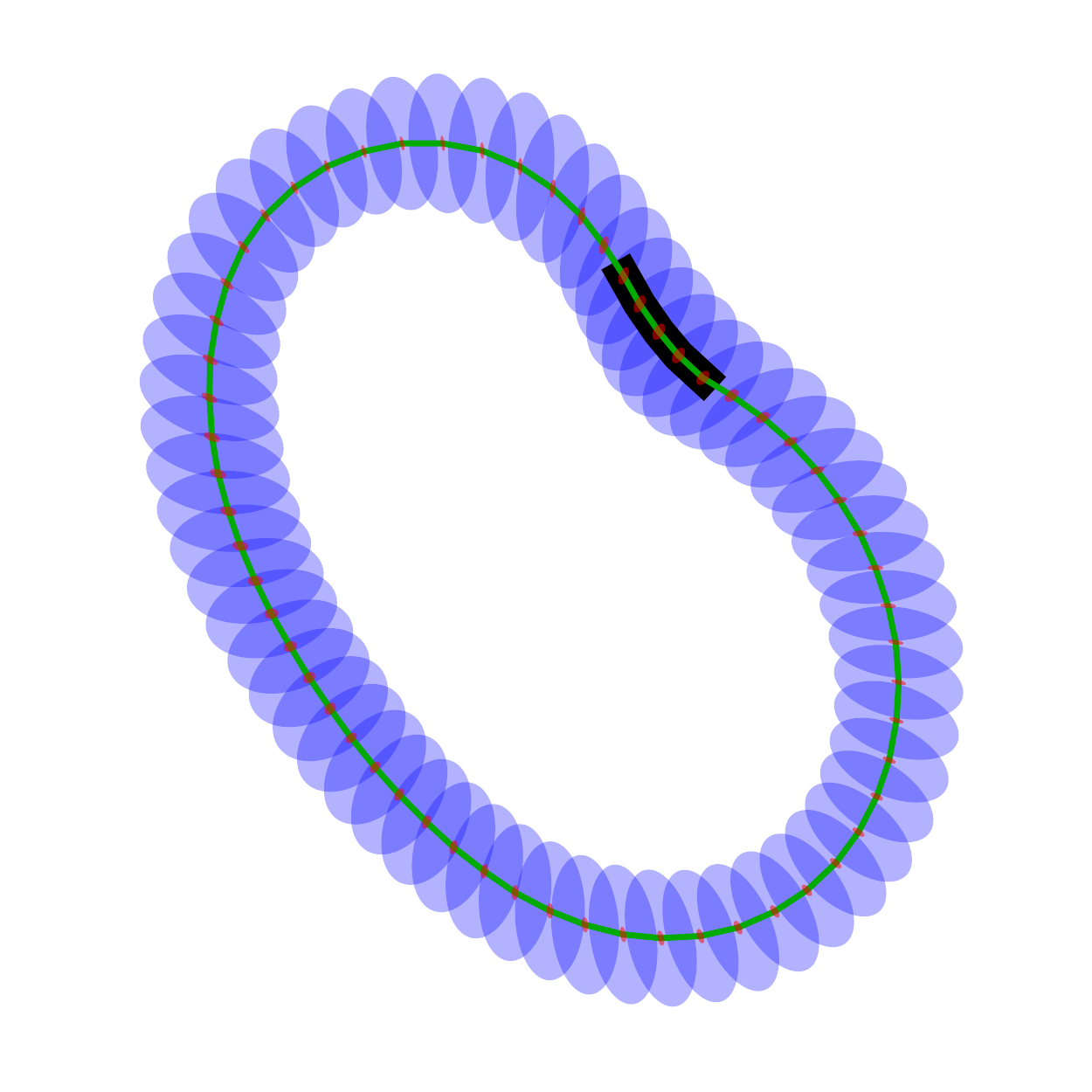}}
		&
		\raisebox{-.5\height}{\includegraphics[width=\figureSize \linewidth]{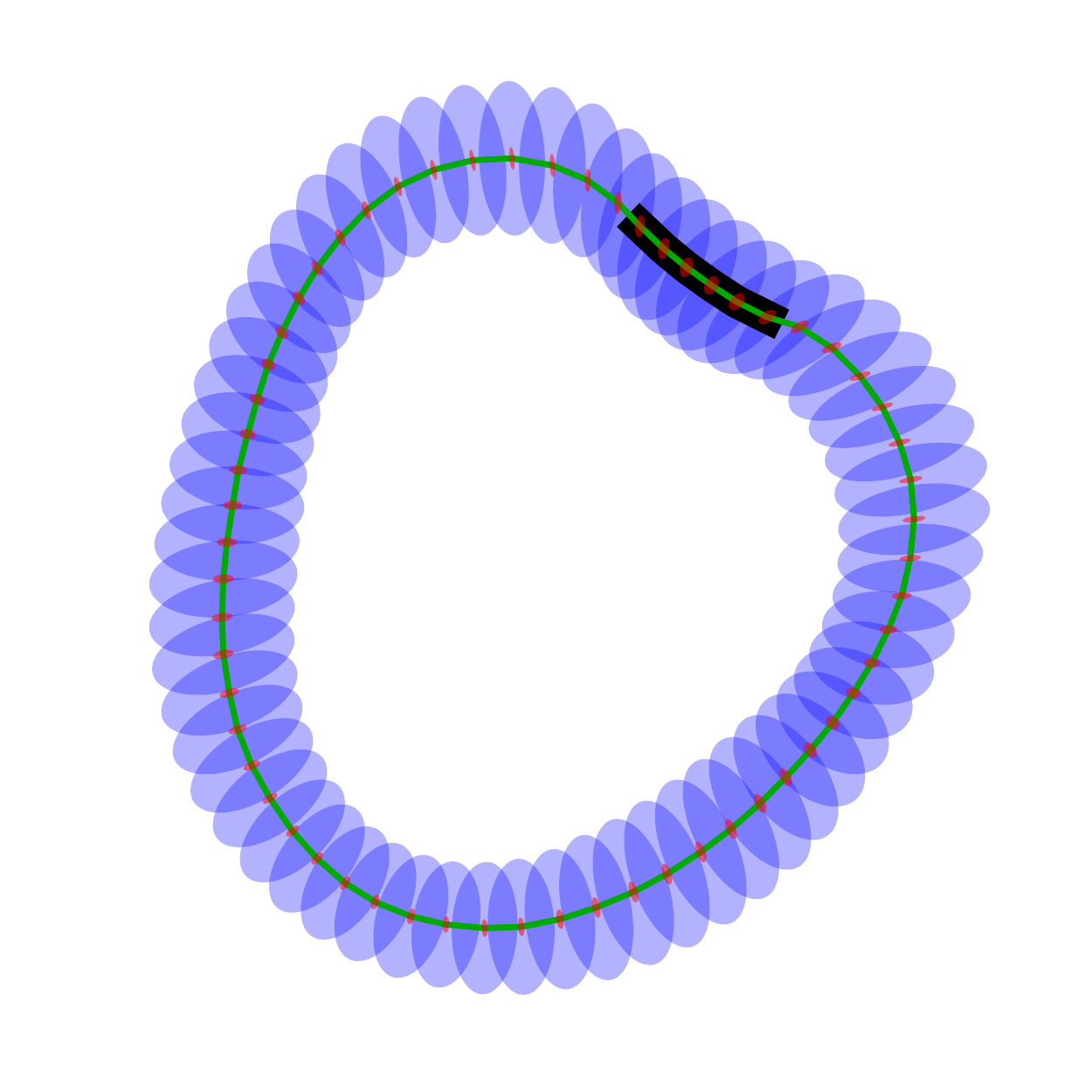}}
		&
		\raisebox{-.5\height}{\includegraphics[width=\figureSize \linewidth]{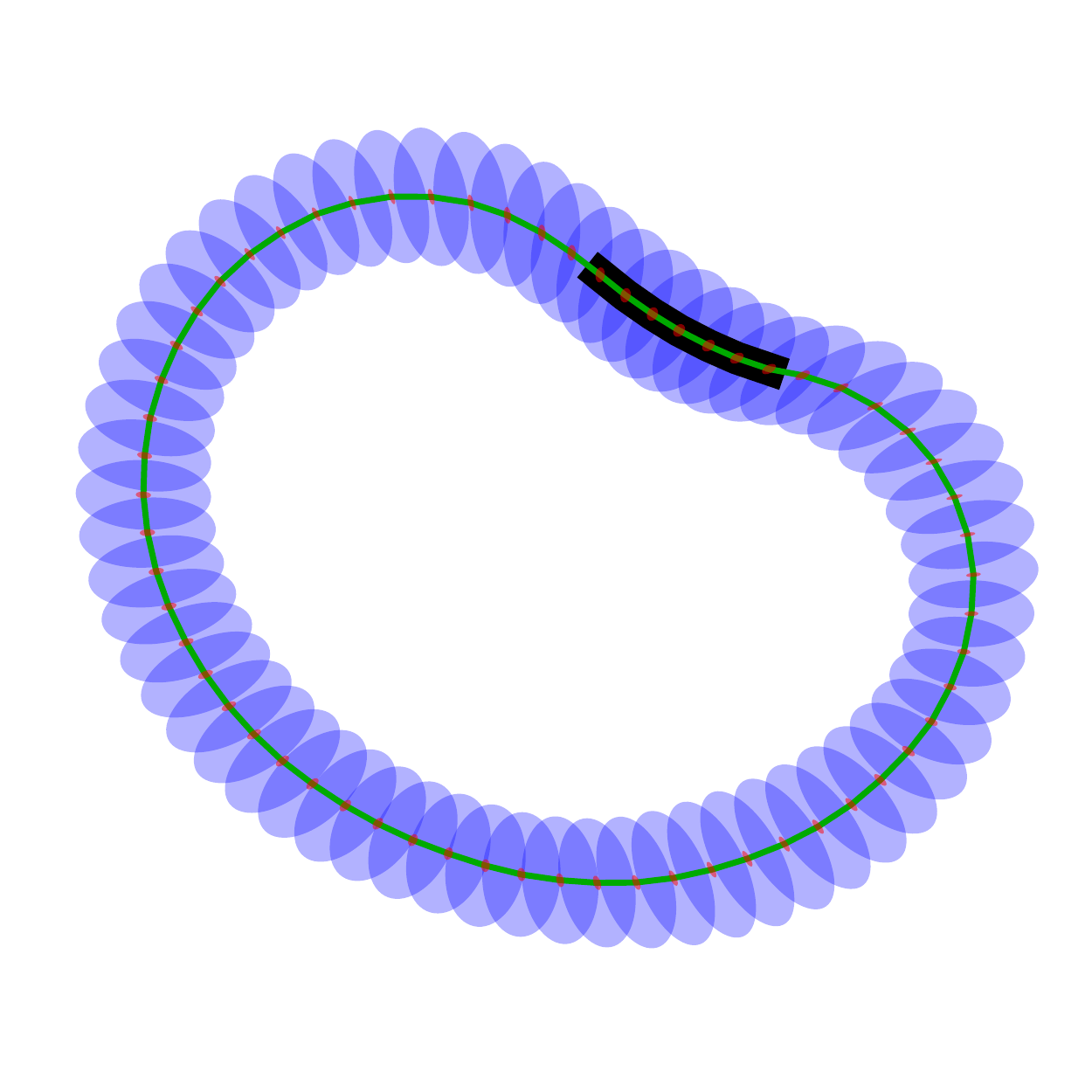}}
		&
		\raisebox{-.5\height}{\includegraphics[width=\figureSize \linewidth]{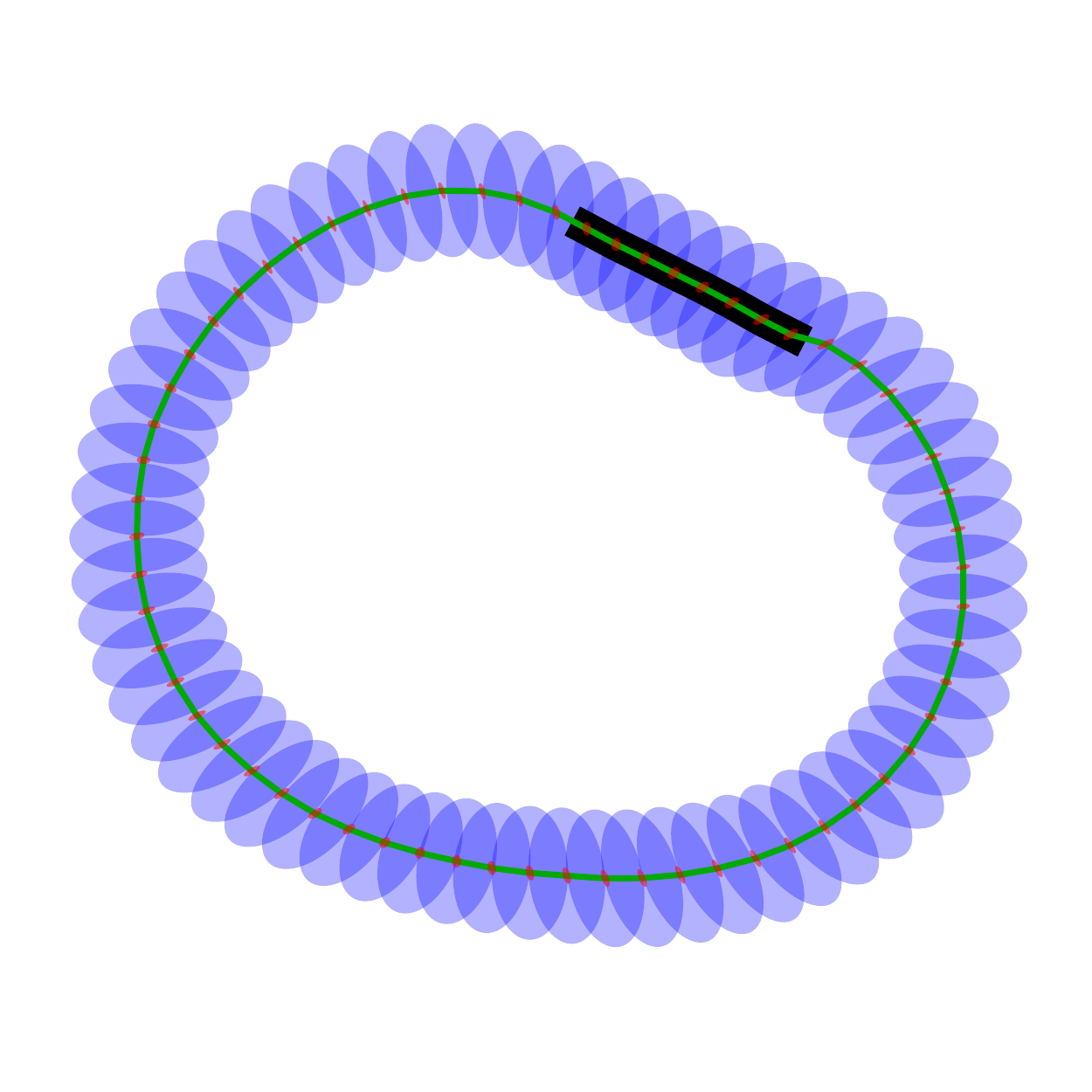}}
		
		\\
		\hline
				$\nch$\T  & \small 171\% &\small 186\% & \small 200\% & \small 214\% & \\
		$\noh$ & 75\% & 83\% & 92\% & 100\%  & \\

		&
		\raisebox{-.5\height}{\includegraphics[width=\figureSize \linewidth]{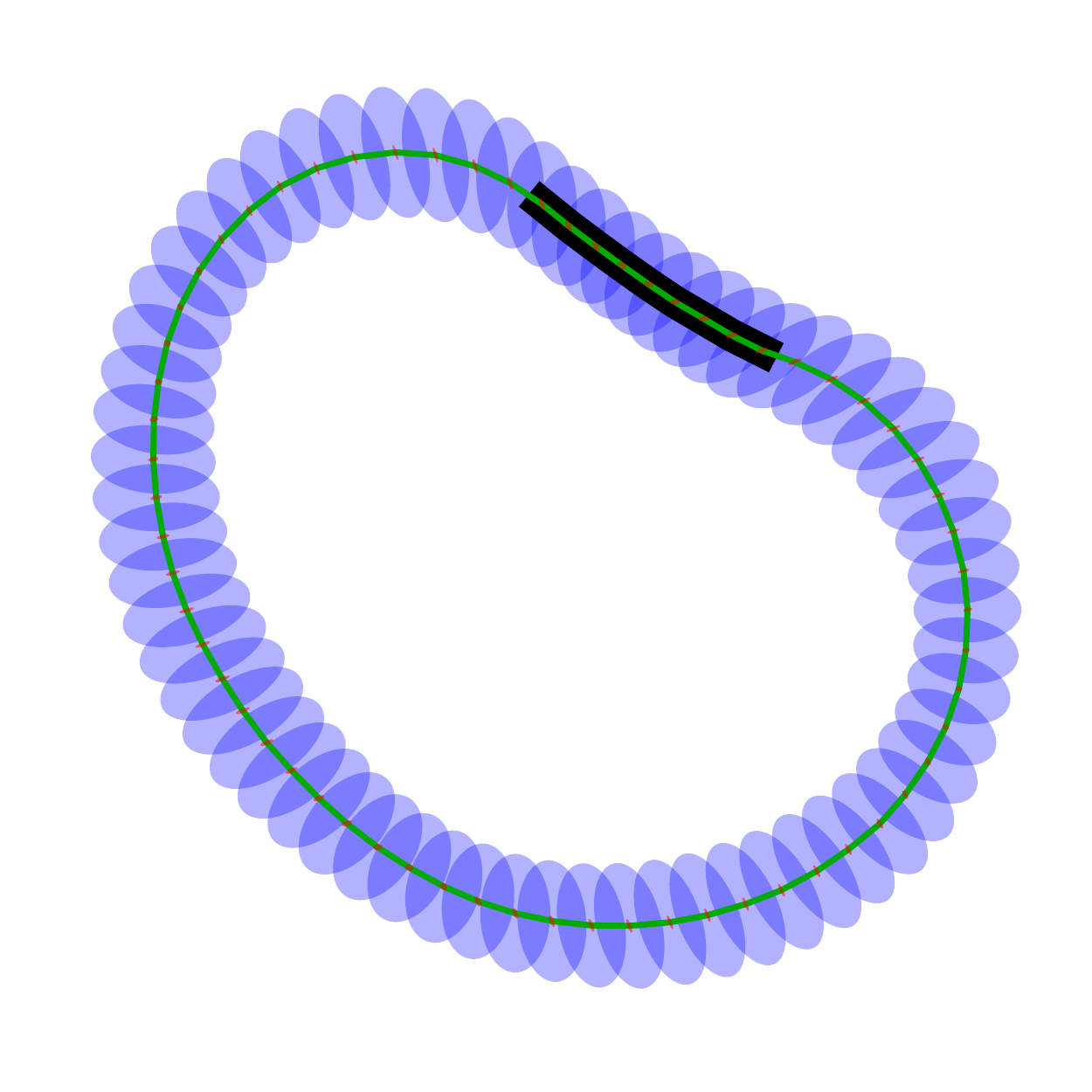}}
		&
		\raisebox{-.5\height}{\includegraphics[width=\figureSize \linewidth]{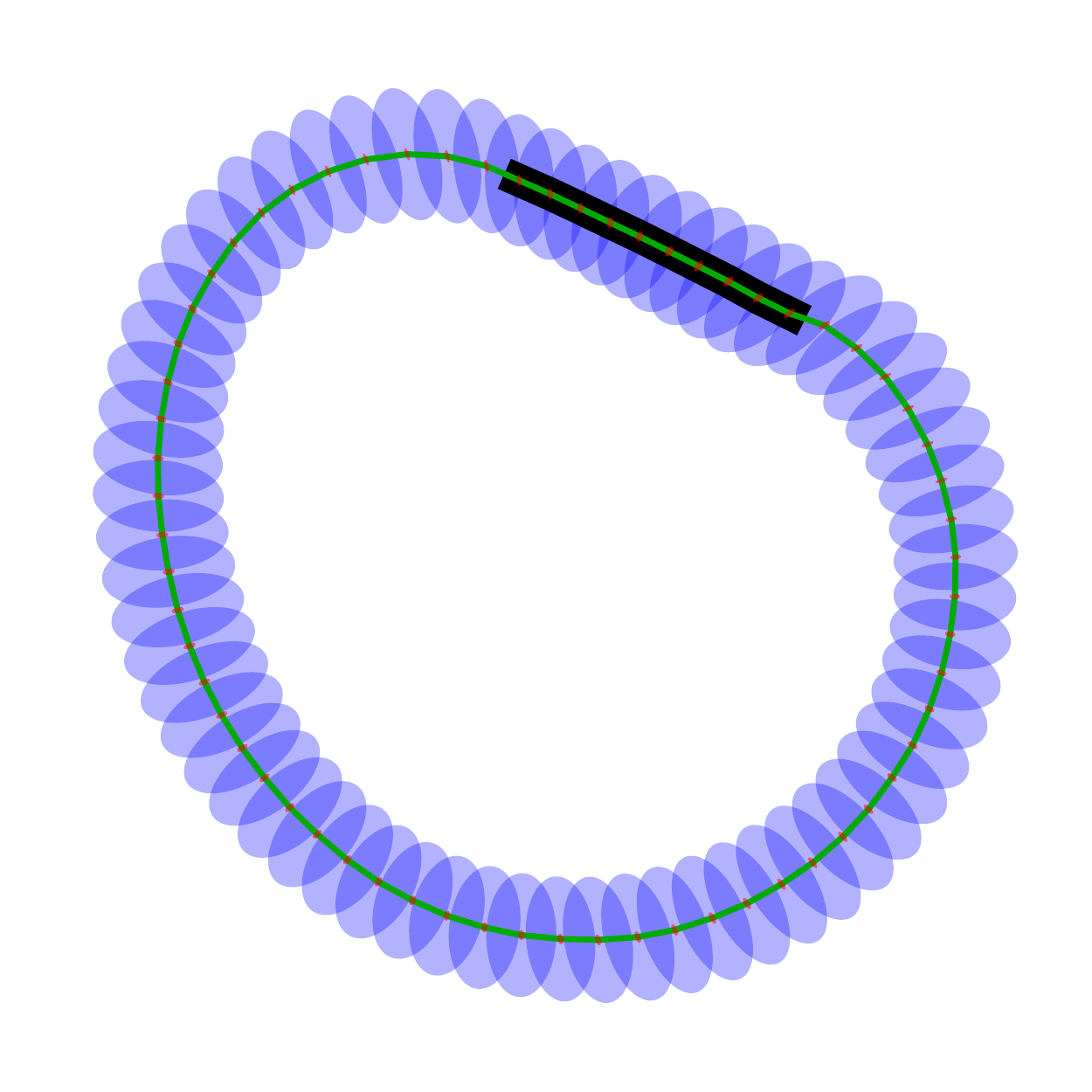}}
		&
		\raisebox{-.5\height}{\includegraphics[width=\figureSize \linewidth]{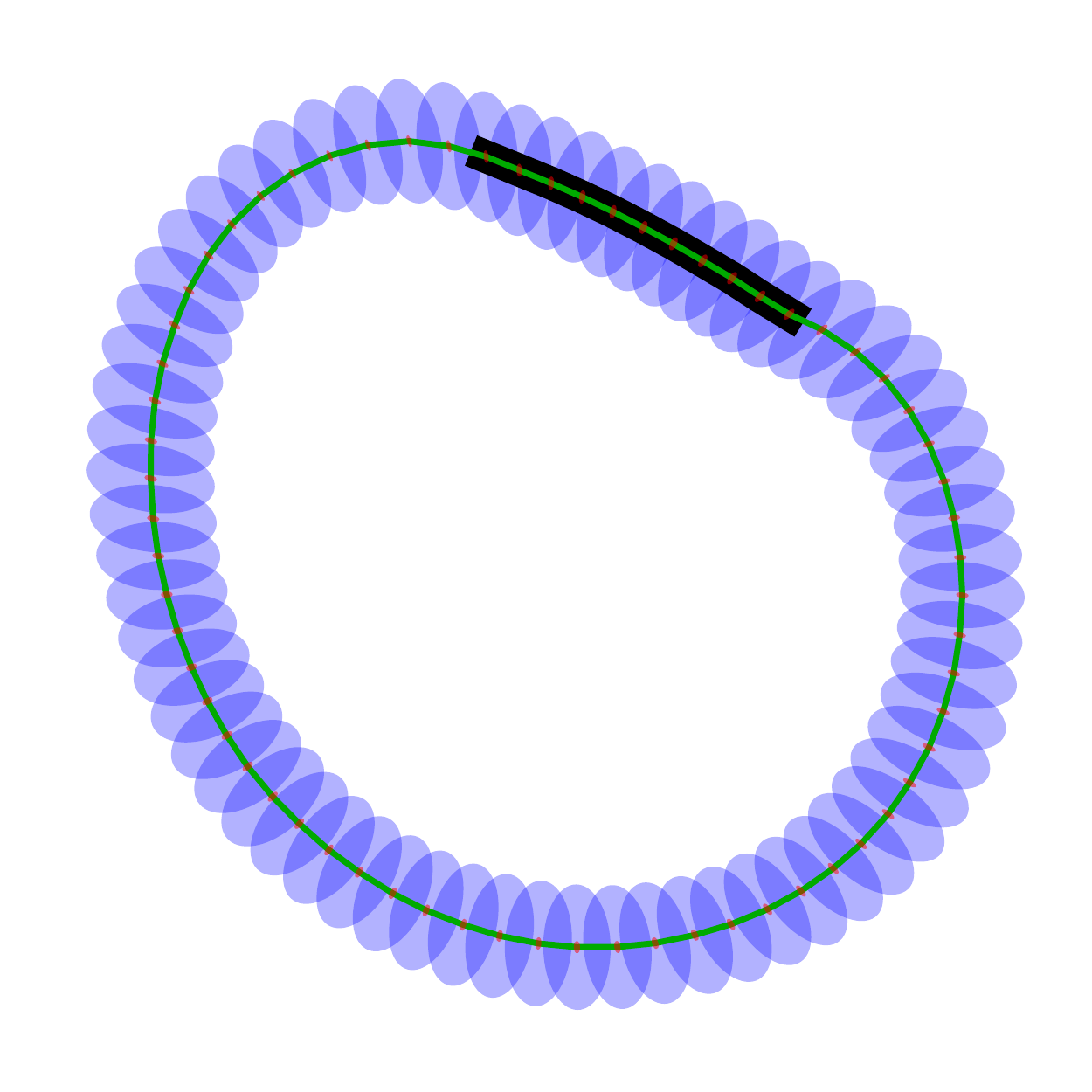}}
		&
		\raisebox{-.5\height}{\includegraphics[width=\figureSize \linewidth]{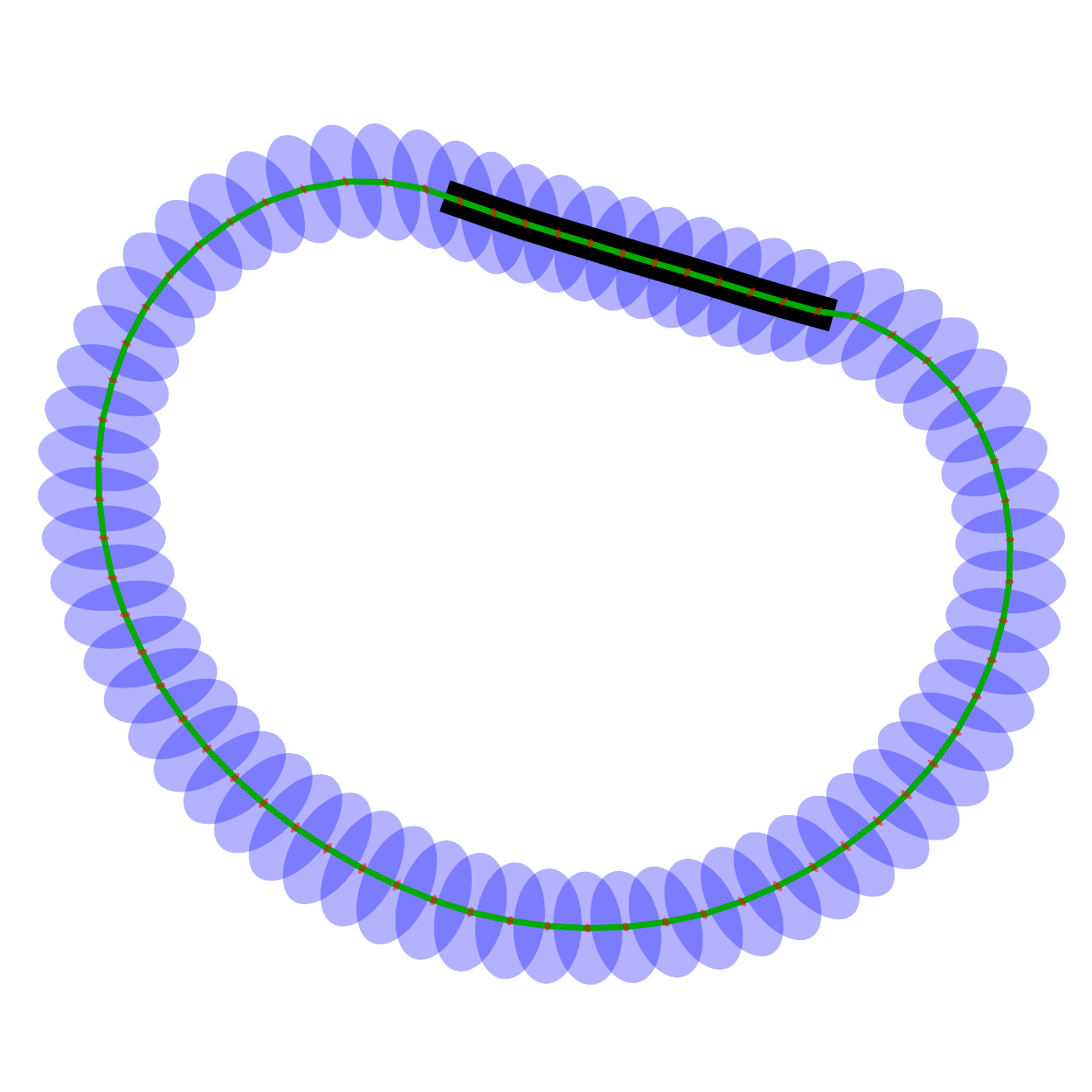}}
		&
		\\
		\hline
	\end{tabular} 
	\caption{
	Mean micelle shapes, thermal fluctuations, and errors in mean micelle shapes (illustrated in the manner of \cref{fig:baseCaseAverage}) for micelles interpolating between the $\nch=214\%$ composition of \cref{tab:coreShapes} and the $\noh=33\%$ composition of \cref{tab:nPhobeShapes}.
	The shape features associated with these data are plotted in \cref{fig:coreNPhobe}.
	}
	\label{tab:coreNPhobeShapes}
\end{table}

\begin{figure}
		\centering
	\includegraphics[width=\linewidth]{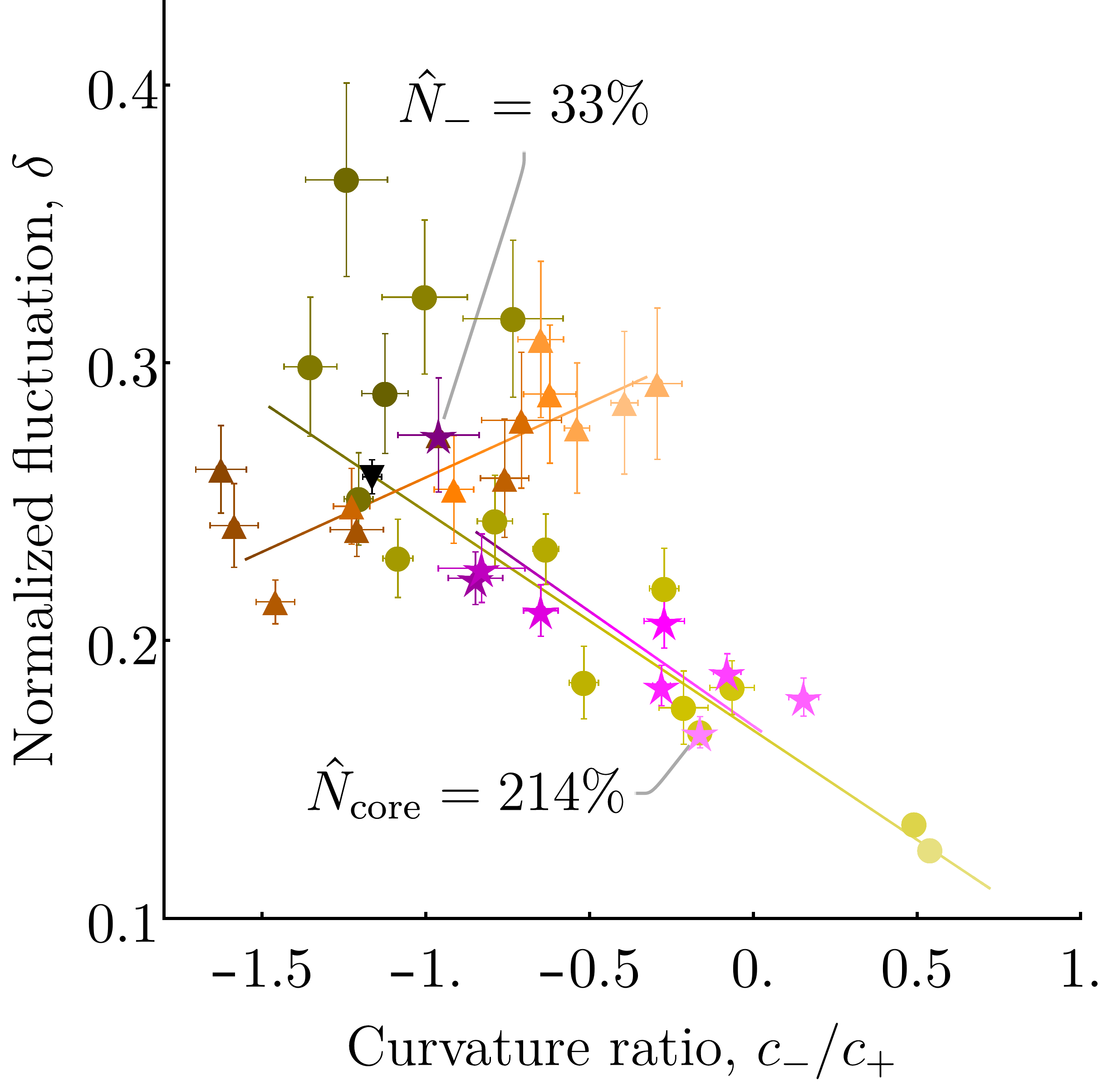}
	\caption{
	Scatter plot showing the curvature ratios and normalized fluctuations of the data shown in \cref{tab:coreNPhobeShapes} (magenta stars).
	The scatter plot and the fit lines are made in the style of \cref{fig:combinedResults}, and the data sets from varying only $\nc$ (yellow circles) and only $\no$ (orange upward-pointing triangles) are reproduced in this figure.
	Notice that the for the data set where only $\no$ is changed, the micelle with the smallest value of $\noh$, namely $33\%$, has a weaker dimple than, and therefore appears to the right of, the reference micelle (downward pointing black triangle).
	By contrast, most other shapes from micelles with $\noh < 100\%$  have stronger dimples than the reference micelle and therefore appear to its left.
	Despite this variation in micelle shape caused by changing $\no$, the magenta points, representing a simultaneous variation in both $\nc$ and $\no$, follow the same trend as the yellow points representing a variation only in $\nc$.
	}
	\label{fig:coreNPhobe}
\end{figure}

To get a closer look on the effect on the shape features, we plot in \cref{fig:coreNPhobe} the curvature ratios and normalized fluctuations for the shapes in \cref{tab:coreNPhobeShapes} as well as the results of individually varying $\nc$ and $\no$.
We see that the interpolating micelle compositions produce shape features that mainly lie along the trend of the data set where just $\nc$ is varied, contrary to the naive expectation that these shape features should be a linear combination both of the shapes feature resulting from varying $\nc$ as well as those resulting from varying $\no$.
To explain this, we hypothesize that the micelles with a large value of $\nc$ have a surface tension so large that the change in spontaneous curvature profile caused by changing $\no$ does not have a noticeable effect on the micelle.

\begin{figure*}
\begin{minipage}{\textwidth}
\centering%
\subfloat[][]{%
\label{fig:nPhilRPhilTriangle}%
\centering%
\includegraphics[width=.5\linewidth]{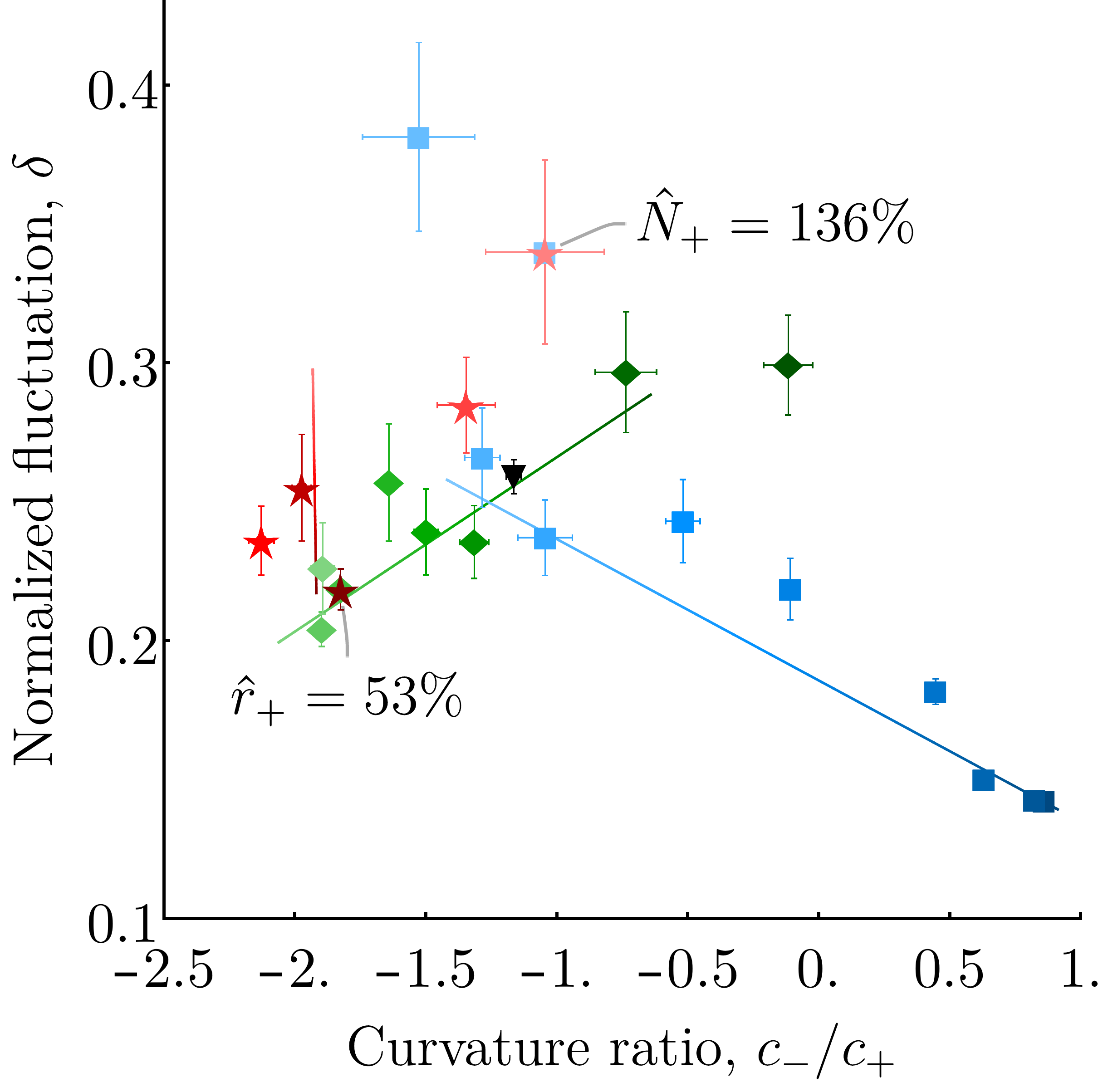}}
\begin{minipage}[b][][c]{.5\textwidth}
\subfloat[][]{%
\label{fig:rPhilnPhilCurvatureRatio}%
\centering%
\includegraphics[width=.55\linewidth]{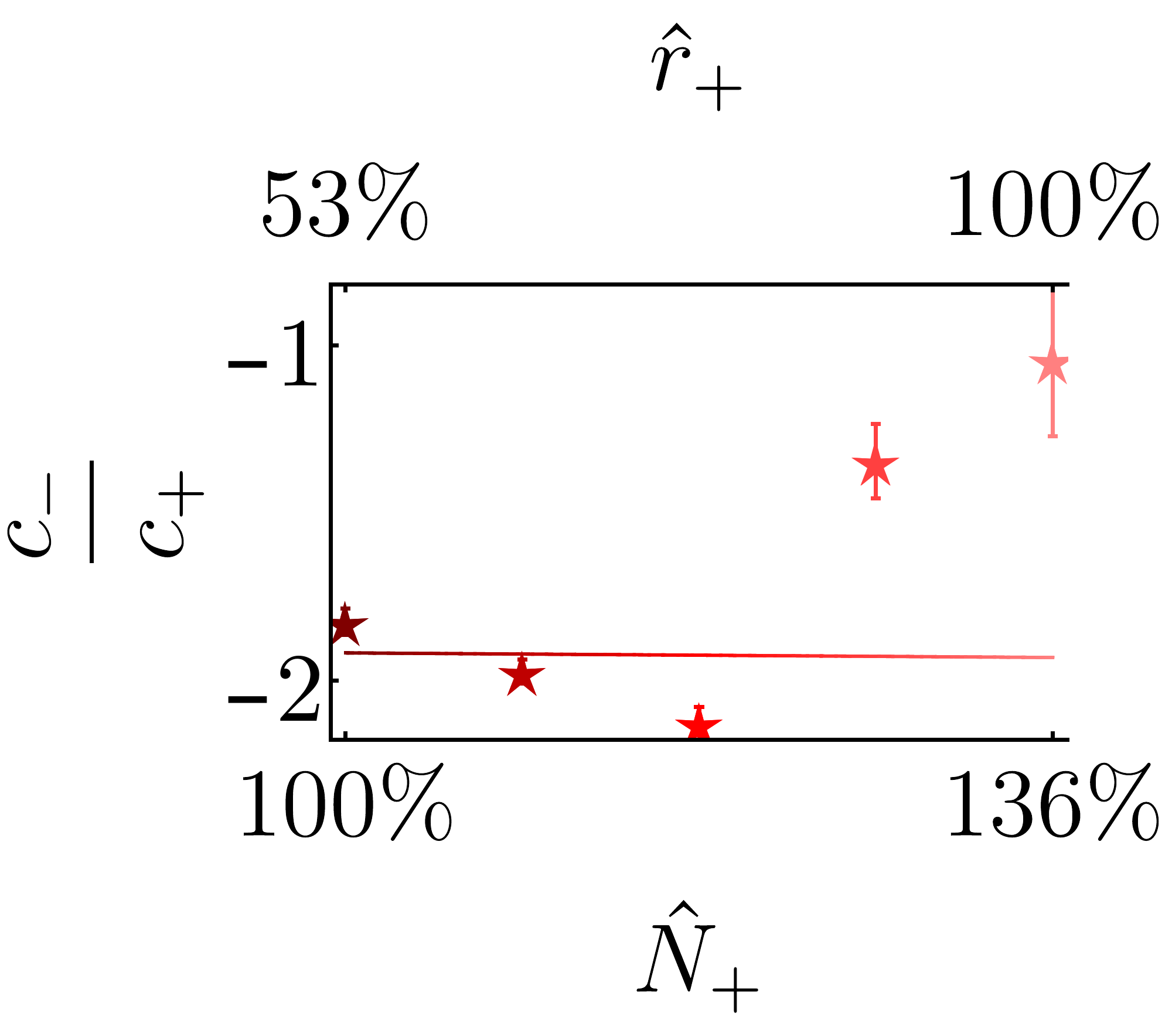}}%

\subfloat[][]{%
\label{fig:nPhilRPhilFluctuation}%
\centering%
\includegraphics[width=.55\linewidth]{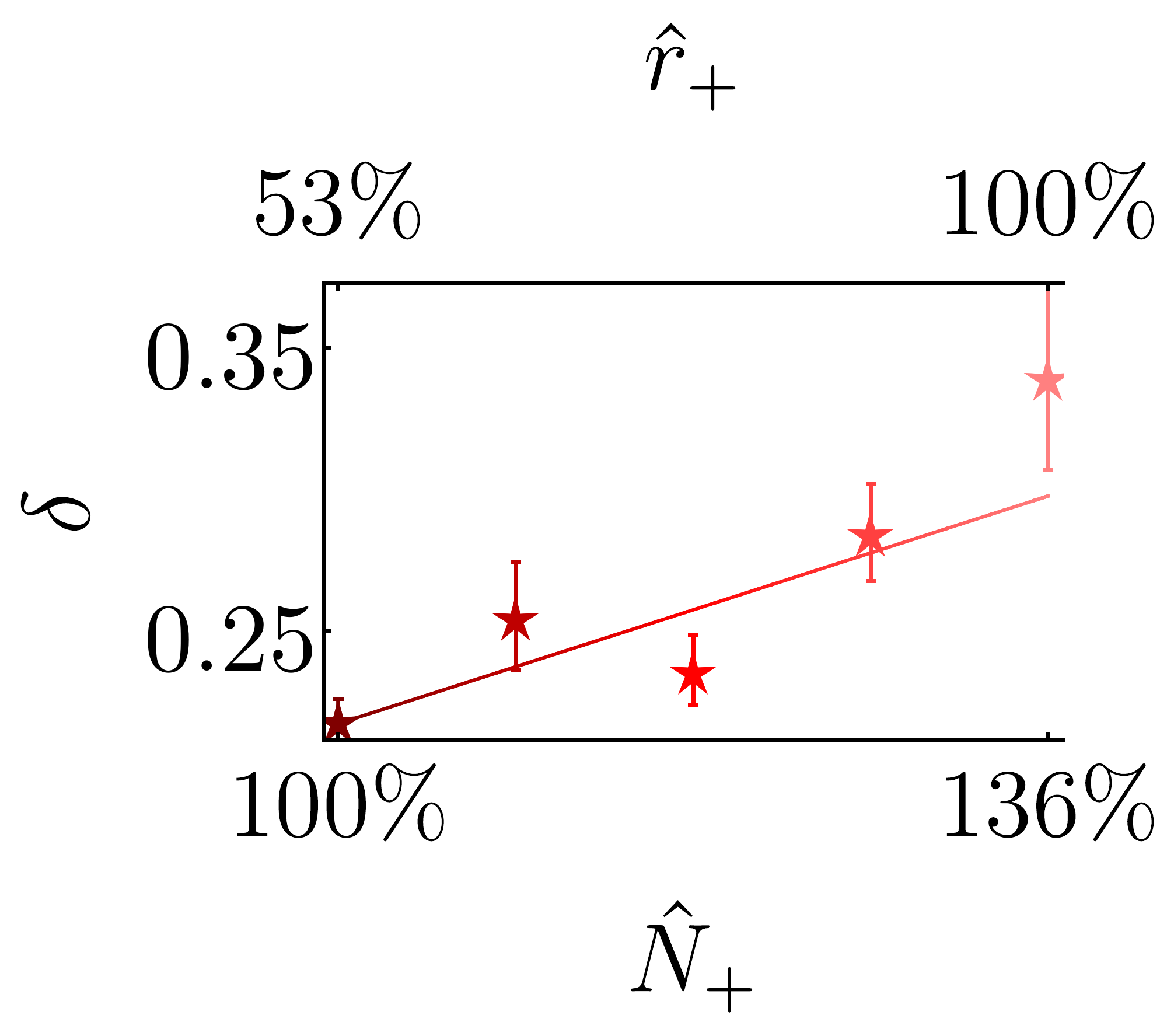}}%
\end{minipage}
\end{minipage}

	\caption{\protect\subref{fig:nPhilRPhilTriangle} Scatter plot showing the curvature ratios and normalized fluctuations of the data shown in \cref{tab:nPhilRPhilShapes} (red stars).
	The scatter plot and the fit lines are made in the style of \cref{fig:combinedResults}, and the data sets from varying only $\ni$ (blue squares) and only $\ri$ (green diamonds) are reproduced in this figure.
	The data from \cref{tab:nPhilRPhilShapes} exhibit a nonmonotonic variation in the curvature ratio, leading to the linear fit of the shape feature dependence on our chosen combination of $\nih$ and $\rih$ to have almost no variation in $\CR$ (resulting in a vertical fit line), despite the significant variation shown by the data.
	\protect\subref{fig:rPhilnPhilCurvatureRatio} and \protect\subref{fig:nPhilRPhilFluctuation} individual dependences of $\CR$ and $\delta$, respectively, on the simultaneous variation in $\nih$ and $\rih$ of \cref{tab:nPhilRPhilShapes} (i.e., red stars of \protect\subref{fig:nPhilRPhilTriangle}), plotted in the style of \cref{fig:corePlots}.
	\protect\subref{fig:rPhilnPhilCurvatureRatio} provides a clear visualization of the nonmonotonicity in $\CR$ and the resulting near constancy of the best fit.
	}
	\label{fig:nPhilRPhil}
\end{figure*}

In the second set of simulations, $\ni$ and $\ri$ are varied to interpolate between the $\nih=136\%$ and $\rih=53\%$ data points of \cref{tab:nPhilShapes} and \cref{tab:rPhilShapes}.
The simulated shapes are shown in \cref{tab:nPhilRPhilShapes}.
\begin{table}
	\newcommand\T{\rule{0pt}{2.6ex}}
	\begin{tabular}{|c|ccccc|}
		\hline
		$\nih$ \T & \small 100\% & \small 109\% &\small 118\% &\small 127\% & 136\%  \\
		$\rih$ & 53\%& 65\% & 76\%& 88\%& 100\%  \\
		&
		\raisebox{-.5\height}{\includegraphics[width=\figureSize \linewidth]{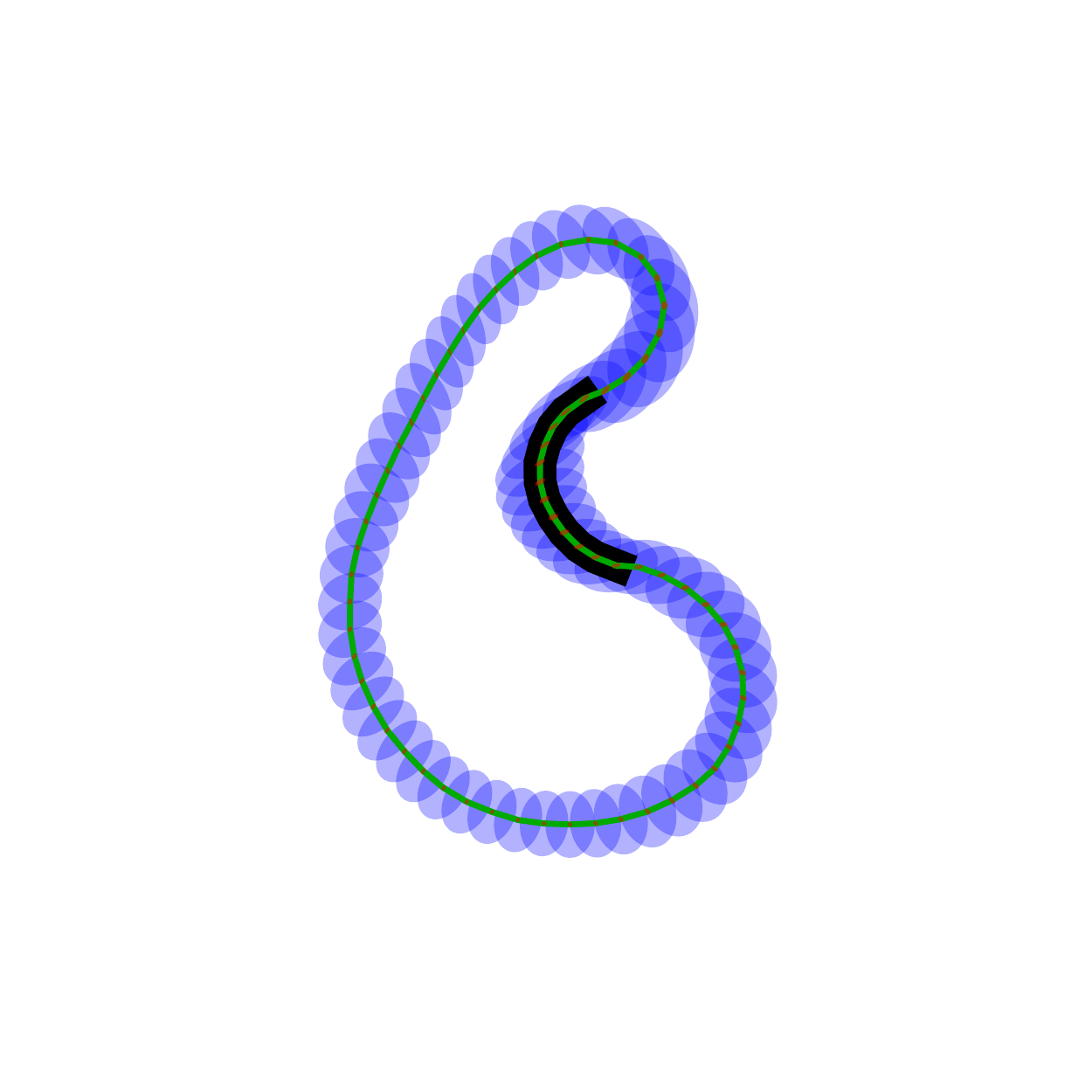}}
		&
		\raisebox{-.5\height}{\includegraphics[width=\figureSize \linewidth]{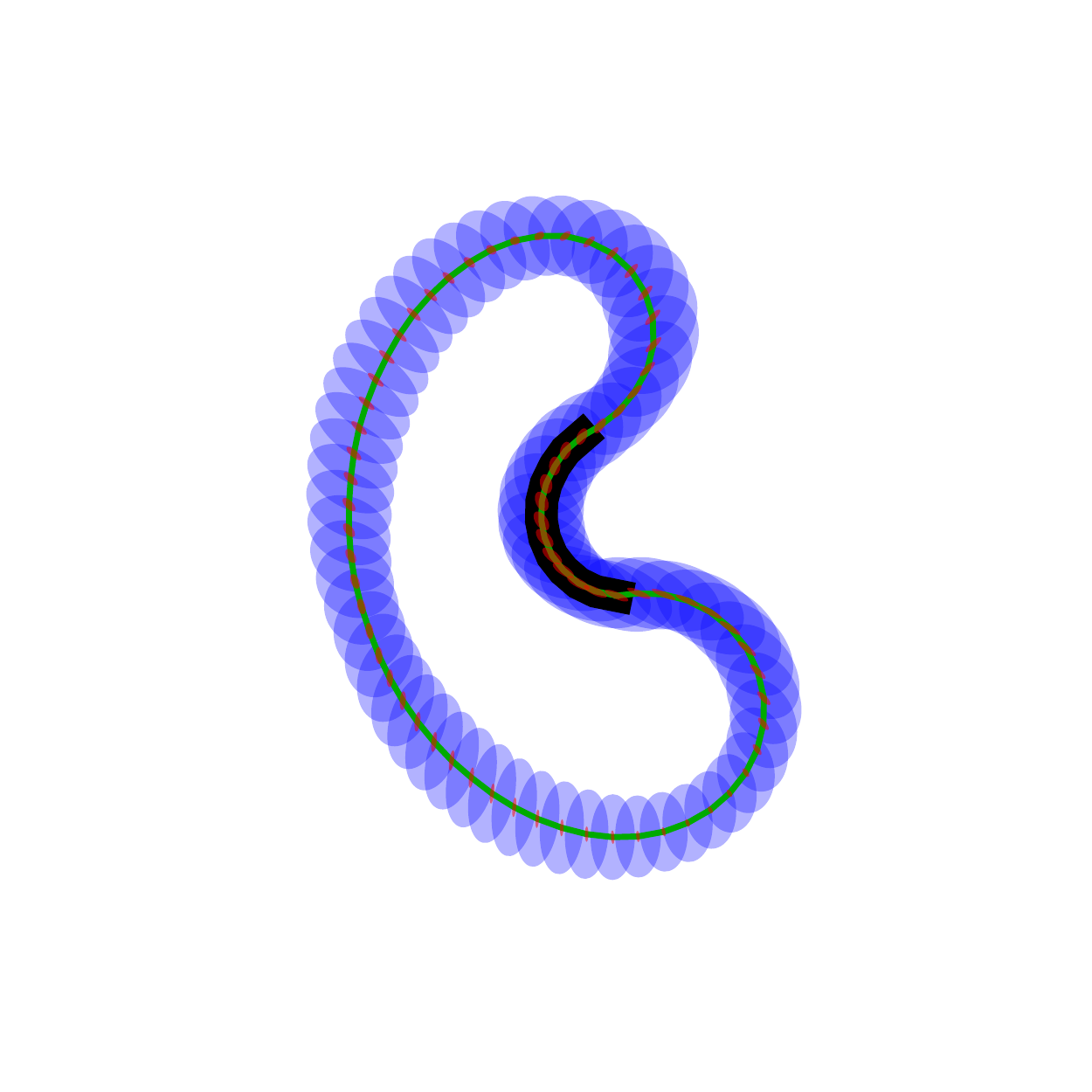}}
		&
		\raisebox{-.5\height}{\includegraphics[width=\figureSize \linewidth]{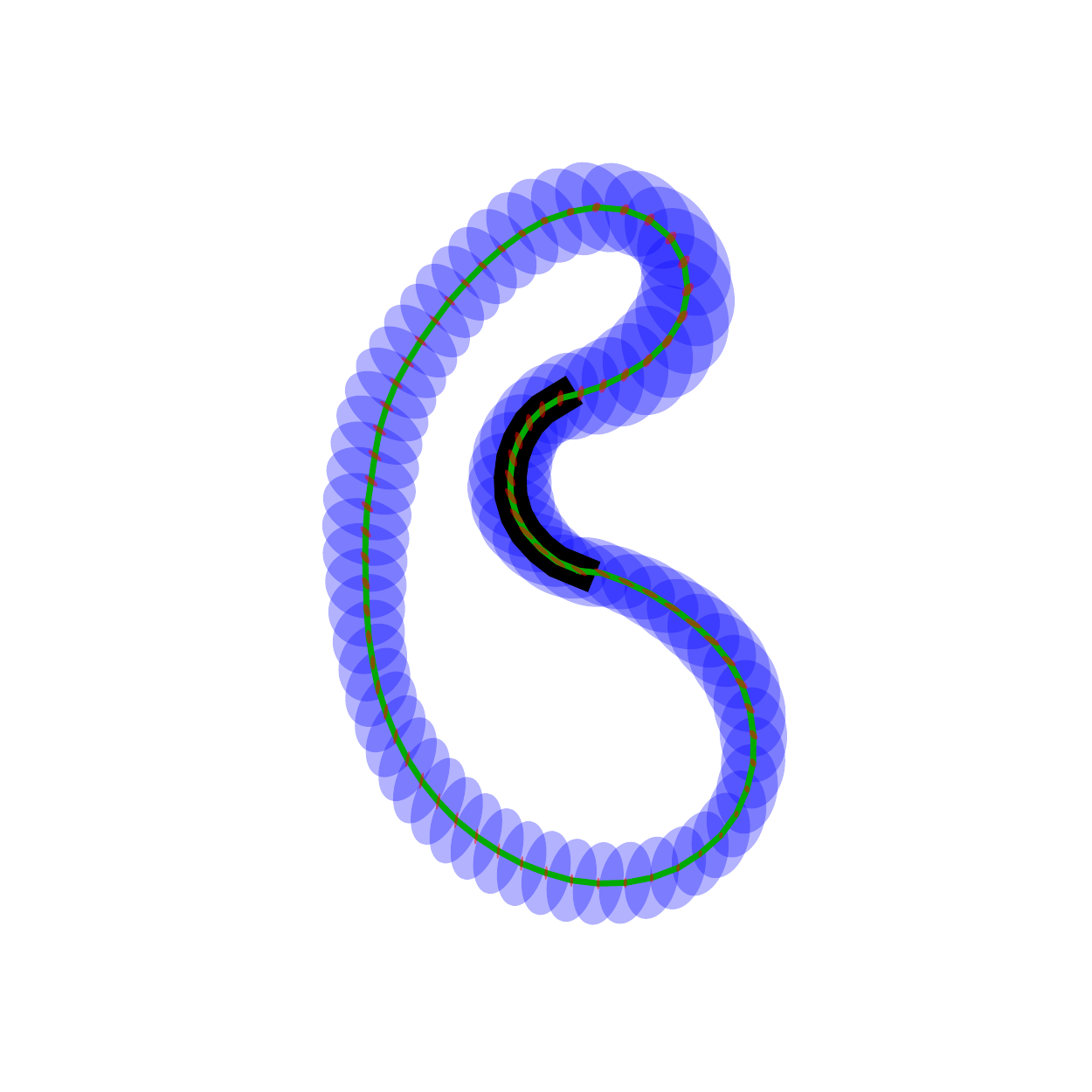}}
		&
		\raisebox{-.5\height}{\includegraphics[width=\figureSize \linewidth]{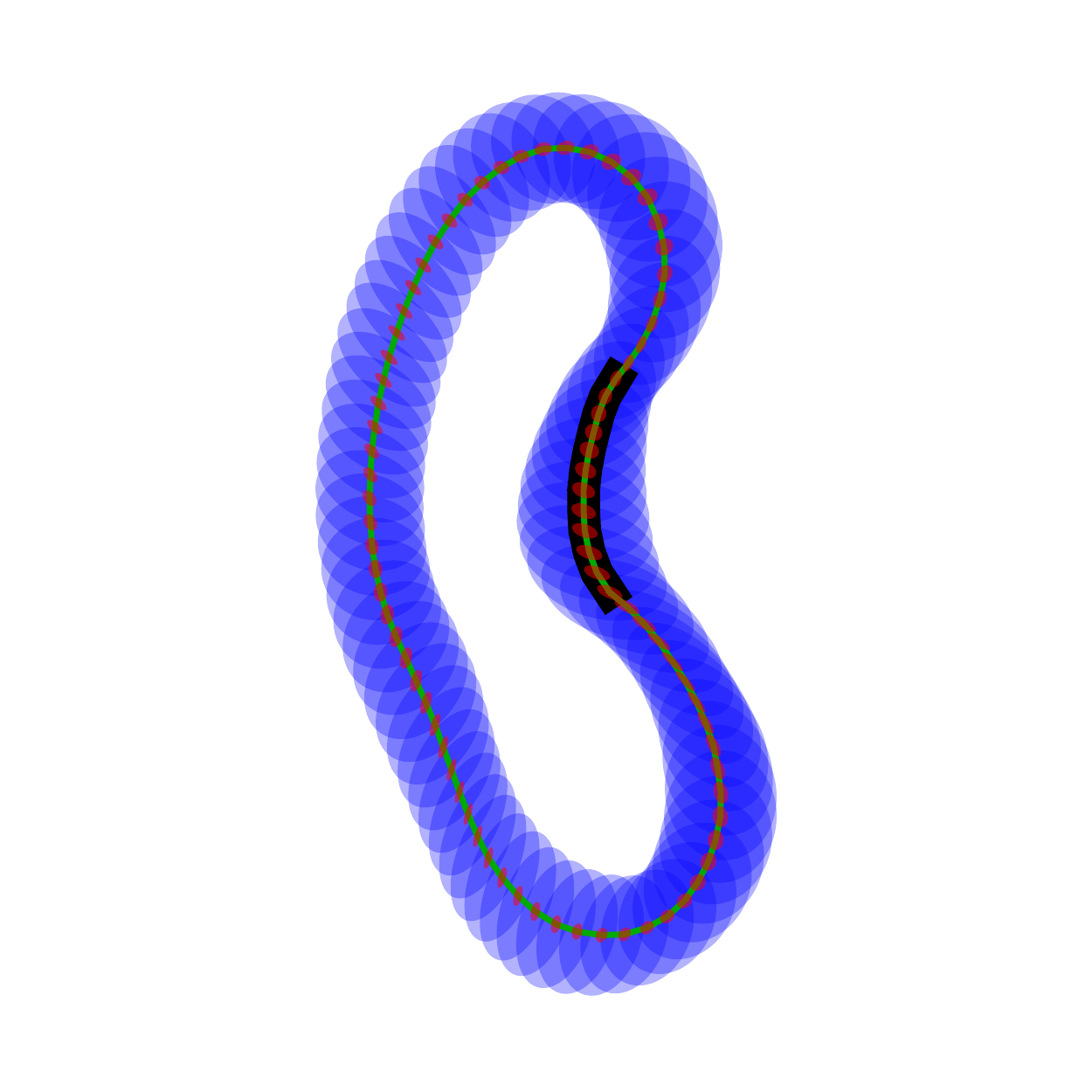}}
		&
		\raisebox{-.5\height}{\includegraphics[width=\figureSize \linewidth]{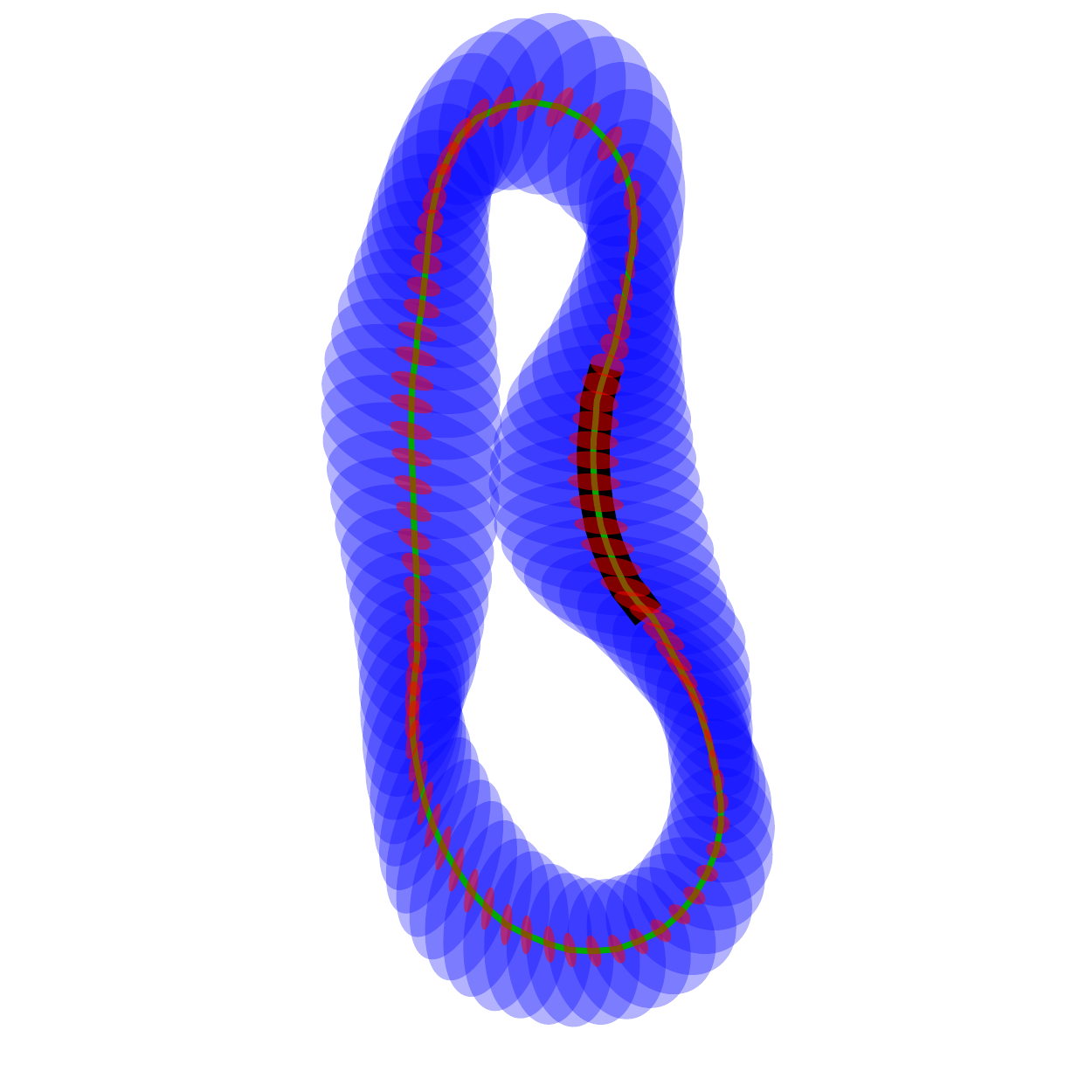}} 
		
		\\
		\hline
	\end{tabular} 
	\caption{
	Mean micelle shapes, thermal fluctuations, and errors in mean micelle shapes (illustrated in the manner of \cref{fig:baseCaseAverage}) for micelles interpolating between the $\nih=136\%$ composition of \cref{tab:nPhilShapes} and the $\rih=53\%$ composition of \cref{tab:rPhilShapes}. 
	The shape features associated with these data are plotted in \cref{fig:nPhilRPhil}.
	}
	\label{tab:nPhilRPhilShapes}
\end{table}
For a more quantitative view of the effect on the shape features, we plot in \cref{fig:nPhilRPhil} the curvature ratios and normalized fluctuations of both the shapes in \cref{tab:coreNPhobeShapes} and the previously discussed shapes of \cref{tab:nPhilShapes} and \cref{tab:rPhilShapes} which resulted from individually varying $\ni$ and $\ri$.

Ideally, the shape features would linearly, or at least monotonically, interpolate between the two extreme cases.
Taking error bars into account, the data are nearly consistent with a monotonic increase in fluctuations from the $\rih=53\%$ data point to the $\nih=136\%$ data point.
However, the curvature ratio dependence is unambiguously nonmonotonic.

To see how the dependence might not be monotonic, consider first the $\rih=53\%$ micelle composition, which has the largest asymmetry contrast of the simulated data shown in \cref{tab:nPhilRPhilShapes} and the fewest diblocks on the micelle perimeter.
On the one hand, the strong asymmetry contrast should lead to a strong dimple, but on the other hand, the decrease in the number of diblocks should lead to a higher surface tension and consequently a smaller dimple.
Next consider the micelle composition at the other extreme, having the largest number of diblocks with $\nih=136\%$.
In this case, there should be a low surface tension, as evidenced by this micelle's large normalized fluctuation, which allows for a larger dimple, but also a low asymmetry contrast which would lead to a smaller dimple.
The nonmonotonicity we observe is that there are intermediate micelles showing a stronger dimple than both of the extreme cases.

We hypothesize that as we move from the first extreme with a large asymmetry contrast and fewer diblocks to the opposite extreme, there comes a critical micelle composition where the increasing preferred perimeter set by the number of diblocks becomes large enough to completely accommodate the preferred curvature of the dimple.
Micelles with more diblocks and less asymmetry contrast than this critical composition do not benefit from the increased perimeter from the diblocks, and instead have a decreasing dimple strength set by the decreasing asymmetry contrast.
On the other hand, micelles with less diblocks and more asymmetry contrast than the critical composition experience both and increased surface tension and an increased preferred curvature.
As evidenced in \cref{fig:coreNPhobe}, the surface tension has a larger effect on the dimple strength, and so we expect dimple strength diminishes.
By this logic, there should be a maximum dimple strength near the critical composition, and so the curvature ratio dependence should be nonmonotonic.
In any event, this example shows that a linear interpolation is insufficient to approximate the behavior of the shape features between two micelle compositions, since nonmonotonic behavior is possible.

\section{Discussion}
\label{sec:discussion}
In \cref{sec:results}, we showed how the micelle shape features depended on the composition parameters.
In this section, we discuss what implications these results have for the central questions of our work, namely whether a micelle may feasibly be designed using the rationale presented in \cref{sec:introduction}.
We begin by discussing if the micelle shape dependence is sufficiently regular to allow for arbitrary shape features to be designed using only a naive strategy.
Next, we address a shortcoming of this work mentioned in \cref{sec:introduction}, which is that the micelles shapes are only metastable.
We explain why the statistics of the metastable shapes examined here are meaningful and discuss what might be done to stabilize the micelles in practice.
Lastly, we justify why the two-dimensional simulations considered here are relevant to practical applications which necessarily have three-dimensions.

In the introduction, we set a goal of identifying good control parameters to design the shape of the micelle. 
Such control parameters ought to have a simple, easily understandable effect.
Indeed, in \cref{sec:results}, we found a smooth variation in the micelle shape features, and we were able to give plausible physical explanations of the observed behavior involving the volume enclosed by the micelle, the surface tension resulting from extension of the micelle perimeter, and the bending energy associated with the curvature of the micelle perimeter.
The explanations were not explicitly verified because it is difficult to define and independently measure the bending energies and surface tensions of the fluctuating, asymmetric micelles considered in this work.
However, we found that the shape dependence was significantly nonlinear and produced nonmonotonicities in some cases, and therefore the dependence cannot be quantitatively explained for the purposes of shape design by simple physical arguments or a naive linear model.
Therefore, if an accurate model of the relationship between the control parameters and the micelle shape is desired for facilitation of shape design, something more must be done.

One approach is to create Hamiltonian whose degrees of freedom are the junction points and which contains terms for the bulk compression of the micelle interior and the stretching and bending of the micelle surface.
It would be necessary to perform a series of simulations to determine a mapping between the micelle composition and the parameters of the simplified Hamiltonian.
Once this mapping is determined, the simplified model, having far fewer degrees of freedom, would give a much simpler and less computationally intensive way of understanding how the micelle shape depends on the micelle composition.

Alternatively, if one desires to design a single specified micelle shape, the required micelle composition could be found by some nonlinear optimization strategy, such as a genetic algorithm.
Such machine learning algorithms have been applied to the design of material properties in a number of contexts \cite{le2015,liu2017,patra2016,miskin2013}.

So far, we have considered only micelles which exhibited the intended positioning of diblocks on the surface, which we call well formed, even though as noted in \cref{sec:methods}, micelles resulting from our simulation often did not have this property.
We now give a justification for considering this seemingly biased sample.
Our justification is based on the fact that the well-formed micelles are metastable, meaning that the micelles have a significant chance of surviving the length of a simulation without forming a defect in the diblock arrangement, but there is a finite probability to form a defect from which the micelle would never recover.
With this in mind, it is natural to investigate the thermodynamic statistical properties of the well-formed micelles, and the appropriate statistical weight to each micelle configuration in this subensemble is found simply giving equal weight to each well-formed micelle while discarding the others. 

However, if this statistical analysis is to be meaningful for practical applications, something must be done to enforce that the micelles be well formed.
We view this problem as separate from the question of how the micelle composition affects the shape of the well-formed micelles, but we believe there are a few promising approaches to solving this problem.
One approach is to change the interaction parameters of the system.
The choice of parameters used in this work was motivated by the desire to have a lower energetic barrier for bead rearrangements allowing shorter simulation times, but this has the drawback of facilitating diblock rearrangements on the micelle surface.
Stronger interactions may increase the energetic penalty for micelle defects, greatly reducing their occurrence.
Another approach is to make the two species of monomers composing the solvophobic-rich diblocks different from the two species composing the solvophilic-rich diblocks.
Such a difference between the two types of diblocks could promote their segregation on the micelle surface, thereby enforcing their intended positioning.

Beyond changing the interaction parameters of the system, a further approach is to alter the polymer architecture with the idea that a different bond topology would better stabilize the well-formed micelles.
Whatever approach is taken to solve this problem, we don't expect it to significantly alter the shape dependences observed in this work, as these are a basic result of the polymer nature of the micelle.

In practical applications, the design problem considered in this work must be solved in three dimensions.
However, we have chosen to conduct two-dimensional simulations, as has often been done~\cite{larson88,miller92,chan93,huopaniemi06,siepmann06}.
We argue that since the physics affecting micelle shape (compressibility, surface tension, spontaneous surface curvature) are qualitatively unchanged, the dependence on the micelle composition of a similar three-dimensional shape, such as a dimpled sphere, should be similar.
In general, one can imagine many more target micelle shapes beyond a dimpled sphere.
In contrast to two-dimensional shapes, the three-dimensional target shapes are described by two principal curvatures at each point on the surface.
The diblock composition on the surface, however, specifies only a mean curvature at each point to lowest order~\cite{helfrich73}.
Therefore we expect that the profile of diblock compositions over the micelle surface is not in general sufficient to completely control the micelle shape in three dimensions, so that full shape control would be harder or perhaps impossible in three dimensions. 
However, some shape control must be possible, and studying the extent of this shape control is an interesting direction for future research.

\section{Conclusion}
\label{sec:conclusion}
We have described a micelle shape-design scheme, and shown its capacity to control the average shape and fluctuations of a micelle in thermal equilibrium.
We began with a reference micelle composition producing a moderately dimpled micelle, and varied, one by one, several aspects of the reference micelle composition to examine the effect on the thermal micelle shape.
We studied two features of the micelle shape in particular, and found that the dependences were somewhat smooth, but significantly nonlinear and sometimes nonmonotonic.
Additionally, simulations were conducted where two aspects of the micelle composition were changed simultaneously, with the result that the combined effect of changing two parameters could not easily be deduced by looking at the individual effects on the micelle.
Plausible rationales were given to explain these results.
Even though the relationship between the micelle composition and shape may not satisfactorily be characterized by a naive linear relationship, we believe more sophisticated methods to characterize the relationship are nonetheless possible, and we proposed examples.
We expect the principles that govern our simple two-dimensional model to extend to three dimensions, and therefore that our results provide evidence that a similar design scheme should work to produce three dimensional shape-designed micelles.

\begin{acknowledgments}
The author thanks Ishanu Chattopadhyay for discussing the applicability of machine learning algorithms to this work.
The author also thanks T. A. Witten for reviewing a manuscript of this paper.
This work is part of a Ph.D. thesis under the supervision of T. A. Witten at the University of Chicago.
This work was completed in part with resources provided by the University of Chicago Research Computing Center.
This work was principally supported by the University of Chicago Materials Research Science and Engineering Center, which is funded by the National Science Foundation under award No. 1420709.
\end{acknowledgments}

\bibliography{references/firstPaper,%
references/artin,%
references/binder11,%
references/bose2017,%
references/chan93,%
references/Decuzzi10,%
references/detcheverry2009CoarseGrained,%
references/Devarajan09,%
references/Dimitrakopoulos04,%
references/dule2015,%
references/Gillies04,%
references/helfrich73,%
references/Hsieh06,%
references/huopaniemi06,%
references/lammps,%
references/larson88,%
references/le2015,%
references/Li2015,%
references/liu2017,%
references/miller92,%
references/miskin2013,%
references/Muro08,%
references/Patil08,%
references/patra2016,%
references/qiu2015,%
references/siepmann06,%
references/Wang91%
} 

\begin{thebibliography}{27}%
\makeatletter
\providecommand \@ifxundefined [1]{%
 \@ifx{#1\undefined}
}%
\providecommand \@ifnum [1]{%
 \ifnum #1\expandafter \@firstoftwo
 \else \expandafter \@secondoftwo
 \fi
}%
\providecommand \@ifx [1]{%
 \ifx #1\expandafter \@firstoftwo
 \else \expandafter \@secondoftwo
 \fi
}%
\providecommand \natexlab [1]{#1}%
\providecommand \enquote  [1]{``#1''}%
\providecommand \bibnamefont  [1]{#1}%
\providecommand \bibfnamefont [1]{#1}%
\providecommand \citenamefont [1]{#1}%
\providecommand \href@noop [0]{\@secondoftwo}%
\providecommand \href [0]{\begingroup \@sanitize@url \@href}%
\providecommand \@href[1]{\@@startlink{#1}\@@href}%
\providecommand \@@href[1]{\endgroup#1\@@endlink}%
\providecommand \@sanitize@url [0]{\catcode `\\12\catcode `\$12\catcode
  `\&12\catcode `\#12\catcode `\^12\catcode `\_12\catcode `\%12\relax}%
\providecommand \@@startlink[1]{}%
\providecommand \@@endlink[0]{}%
\providecommand \url  [0]{\begingroup\@sanitize@url \@url }%
\providecommand \@url [1]{\endgroup\@href {#1}{\urlprefix }}%
\providecommand \urlprefix  [0]{URL }%
\providecommand \Eprint [0]{\href }%
\providecommand \doibase [0]{http://dx.doi.org/}%
\providecommand \selectlanguage [0]{\@gobble}%
\providecommand \bibinfo  [0]{\@secondoftwo}%
\providecommand \bibfield  [0]{\@secondoftwo}%
\providecommand \translation [1]{[#1]}%
\providecommand \BibitemOpen [0]{}%
\providecommand \bibitemStop [0]{}%
\providecommand \bibitemNoStop [0]{.\EOS\space}%
\providecommand \EOS [0]{\spacefactor3000\relax}%
\providecommand \BibitemShut  [1]{\csname bibitem#1\endcsname}%
\let\auto@bib@innerbib\@empty
\bibitem [{\citenamefont {Decuzzi}\ \emph {et~al.}(2010)\citenamefont
  {Decuzzi}, \citenamefont {Godin}, \citenamefont {Tanaka}, \citenamefont
  {Lee}, \citenamefont {Chiappini}, \citenamefont {Liu},\ and\ \citenamefont
  {Ferrari}}]{Decuzzi10}%
  \BibitemOpen
  \bibfield  {author} {\bibinfo {author} {\bibfnamefont {P.}~\bibnamefont
  {Decuzzi}}, \bibinfo {author} {\bibfnamefont {B.}~\bibnamefont {Godin}},
  \bibinfo {author} {\bibfnamefont {T.}~\bibnamefont {Tanaka}}, \bibinfo
  {author} {\bibfnamefont {S.-Y.}\ \bibnamefont {Lee}}, \bibinfo {author}
  {\bibfnamefont {C.}~\bibnamefont {Chiappini}}, \bibinfo {author}
  {\bibfnamefont {X.}~\bibnamefont {Liu}}, \ and\ \bibinfo {author}
  {\bibfnamefont {M.}~\bibnamefont {Ferrari}},\ }\href {\doibase
  https://doi.org/10.1016/j.jconrel.2009.10.014} {\bibfield  {journal}
  {\bibinfo  {journal} {Journal of Controlled Release}\ }\textbf {\bibinfo
  {volume} {141}},\ \bibinfo {pages} {320 } (\bibinfo {year}
  {2010})}\BibitemShut {NoStop}%
\bibitem [{\citenamefont {Devarajan}\ \emph {et~al.}(2010)\citenamefont
  {Devarajan}, \citenamefont {Jindal}, \citenamefont {Patil}, \citenamefont
  {Mulla}, \citenamefont {Gaikwad},\ and\ \citenamefont {Samad}}]{Devarajan09}%
  \BibitemOpen
  \bibfield  {author} {\bibinfo {author} {\bibfnamefont {P.~V.}\ \bibnamefont
  {Devarajan}}, \bibinfo {author} {\bibfnamefont {A.~B.}\ \bibnamefont
  {Jindal}}, \bibinfo {author} {\bibfnamefont {R.~R.}\ \bibnamefont {Patil}},
  \bibinfo {author} {\bibfnamefont {F.}~\bibnamefont {Mulla}}, \bibinfo
  {author} {\bibfnamefont {R.~V.}\ \bibnamefont {Gaikwad}}, \ and\ \bibinfo
  {author} {\bibfnamefont {A.}~\bibnamefont {Samad}},\ }\href {\doibase
  http://dx.doi.org/10.1002/jps.22052} {\bibfield  {journal} {\bibinfo
  {journal} {Journal of Pharmaceutical Sciences}\ }\textbf {\bibinfo {volume}
  {99}},\ \bibinfo {pages} {2576 } (\bibinfo {year} {2010})}\BibitemShut
  {NoStop}%
\bibitem [{\citenamefont {Muro}\ \emph {et~al.}(2008)\citenamefont {Muro},
  \citenamefont {Garnacho}, \citenamefont {Champion}, \citenamefont
  {Leferovich}, \citenamefont {Gajewski}, \citenamefont {Schuchman},
  \citenamefont {Mitragotri},\ and\ \citenamefont {Muzykantov}}]{Muro08}%
  \BibitemOpen
  \bibfield  {author} {\bibinfo {author} {\bibfnamefont {S.}~\bibnamefont
  {Muro}}, \bibinfo {author} {\bibfnamefont {C.}~\bibnamefont {Garnacho}},
  \bibinfo {author} {\bibfnamefont {J.~A.}\ \bibnamefont {Champion}}, \bibinfo
  {author} {\bibfnamefont {J.}~\bibnamefont {Leferovich}}, \bibinfo {author}
  {\bibfnamefont {C.}~\bibnamefont {Gajewski}}, \bibinfo {author}
  {\bibfnamefont {E.~H.}\ \bibnamefont {Schuchman}}, \bibinfo {author}
  {\bibfnamefont {S.}~\bibnamefont {Mitragotri}}, \ and\ \bibinfo {author}
  {\bibfnamefont {V.~R.}\ \bibnamefont {Muzykantov}},\ }\href {\doibase
  http://dx.doi.org/10.1038/mt.2008.127} {\bibfield  {journal} {\bibinfo
  {journal} {Molecular Therapy}\ }\textbf {\bibinfo {volume} {16}},\ \bibinfo
  {pages} {1450 } (\bibinfo {year} {2008})}\BibitemShut {NoStop}%
\bibitem [{\citenamefont {Patil}\ \emph {et~al.}(2008)\citenamefont {Patil},
  \citenamefont {Gaikwad}, \citenamefont {Samad},\ and\ \citenamefont
  {Devarajan}}]{Patil08}%
  \BibitemOpen
  \bibfield  {author} {\bibinfo {author} {\bibfnamefont {R.~R.}\ \bibnamefont
  {Patil}}, \bibinfo {author} {\bibfnamefont {R.~V.}\ \bibnamefont {Gaikwad}},
  \bibinfo {author} {\bibfnamefont {A.}~\bibnamefont {Samad}}, \ and\ \bibinfo
  {author} {\bibfnamefont {P.~V.}\ \bibnamefont {Devarajan}},\ }\href {\doibase
  doi:10.1166/jbn.2008.320} {\bibfield  {journal} {\bibinfo  {journal} {Journal
  of Biomedical Nanotechnology}\ }\textbf {\bibinfo {volume} {4}},\ \bibinfo
  {pages} {359} (\bibinfo {year} {2008})}\BibitemShut {NoStop}%
\bibitem [{\citenamefont {Gillies}\ \emph {et~al.}(2005)\citenamefont
  {Gillies}, \citenamefont {Dy}, \citenamefont {Fr\'echet},\ and\ \citenamefont
  {Szoka}}]{Gillies04}%
  \BibitemOpen
  \bibfield  {author} {\bibinfo {author} {\bibfnamefont {E.~R.}\ \bibnamefont
  {Gillies}}, \bibinfo {author} {\bibfnamefont {E.}~\bibnamefont {Dy}},
  \bibinfo {author} {\bibfnamefont {J.~M.~J.}\ \bibnamefont {Fr\'echet}}, \
  and\ \bibinfo {author} {\bibfnamefont {F.~C.}\ \bibnamefont {Szoka}},\ }\href
  {\doibase 10.1021/mp049886u} {\bibfield  {journal} {\bibinfo  {journal}
  {Molecular Pharmaceutics}\ }\textbf {\bibinfo {volume} {2}},\ \bibinfo
  {pages} {129} (\bibinfo {year} {2005})}\BibitemShut {NoStop}%
\bibitem [{\citenamefont {Bose}\ \emph {et~al.}(2017)\citenamefont {Bose},
  \citenamefont {Jana}, \citenamefont {Saha},\ and\ \citenamefont
  {Mandal}}]{Bose2017}%
  \BibitemOpen
  \bibfield  {author} {\bibinfo {author} {\bibfnamefont {A.}~\bibnamefont
  {Bose}}, \bibinfo {author} {\bibfnamefont {S.}~\bibnamefont {Jana}}, \bibinfo
  {author} {\bibfnamefont {A.}~\bibnamefont {Saha}}, \ and\ \bibinfo {author}
  {\bibfnamefont {T.~K.}\ \bibnamefont {Mandal}},\ }\href {\doibase
  10.1016/j.polymer.2016.12.068} {\bibfield  {journal} {\bibinfo  {journal}
  {Polymer}\ }\textbf {\bibinfo {volume} {110}},\ \bibinfo {pages} {12}
  (\bibinfo {year} {2017})}\BibitemShut {NoStop}%
\bibitem [{\citenamefont {Dule}\ \emph {et~al.}(2015)\citenamefont {Dule},
  \citenamefont {Biswas}, \citenamefont {Paira},\ and\ \citenamefont
  {Mandal}}]{dule2015}%
  \BibitemOpen
  \bibfield  {author} {\bibinfo {author} {\bibfnamefont {M.}~\bibnamefont
  {Dule}}, \bibinfo {author} {\bibfnamefont {M.}~\bibnamefont {Biswas}},
  \bibinfo {author} {\bibfnamefont {T.~K.}\ \bibnamefont {Paira}}, \ and\
  \bibinfo {author} {\bibfnamefont {T.~K.}\ \bibnamefont {Mandal}},\ }\href
  {\doibase https://doi.org/10.1016/j.polymer.2015.09.020} {\bibfield
  {journal} {\bibinfo  {journal} {Polymer}\ }\textbf {\bibinfo {volume} {77}},\
  \bibinfo {pages} {32 } (\bibinfo {year} {2015})}\BibitemShut {NoStop}%
\bibitem [{\citenamefont {Li}\ \emph {et~al.}(2015)\citenamefont {Li},
  \citenamefont {Gao}, \citenamefont {Boott}, \citenamefont {Winnik},\ and\
  \citenamefont {Manners}}]{Li2015}%
  \BibitemOpen
  \bibfield  {author} {\bibinfo {author} {\bibfnamefont {X.}~\bibnamefont
  {Li}}, \bibinfo {author} {\bibfnamefont {Y.}~\bibnamefont {Gao}}, \bibinfo
  {author} {\bibfnamefont {C.~E.}\ \bibnamefont {Boott}}, \bibinfo {author}
  {\bibfnamefont {M.~A.}\ \bibnamefont {Winnik}}, \ and\ \bibinfo {author}
  {\bibfnamefont {I.}~\bibnamefont {Manners}},\ }\href
  {http://dx.doi.org/10.1038/ncomms9127} {\bibfield  {journal} {\bibinfo
  {journal} {Nature Communications}\ }\textbf {\bibinfo {volume} {6}},\
  \bibinfo {pages} {8127} (\bibinfo {year} {2015})}\BibitemShut {NoStop}%
\bibitem [{\citenamefont {Qiu}\ \emph {et~al.}(2015)\citenamefont {Qiu},
  \citenamefont {Hudson}, \citenamefont {Winnik},\ and\ \citenamefont
  {Manners}}]{Qiu2015}%
  \BibitemOpen
  \bibfield  {author} {\bibinfo {author} {\bibfnamefont {H.}~\bibnamefont
  {Qiu}}, \bibinfo {author} {\bibfnamefont {Z.~M.}\ \bibnamefont {Hudson}},
  \bibinfo {author} {\bibfnamefont {M.~A.}\ \bibnamefont {Winnik}}, \ and\
  \bibinfo {author} {\bibfnamefont {I.}~\bibnamefont {Manners}},\ }\href
  {\doibase 10.1126/science.1261816} {\bibfield  {journal} {\bibinfo  {journal}
  {Science}\ }\textbf {\bibinfo {volume} {347}},\ \bibinfo {pages} {1329}
  (\bibinfo {year} {2015})}\BibitemShut {NoStop}%
\bibitem [{\citenamefont {Moths}\ and\ \citenamefont
  {Witten}(2018)}]{firstPaper}%
  \BibitemOpen
  \bibfield  {author} {\bibinfo {author} {\bibfnamefont {B.}~\bibnamefont
  {Moths}}\ and\ \bibinfo {author} {\bibfnamefont {T.~A.}\ \bibnamefont
  {Witten}},\ }\href {\doibase 10.1103/PhysRevE.97.032505} {\bibfield
  {journal} {\bibinfo  {journal} {Phys. Rev. E}\ }\textbf {\bibinfo {volume}
  {97}},\ \bibinfo {pages} {032505} (\bibinfo {year} {2018})}\BibitemShut
  {NoStop}%
\bibitem [{\citenamefont {Wang}\ and\ \citenamefont {Safran}(1991)}]{Wang91}%
  \BibitemOpen
  \bibfield  {author} {\bibinfo {author} {\bibfnamefont {Z.}~\bibnamefont
  {Wang}}\ and\ \bibinfo {author} {\bibfnamefont {S.~A.}\ \bibnamefont
  {Safran}},\ }\href {\doibase 10.1063/1.460334} {\bibfield  {journal}
  {\bibinfo  {journal} {The Journal of Chemical Physics}\ }\textbf {\bibinfo
  {volume} {94}},\ \bibinfo {pages} {679} (\bibinfo {year} {1991})}\BibitemShut
  {NoStop}%
\bibitem [{\citenamefont {Detcheverry}\ \emph {et~al.}(2009)\citenamefont
  {Detcheverry}, \citenamefont {Pike}, \citenamefont {Nealey}, \citenamefont
  {M\"uller},\ and\ \citenamefont {de~Pablo}}]{detcheverry2009}%
  \BibitemOpen
  \bibfield  {author} {\bibinfo {author} {\bibfnamefont {F.~A.}\ \bibnamefont
  {Detcheverry}}, \bibinfo {author} {\bibfnamefont {D.~Q.}\ \bibnamefont
  {Pike}}, \bibinfo {author} {\bibfnamefont {P.~F.}\ \bibnamefont {Nealey}},
  \bibinfo {author} {\bibfnamefont {M.}~\bibnamefont {M\"uller}}, \ and\
  \bibinfo {author} {\bibfnamefont {J.~J.}\ \bibnamefont {de~Pablo}},\ }\href
  {\doibase 10.1103/PhysRevLett.102.197801} {\bibfield  {journal} {\bibinfo
  {journal} {Phys. Rev. Lett.}\ }\textbf {\bibinfo {volume} {102}},\ \bibinfo
  {pages} {197801} (\bibinfo {year} {2009})}\BibitemShut {NoStop}%
\bibitem [{\citenamefont {Binder}\ \emph {et~al.}(2011)\citenamefont {Binder},
  \citenamefont {Mognetti}, \citenamefont {Paul}, \citenamefont {Virnau},\ and\
  \citenamefont {Yelash}}]{Binder11}%
  \BibitemOpen
  \bibfield  {author} {\bibinfo {author} {\bibfnamefont {K.}~\bibnamefont
  {Binder}}, \bibinfo {author} {\bibfnamefont {B.}~\bibnamefont {Mognetti}},
  \bibinfo {author} {\bibfnamefont {W.}~\bibnamefont {Paul}}, \bibinfo {author}
  {\bibfnamefont {P.}~\bibnamefont {Virnau}}, \ and\ \bibinfo {author}
  {\bibfnamefont {L.}~\bibnamefont {Yelash}},\ }in\ \href {\doibase
  10.1007/12_2010_82} {\emph {\bibinfo {booktitle} {Polymer Thermodynamics:
  Liquid Polymer-Containing Mixtures}}},\ \bibinfo {editor} {edited by\
  \bibinfo {editor} {\bibfnamefont {B.~A.}\ \bibnamefont {Wolf}}\ and\ \bibinfo
  {editor} {\bibfnamefont {S.}~\bibnamefont {Enders}}}\ (\bibinfo  {publisher}
  {Springer},\ \bibinfo {address} {Berlin, Heidelberg},\ \bibinfo {year}
  {2011})\ pp.\ \bibinfo {pages} {329--387}\BibitemShut {NoStop}%
\bibitem [{\citenamefont {Dimitrakopoulos}(2004)}]{Dimitrakopoulos04}%
  \BibitemOpen
  \bibfield  {author} {\bibinfo {author} {\bibfnamefont {P.}~\bibnamefont
  {Dimitrakopoulos}},\ }\href {\doibase 10.1103/PhysRevLett.93.217801}
  {\bibfield  {journal} {\bibinfo  {journal} {Phys. Rev. Lett.}\ }\textbf
  {\bibinfo {volume} {93}},\ \bibinfo {pages} {217801} (\bibinfo {year}
  {2004})}\BibitemShut {NoStop}%
\bibitem [{\citenamefont {Hsieh}\ \emph {et~al.}(2006)\citenamefont {Hsieh},
  \citenamefont {Jain},\ and\ \citenamefont {Larson}}]{Hsieh06}%
  \BibitemOpen
  \bibfield  {author} {\bibinfo {author} {\bibfnamefont {C.-C.}\ \bibnamefont
  {Hsieh}}, \bibinfo {author} {\bibfnamefont {S.}~\bibnamefont {Jain}}, \ and\
  \bibinfo {author} {\bibfnamefont {R.~G.}\ \bibnamefont {Larson}},\ }\href
  {\doibase 10.1063/1.2161210} {\bibfield  {journal} {\bibinfo  {journal} {The
  Journal of Chemical Physics}\ }\textbf {\bibinfo {volume} {124}},\ \bibinfo
  {pages} {044911} (\bibinfo {year} {2006})}\BibitemShut {NoStop}%
\bibitem [{\citenamefont {Plimpton}(1995)}]{lammps95}%
  \BibitemOpen
  \bibfield  {author} {\bibinfo {author} {\bibfnamefont {S.}~\bibnamefont
  {Plimpton}},\ }\href {\doibase http://dx.doi.org/10.1006/jcph.1995.1039}
  {\bibfield  {journal} {\bibinfo  {journal} {Journal of Computational
  Physics}\ }\textbf {\bibinfo {volume} {117}},\ \bibinfo {pages} {1 }
  (\bibinfo {year} {1995})}\BibitemShut {NoStop}%
\bibitem [{\citenamefont {Artin}(2011)}]{artin}%
  \BibitemOpen
  \bibfield  {author} {\bibinfo {author} {\bibfnamefont {M.}~\bibnamefont
  {Artin}},\ }\href {https://books.google.com/books?id=S6GSAgAAQBAJ} {\emph
  {\bibinfo {title} {Algebra}}}\ (\bibinfo  {publisher} {Boston, Prentice
  Hall},\ \bibinfo {year} {2011})\ pp.\ \bibinfo {pages} {102--103}\BibitemShut
  {NoStop}%
\bibitem [{\citenamefont {Le}\ and\ \citenamefont {Winkler}(2016)}]{le2015}%
  \BibitemOpen
  \bibfield  {author} {\bibinfo {author} {\bibfnamefont {T.~C.}\ \bibnamefont
  {Le}}\ and\ \bibinfo {author} {\bibfnamefont {D.~A.}\ \bibnamefont
  {Winkler}},\ }\href {\doibase 10.1021/acs.chemrev.5b00691} {\bibfield
  {journal} {\bibinfo  {journal} {Chemical Reviews}\ }\textbf {\bibinfo
  {volume} {116}},\ \bibinfo {pages} {6107} (\bibinfo {year}
  {2016})}\BibitemShut {NoStop}%
\bibitem [{\citenamefont {Liu}\ \emph {et~al.}(2017)\citenamefont {Liu},
  \citenamefont {Zhao}, \citenamefont {Ju},\ and\ \citenamefont
  {Shi}}]{liu2017}%
  \BibitemOpen
  \bibfield  {author} {\bibinfo {author} {\bibfnamefont {Y.}~\bibnamefont
  {Liu}}, \bibinfo {author} {\bibfnamefont {T.}~\bibnamefont {Zhao}}, \bibinfo
  {author} {\bibfnamefont {W.}~\bibnamefont {Ju}}, \ and\ \bibinfo {author}
  {\bibfnamefont {S.}~\bibnamefont {Shi}},\ }\href {\doibase
  https://doi.org/10.1016/j.jmat.2017.08.002} {\bibfield  {journal} {\bibinfo
  {journal} {Journal of Materiomics}\ }\textbf {\bibinfo {volume} {3}},\
  \bibinfo {pages} {159 } (\bibinfo {year} {2017})}\BibitemShut {NoStop}%
\bibitem [{\citenamefont {Patra}\ \emph {et~al.}(2017)\citenamefont {Patra},
  \citenamefont {Meenakshisundaram}, \citenamefont {Hung},\ and\ \citenamefont
  {Simmons}}]{patra2016}%
  \BibitemOpen
  \bibfield  {author} {\bibinfo {author} {\bibfnamefont {T.~K.}\ \bibnamefont
  {Patra}}, \bibinfo {author} {\bibfnamefont {V.}~\bibnamefont
  {Meenakshisundaram}}, \bibinfo {author} {\bibfnamefont {J.-H.}\ \bibnamefont
  {Hung}}, \ and\ \bibinfo {author} {\bibfnamefont {D.~S.}\ \bibnamefont
  {Simmons}},\ }\href {\doibase 10.1021/acscombsci.6b00136} {\bibfield
  {journal} {\bibinfo  {journal} {ACS Combinatorial Science}\ }\textbf
  {\bibinfo {volume} {19}},\ \bibinfo {pages} {96} (\bibinfo {year}
  {2017})}\BibitemShut {NoStop}%
\bibitem [{\citenamefont {Miskin}\ and\ \citenamefont
  {Jaeger}(2013)}]{miskin2013}%
  \BibitemOpen
  \bibfield  {author} {\bibinfo {author} {\bibfnamefont {M.~Z.}\ \bibnamefont
  {Miskin}}\ and\ \bibinfo {author} {\bibfnamefont {H.~M.}\ \bibnamefont
  {Jaeger}},\ }\href {http://dx.doi.org/10.1038/nmat3543} {\bibfield  {journal}
  {\bibinfo  {journal} {Nature Materials}\ }\textbf {\bibinfo {volume} {12}},\
  \bibinfo {pages} {326 EP } (\bibinfo {year} {2013})}\BibitemShut {NoStop}%
\bibitem [{\citenamefont {Larson}(1988)}]{larson88}%
  \BibitemOpen
  \bibfield  {author} {\bibinfo {author} {\bibfnamefont {R.~G.}\ \bibnamefont
  {Larson}},\ }\href {\doibase 10.1063/1.455110} {\bibfield  {journal}
  {\bibinfo  {journal} {The Journal of Chemical Physics}\ }\textbf {\bibinfo
  {volume} {89}},\ \bibinfo {pages} {1642} (\bibinfo {year}
  {1988})}\BibitemShut {NoStop}%
\bibitem [{\citenamefont {Miller}\ \emph {et~al.}(1992)\citenamefont {Miller},
  \citenamefont {Danko}, \citenamefont {Fasolka}, \citenamefont {Balazs},
  \citenamefont {Chan},\ and\ \citenamefont {Dill}}]{miller92}%
  \BibitemOpen
  \bibfield  {author} {\bibinfo {author} {\bibfnamefont {R.}~\bibnamefont
  {Miller}}, \bibinfo {author} {\bibfnamefont {C.~A.}\ \bibnamefont {Danko}},
  \bibinfo {author} {\bibfnamefont {M.~J.}\ \bibnamefont {Fasolka}}, \bibinfo
  {author} {\bibfnamefont {A.~C.}\ \bibnamefont {Balazs}}, \bibinfo {author}
  {\bibfnamefont {H.~S.}\ \bibnamefont {Chan}}, \ and\ \bibinfo {author}
  {\bibfnamefont {K.~A.}\ \bibnamefont {Dill}},\ }\href {\doibase
  10.1063/1.462462} {\bibfield  {journal} {\bibinfo  {journal} {The Journal of
  Chemical Physics}\ }\textbf {\bibinfo {volume} {96}},\ \bibinfo {pages} {768}
  (\bibinfo {year} {1992})}\BibitemShut {NoStop}%
\bibitem [{\citenamefont {Chan}\ and\ \citenamefont {Dill}(1993)}]{chan93}%
  \BibitemOpen
  \bibfield  {author} {\bibinfo {author} {\bibfnamefont {H.~S.}\ \bibnamefont
  {Chan}}\ and\ \bibinfo {author} {\bibfnamefont {K.~A.}\ \bibnamefont
  {Dill}},\ }\href {\doibase 10.1063/1.465277} {\bibfield  {journal} {\bibinfo
  {journal} {The Journal of Chemical Physics}\ }\textbf {\bibinfo {volume}
  {99}},\ \bibinfo {pages} {2116} (\bibinfo {year} {1993})}\BibitemShut
  {NoStop}%
\bibitem [{\citenamefont {Huopaniemi}\ \emph {et~al.}(2006)\citenamefont
  {Huopaniemi}, \citenamefont {Luo}, \citenamefont {Ala-Nissila},\ and\
  \citenamefont {Ying}}]{huopaniemi06}%
  \BibitemOpen
  \bibfield  {author} {\bibinfo {author} {\bibfnamefont {I.}~\bibnamefont
  {Huopaniemi}}, \bibinfo {author} {\bibfnamefont {K.}~\bibnamefont {Luo}},
  \bibinfo {author} {\bibfnamefont {T.}~\bibnamefont {Ala-Nissila}}, \ and\
  \bibinfo {author} {\bibfnamefont {S.-C.}\ \bibnamefont {Ying}},\ }\href
  {\doibase 10.1063/1.2357118} {\bibfield  {journal} {\bibinfo  {journal} {The
  Journal of Chemical Physics}\ }\textbf {\bibinfo {volume} {125}},\ \bibinfo
  {pages} {124901} (\bibinfo {year} {2006})}\BibitemShut {NoStop}%
\bibitem [{\citenamefont {Siepmann}\ and\ \citenamefont
  {Frenkel}(1992)}]{siepmann06}%
  \BibitemOpen
  \bibfield  {author} {\bibinfo {author} {\bibfnamefont {J.~I.}\ \bibnamefont
  {Siepmann}}\ and\ \bibinfo {author} {\bibfnamefont {D.}~\bibnamefont
  {Frenkel}},\ }\href {\doibase 10.1080/00268979200100061} {\bibfield
  {journal} {\bibinfo  {journal} {Molecular Physics}\ }\textbf {\bibinfo
  {volume} {75}},\ \bibinfo {pages} {59} (\bibinfo {year} {1992})}\BibitemShut
  {NoStop}%
\bibitem [{\citenamefont {Helfrich}(1973)}]{helfrich73}%
  \BibitemOpen
  \bibfield  {author} {\bibinfo {author} {\bibfnamefont {W.}~\bibnamefont
  {Helfrich}},\ }\href
  {http://ludfc39.u-strasbg.fr/pdflib/membranes/Elasticity/1973\_Helfrich\_b.pdf}
  {\bibfield  {journal} {\bibinfo  {journal} {Z. Naturforsch. C}\ }\textbf
  {\bibinfo {volume} {28}},\ \bibinfo {pages} {693} (\bibinfo {year}
  {1973})}\BibitemShut {NoStop}%
\end{thebibliography}%
\end{document}